\documentclass[times,twocolumn]{aastex631}

\usepackage{CJK}
\usepackage{amsmath, amssymb, bm}
\usepackage{mathrsfs}
\usepackage{color}
\usepackage[nodayofweek]{datetime}
\newdateformat{monthyearday}{%
  \THEYEAR\ \monthname[\THEMONTH] \twodigit{\THEDAY}}
\usepackage{multirow}

\makeatletter
\patchcmd{\NAT@citex}
  {\@citea\NAT@hyper@{%
     \NAT@nmfmt{\NAT@nm}%
     \hyper@natlinkbreak{\NAT@aysep\NAT@spacechar}{\@citeb\@extra@b@citeb}%
     \NAT@date}}
  {\@citea\NAT@nmfmt{\NAT@nm}%
   \NAT@aysep\NAT@spacechar\NAT@hyper@{\NAT@date}}{}{}

\patchcmd{\NAT@citex}
  {\@citea\NAT@hyper@{%
     \NAT@nmfmt{\NAT@nm}%
     \hyper@natlinkbreak{\NAT@spacechar\NAT@@open\if*#1*\else#1\NAT@spacechar\fi}%
       {\@citeb\@extra@b@citeb}%
     \NAT@date}}
  {\@citea\NAT@nmfmt{\NAT@nm}%
   \NAT@spacechar\NAT@@open\if*#1*\else#1\NAT@spacechar\fi\NAT@hyper@{\NAT@date}}
  {}{}
\makeatother
\newcommand{\uat}[2]{\href{http://astrothesaurus.org/uat/#1}{#2  (#1)}}
\newcommand{\hi}{H\,{\textsc{\romannumeral 1}}}

\newcommand{\mhi}{{M_{\rm H\,{\textsc{\romannumeral 1}}}}}

\newcommand{\hh}{$\rm H_2$} 
\newcommand\fhh{$f_{\rm H_2}$}
\newcommand\mhh{$M_{\rm H_2}$} 
\newcommand{\ha}{H{\sc $\alpha$}}
\newcommand{\sfe}{$\rm SFE_{\text{\hi}}$}
\received{March 5, 2024}
\revised{April 19, 2024}
\accepted{April 28, 2024}
\published{June 20, 2024}
\shorttitle{The Astrophysical Journal Supplement Series, 273:2 (22pp), 2024 July}
\shortauthors{Yu et al.}
\graphicspath{{./}{figures/}}
\begin{document}
\begin{CJK*}{UTF8}{gbsn}
\title{CO Observations of Early-mid Stage Major Mergers in the MaNGA Survey}

% \correspondingauthor{Qingzheng Yu, Taotao Fang}
% \email{yuqingzheng@stu.xmu.edu.cn, fangt@xmu.edu.cn}

\author[0000-0003-3230-3981]{Qingzheng Yu (余清正)}
\affiliation{Department of Astronomy, Xiamen University, Xiamen 361005, People's Republic of China; \url{yuqingzheng@stu.xmu.edu.cn, fangt@xmu.edu.cn}}

\author[0000-0002-2853-3808]{Taotao Fang (方陶陶)}
\affiliation{Department of Astronomy, Xiamen University, Xiamen 361005, People's Republic of China; \url{yuqingzheng@stu.xmu.edu.cn, fangt@xmu.edu.cn}}

\author[0000-0002-1588-6700]{Cong Kevin Xu (徐聪)}
\affiliation{Chinese Academy of Sciences South America Center for Astronomy, National Astronomical Observatories, CAS, Beijing 100101, People's Republic of China}
\affiliation{National Astronomical Observatories, Chinese Academy of Sciences, 20A Datun Road, Chaoyang District, Beijing 100101, People's Republic of China.}

\author[0000-0002-9767-9237]{Shuai Feng (冯帅)}
\affiliation{College of Physics, Hebei Normal University, 20 South Erhuan Road, Shijiazhuang, 050024, People's Republic of China}
\affiliation{Hebei Key Laboratory of Photophysics Research and Application, Shijiazhuang 050024, People's Republic of China}

\author[0000-0002-4707-8409]{Siyi Feng (冯思轶)}
\affiliation{Department of Astronomy, Xiamen University, Xiamen 361005, People's Republic of China; \url{yuqingzheng@stu.xmu.edu.cn, fangt@xmu.edu.cn}}

\author[0000-0002-4707-8409]{Yu Gao (高煜)}
\affiliation{Department of Astronomy, Xiamen University, Xiamen 361005, People's Republic of China; \url{yuqingzheng@stu.xmu.edu.cn, fangt@xmu.edu.cn}}

\author[0000-0002-8899-4673]{Xue-Jian Jiang (蒋雪健)}
\affiliation{Research Center for Astronomical Computing, Zhejiang Laboratory, Hangzhou 311100, People's Republic of China}

\author[0000-0002-9471-5423]{Ute Lisenfeld}
\affiliation{Dept. F\'isica Te\'orica y del Cosmos, Campus de Fuentenueva, Edificio Mecenas, Universidad de Granada, E-18071 Granada, Spain}
\affiliation{Instituto Carlos I de F\'sica T\'orica y Computacional, Facultad de Ciencias, Granada E-18071, Spain}



\begin{abstract}
We present a study of the molecular gas in early-mid stage major mergers, with a sample of 43 major-merger galaxy pairs selected from the Mapping Nearby Galaxies at Apache Point Observatory survey and a control sample of 195 isolated galaxies selected from the xCOLD GASS survey. Adopting kinematic asymmetry as a new effective indicator to describe the merger stage, we aim to study the role of molecular gas in the merger-induced star formation enhancement along the merger sequence of galaxy pairs. We obtain the molecular gas properties from CO observations with the James Clerk Maxwell Telescope, Institut de Radioastronomie Milimetrique 30 m telescope, and the MaNGA-ARO Survey of CO Targets survey. Using these data, we investigate the differences in molecular gas fraction (\fhh), star formation rate (SFR), star formation efficiency (SFE), molecular-to-atomic gas ratio (\mhh$/\mhi$), total gas fraction ($f_{\rm gas}$), and the SFE of total gas (${\rm SFE}_{\rm gas}$) between the pair and control samples. In the full pair sample, our results suggest the \fhh\ of paired galaxies is significantly enhanced, while the SFE is comparable to that of isolated galaxies. We detect significantly increased \fhh\ and \mhh$/\mhi$ in paired galaxies at the pericenter stage, indicating an accelerated transition from atomic gas to molecular gas due to interactions. Our results indicate that the elevation of \fhh\ plays a major role in the enhancement of global SFR in paired galaxies at the pericenter stage, while the contribution of enhanced SFE in specific regions requires further explorations through spatially resolved observations of a larger sample spanning a wide range of merger stages.

\end{abstract}


\section{Introduction} \label{sec:intro}

Merging galaxies are believed to play a fundamental role in regulating hierarchical galaxy formation and evolution \citep[e.g.,][]{Toomre1972,Toomre1977,Barnes1992}. Numerous observations and numerical simulations have revealed the drastic influences of galaxy interactions and mergers, such as the enhancement of star formation \citep[e.g.,][]{Xu1991,Mihos1996,Ellison2008,Li2008,Xu2010}, regulation of gas \citep[e.g.,][]{Hibbard1996,Ellison2018,Hani2018,Lisenfeld2019,Moreno2019}, triggering of active galactic nucleus \citep[e.g.,][]{Kennicutt1984,Ellison2011,Satyapal2014,Goulding2018}, and redistribution of metallicities \citep[e.g.,][]{Kewley2006,Rupke2010,Scudder2012,Torrey2012}. As a crucial and evident effect, the enhancement of star formation rate (SFR) depends on multiple factors, such as the properties of the progenitor galaxies \citep[e.g.,][]{Mihos1996,Cox2008,Xu2010}, orbital parameters of the interacting
galaxies \citep[e.g.,][]{Kennicutt1987,Keel1993,Sparre2016,Xu2021}, and the interacting phase of galaxies \citep{DiMatteo2007,Scudder2012,Patton2013}. The strongest SFR enhancement or starbursts are believed to occur after first pericenter and during coalescence, supported by studies of major-merger galaxy pairs \citep{Nikolic2004,Scudder2012,Moreno2015,Pan2019} and ultraluminous infrared galaxies (ULIRGs) in interacting systems \citep[e.g.,][]{Sanders1996,Veilleux2002,Haan2011,Ellison2013,Thorp2019}.

Theoretical models and numerical simulations have predicted that the gravitational tidal torques induced by galaxy interactions can trigger strong gas inflows in the interstellar medium (ISM), resulting in the enhancement of SFR and nuclear starbursts \citep{Barnes1991,Barnes1996,Mihos1996,Iono2004,Hopkins2009,Hopkins2013}. Hence, studies of the cold gas in the ISM provide a crucial approach for investigating how star formation is enhanced during galaxy interactions, and two major mechanisms have been commonly proposed by observational studies. The first scenario suggests that the merger-induced external pressure can accelerate the transition from atomic (\hi) to molecular (\hh) gas \citep[e.g.,][]{Braine1993,Elmegreen1993,Kaneko2017}, leading to the observed enrichment of the molecular gas reservoir (traced by CO) and enhancement of SFR \citep{Braine1993,Combes1994,Casasola2004,Wilson2008,Lisenfeld2019,Shangguan2019}.
The second perspective attributes enhanced star formation to the increased density of molecular gas in the galactic center, thereby boosting the efficiency of star formation \citep{Solomon1988,Sofue1993,Gao1999,Michiyama2016,Yamashita2017}. Recent observations of galaxy pairs have found enhancement of both \hh\ gas fraction (\fhh) and star formation efficiency \citep[SFE;][]{Pan2018,Violino2018}. Thus, it remains unclear whether the enhancement of SFR is driven by the total amount of molecular gas, SFE, or both.  

To test the theoretical predictions and investigate the mutual impact of molecular gas and SFR along the evolutionary merger sequence, several factors that may affect the results should be addressed for observational studies. First, the selection and comparison of galaxy pairs and control galaxies could introduce biased results if the global properties are not carefully controlled between samples \citep{Braine1993,Michiyama2016}. Furthermore, single-dish CO observations provide a crucial and efficient approach to the evaluation and analysis of global galactic quantities. However, results from single-dish observations should be taken with caution, considering the limitation of spatial resolution. Previous CO observations performed with interferometers have found consistent results relative to the single-dish observations but often suffered from limited numbers and types of targets \citep[e.g.,][]{Iono2009,Ueda2014,Xu2021,Hou2023}. A recent study used the Atacama Large Millimeter Array to observe 31 mergers selected from the  Mapping Nearby Galaxies at Apache Point Observatory (MaNGA) survey, and they have found that merger-induced star formation can be driven by a variety of mechanisms, with comparable contributions from both \fhh\ and SFE \citep{Thorp2022}. Another less explored but crucial factor pertains to the lack of rigorously defined merger stages, particularly in studies involving galaxy pairs \citep{Pan2019}.

The commonly adopted definition of the merger stage in galaxy pairs highly relies on the projected separation ($d_{\rm p}$) and the difference between line-of-sight velocities  ($\Delta v$) \citep{Ellison2008,Zuo2018,Feng2019,Zhang2020}, which has been hampered by several shortcomings. Considering the projection effect, relying solely on the projected separation ($d_{\rm P}$) may introduce large uncertainty in deriving the physical separation of member galaxies  \citep{Soares2007}. Furthermore, galaxy pairs with the same separation can be in different merger stages. Although the degree of interactions has been commonly characterized by galaxy morphology \citep[e.g.,][]{Toomre1977,Barrera2015,Smith2018,Pan2019}, a more sensitive and quantitative indicator is required to study the early-stage galaxy pairs \citep{Feng2020}. With the emergence of integral field unit (IFU) surveys on nearby galaxies, such as the Calar Alto Legacy Integral Field Area survey \citep[][]{Sanchez2012} and the Sydney-AAO Multi-object Integral field spectrograph galaxy survey \citep[][]{Croom2012}, recent studies have exploited unique data of gas kinematics and revealed connections between the asymmetry of gas kinematics and galaxy interactions \citep{Barrera2015,Bloom2018}. Using the IFU data from Mapping Nearby Galaxies at Apache Point Observatory (MaNGA) survey \citep{Bundy2015}, \cite{Feng2020} investigated the kinematic asymmetry of the ionized gas in a large sample of paired galaxies. By fitting the velocity maps of \ha\ gas, \cite{Feng2020} quantified the degree of kinematic asymmetry ($\overline{v}_{\rm asym}$), which depicts the asymmetry of velocity field contributed by the interaction-induced nonrotating motion. The value of $\overline{v}_{\rm asym}$ serves as an indicator of galaxy interaction strength in statistical analyses \citep[see details in][]{Feng2020}. They find a significant enhancement in the SFR of paired galaxies with high kinematic asymmetries, while those with low kinematic asymmetries show no enhancement of SFR whether at small or large projected separations. For paired galaxies with high kinematic asymmetries, the enhancement of SFR presents a tight anticorrelation with projected separation. These findings suggest that the combination of kinematic asymmetry and projected separation is effective in quantifying the stage of galaxy pairs, consistent with previous findings that used projected separation and morphology to determine merger stages \citep[e.g.,][]{Pan2019}. By employing 24 high-resolution idealized hydrodynamical galaxy merger simulations based on the Feedback In Realistic Environment (FIRE-2) model, this new method has been confirmed to be highly effective in distinguishing mergers in different stages, e.g., mergers during pericentric passages and in/after the final coalescence \citep{McElroy2022}.

Given that the enhancement of star formation during galaxy--galaxy interactions requires sustaining gas supply, we conduct studies on the cold gas properties of galaxy pairs at different merger stages. \cite{Yu2022} compiled a major-merger galaxy pair sample selected from the MaNGA survey to study the \hi\ content of merging galaxies \citep{Yu2022}, adopting the kinematic asymmetry and projected separation as indicators of the merging stage \citep{Feng2020}. Their results suggest that the \hi\ gas fraction of major-merger galaxy pairs, on average, is marginally decreased by $\sim$15\% relative to isolated galaxies, implying mild \hi\ depletion during galaxy interactions \citep{Yu2022}. In particular, galaxy pairs during the pericenter passage present the largest \hi\ deficiency \citep[$\sim$26\%,][]{Yu2022}. Based on these previous findings, it is necessary to observe the molecular gas to test whether the detection of \hi\ deficiency is attributed to the accelerated conversion from atomic gas to molecular gas in galaxy pairs.

To study the interplay between galaxy--galaxy interactions and the molecular gas properties along the merger sequence, we conducted CO observations for a sample of 43 major-merger paired galaxies selected from the MaNGA survey \citep{Bundy2015}, with ancillary \hi\ data from the HI-MaNGA survey \citep{Stark2021} and \cite{Yu2022}. This paper is organized as follows. In Section \ref{sec:obs}, we introduce the sample selection of galaxy pairs, observation setup, and data reduction in this work. We then present the main results of the molecular gas properties and star formation in Section \ref{sec:results}, with a further discussion presented in Section \ref{sec:discussions}. Finally, we summarize the main results in Section \ref{sec:Summary}. Throughout the whole paper, we adopt the standard $\Lambda$CDM cosmology with $H_0 = 70$ km s$^{-1}$ Mpc$^{-1}$, $\Omega_M = 0.3$, and $\Omega_{\Lambda} = 0.7$.

\section{Samples and Data} \label{sec:obs}

\subsection{Pair Sample}\label{subsec:sample}
In our parent sample \citep{Feng2019,Feng2020}, we selected the isolated galaxy pairs based on the following criteria: (1) the projected separation for member galaxies: 5 $h^{-1}$ kpc $\leqslant d_{\rm p} \leqslant$  200 $h^{-1}$ kpc, (2) the line-of-sight velocity difference: $|\Delta v|\leqslant$ 500 km $\rm s^{-1}$, (3) each member of the pair has only one neighbor satisfying the above criteria, (4) at least one member galaxy of each pair has been observed in the MaNGA survey \citep{Bundy2015}, and the member galaxy has more than 70\% spaxels with \ha\ emission at S/N $>$ 5 within 1.5 effective radius ($R_e$), and (5) we only study star-forming galaxies (log$(\rm{sSFR/yr}^{-1})>-11$) in this work. To study major-merger pairs, we constrain the mass ratio as $M_1/M_2< 3$, where $M_1$ and $M_2$ represent the stellar masses of primary galaxies and companions, respectively. Among the selected major-merger galaxy pairs, we adopt a mass cut of $\text{log}(M_{\star}/M_{\odot}) \geqslant 9$ to match the stellar mass range of the control sample \citep[][]{Saintonge2017} used in this work (see section \ref{subsec:subsample}). 

The final pair sample consists of 43 major-merger galaxies in pairs, of which 36 paired galaxies are observed through our PI programs, and seven sources are adopted from the MaNGA-ARO Survey of CO Targets \citep[MASCOT;][]{Wylezalek2022}. Based on the previous morphological classification \citep{Yu2022}, this sample comprises 36 spiral+spiral (S+S) and seven spiral+elliptical (S+E) pairs, and all the observed sources in the sample are spiral galaxies. The redshift range of this sample spans from 0.017--0.062. We adopt the SFR and stellar mass ($M_{\star}$) for each pair from the public catalog of MPA-JHU DR7\footnote{\url{https://wwwmpa.mpa-garching.mpg.de/SDSS/DR7/\#derived}}. For all the galaxies in the pair sample, the following key parameters are presented in Table \ref{tab:major_pair}: 

\begin{enumerate}
\setlength{\itemsep}{0pt}
\setlength{\parsep}{0pt}
\setlength{\parskip}{0pt}
\item {\tt MaNGA ID}: Source name presented by MaNGA plate-ifu, with B representing the companion galaxy of the primary MaNGA galaxy.
\item {\tt R.A.} : right ascension in degrees. 
\item {\tt Decl.} : Declination in degrees.
\item {\tt $z$}: Sloan Digital Sky Survey (SDSS) spectroscopic redshift. 
\item {\tt log($ M_{\star}$)} :  stellar mass from the MPA-JHU Catalog.
\item {\tt log(SFR)} :  SFR from the MPA-JHU Catalog. 
\item {\tt $\overline{v}_{\rm asym}$}:  kinematic asymmetry measured from \ha\ velocity maps. The velocity map is divided into a sequence of concentric elliptical rings and fitted with the Fourier series, and the kinematic asymmetry is defined as the ratio between high-order (sum from 2--5 orders) and first-order coefficients: $\overline{v}_{\rm asym}=(k_{2}+k_{3}+k_{4}+k_{5})/4k_{1}$, where $k_n$ is the coefficient of the $n$th-order Fourier component \citep{Feng2020}.
\item {\tt $d_{\rm p}$}: projected separation. 
\item {\tt log($\mhi$)} : \hi\ gas mass from HI-MaNGA \citep{Stark2021} and \cite{Yu2022}.
\item {\tt $r_{25}$}:  isophotal optical major radius at the isophotal level 25 mag arcsec$^{-1}$ in the $r$ band from the NASA/IPAC Extragalactic Database (NED)\footnote{\url{https://ned.ipac.caltech.edu/}}.
\item {\tt $i$}:  inclination from NED.
\item {\tt $f_{\rm aper}$}:  aperture correction calculated with Equation (\ref{eq:f_aper}).
\item {\tt log(\mhh)}:  molecular gas mass calculated from the total CO line luminosity after aperture correction, details in Section \ref{subsec:aperture_correct}.
\item {\tt Stage}:  Merging stage based on $\overline{v}_{\rm asym}$ and $d_{\rm p}$, see Section \ref{subsec:subsample} for the definition; 1, 2, and 3 represent pre-passage, pericenter, and apocenter, respectively.
\item {\tt Type}:  Type of galaxy pair, with SS and SE representing spiral$+$spiral and spiral$+$elliptical pairs, respectively. 
\item {\tt Blending}:  Source blending in \hi\ observations, with Y and N representing yes and no for source blending problems, respectively.
\end{enumerate}

%\startlongtable
% \begin{rotatetable*}
% \centerwidetable
\begin{deluxetable*}{lrrcDDlDDrccDccc}
\setlength\tabcolsep{4pt}
\tabletypesize{\scriptsize}
\tablenum{1}
\tablecaption{Identifiers and Properties of the Major-merger Galaxies in Pairs\label{tab:major_pair}}
\centering
%\tablewidth{0pt}
\tablehead{
\colhead{MaNGA ID} & \colhead{R.A.} & \colhead{Decl.} & \colhead{$z$} & \twocolhead{log($ M_{\star}$)} & \twocolhead{log(SFR)} & \colhead{$\overline{v}_{\rm asym}$} & \twocolhead{$d_{\rm p}$} & \twocolhead{log($\mhi$)} & \colhead{$r_{25}$} & \colhead{$i$} & \colhead{$f_{\rm aper}$} & \twocolhead{log(\mhh)} & \colhead{Stage} & \colhead{Type} & \colhead{Blending}\\
\colhead{}  & \colhead{(deg)} & \colhead{(deg)} & \colhead{} & \twocolhead{($M_{\odot}$)} & \twocolhead{($M_{\odot}\ \rm{yr^{-1}}$)} & \colhead{} & \twocolhead{($h^{-1}$ kpc)} & \twocolhead{($M_{\odot}$)} & \colhead{(\arcsec)} & \colhead{(\arcdeg)} & \colhead{} & \twocolhead{($M_{\sun}$)} & \colhead{} & \colhead{} & \colhead{}
}
\decimalcolnumbers
\decimals
\startdata
8078-6104	&	42.73943	&	0.36941 	&	0.04421	&	10.03	&	0.66	&	0.179	&	18.35	&	9.88	&	2.9 	&	58	&	1.52	&	9.39	&	2	&	S+S	&	Y	\\
8078-6104B	&	42.73145	&	0.36689 	&	0.04391	&	10.21	&	0.49	&	0.179	&	18.35	&	9.88	&	2.8 	&	34	&	1.67	&	9.59	&	2	&	S+S	&	Y	\\
8082-12703	&	49.51165	&	-0.53896	&	0.02104	&	10.33	&	-0.11	&	0.0295	&	63.14	&	9.50	&	5.3 	&	69	&	1.73	&	9.65	&	3	&	S+S	&	Y	\\
8085-12704	&	52.61990	&	0.81150 	&	0.03080	&	10.55	&	-0.15	&	0.0221	&	98.10	&	10.07	&	4.6 	&	65	&	1.69	&	$<$9.20	&	1	&	S+S	&	Y	\\
8153-12701	&	39.63722	&	-0.86747	&	0.03923	&	9.72	&	-0.35	&	0.0422	&	158.83	&	10.00	&	3.2 	&	48	&	1.65	&	$<$8.79	&	3	&	S+S	&	Y	\\
8250-6101	&	138.75315	&	42.02439	&	0.02790	&	10.28	&	0.89	&	0.0396	&	44.97	&	9.88	&	4.1 	&	53	&	1.79	&	9.75	&	2	&	S+S	&	Y	\\
8260-6101	&	182.40876	&	42.00967	&	0.02288	&	9.89	&	0.15	&	0.0293	&	28.93	&	9.50	&	3.7 	&	57	&	1.69	&	8.86	&	2	&	S+S	&	Y	\\
8338-6102	&	172.68267	&	22.36354	&	0.02236	&	9.45	&	-0.09	&	0.0360	&	76.13	&	9.29	&	3.0 	&	41	&	1.68	&	8.34	&	3	&	S+S	&	N	\\
8450-6102	&	171.74883	&	21.14168	&	0.04177	&	10.17	&	0.67	&	0.0516	&	57.37	&	9.72	&	3.1 	&	34	&	1.77	&	9.37	&	3	&	S+E	&	N	\\
8456-12702	&	149.96826	&	45.28310	&	0.02349	&	9.39	&	-0.04	&	0.0813	&	13.35	&	9.98	&	6.5 	&	76	&	1.76	&	8.55	&	2	&	S+S	&	Y	\\
8547-12702	&	217.91070	&	52.74946	&	0.04566	&	10.62	&	0.02	&	0.0187	&	137.28	&	10.40	&	5.8 	&	81	&	1.64	&	9.31	&	1	&	S+E	&	N	\\
8552-12702	&	227.92840	&	43.97044	&	0.02758	&	9.25	&	0.05	&	0.0558	&	90.60	&	9.73	&	4.0 	&	73	&	1.55	&	8.92	&	3	&	S+S	&	Y	\\
8588-12702	&	250.31305	&	39.29009	&	0.03054	&	9.71	&	-0.08	&	0.0648	&	88.57	&	9.95	&	5.2 	&	54	&	1.91	&	8.93	&	3	&	S+S	&	N	\\
8656-1901	&	7.71740  	&	0.52876 	&	0.01914	&	9.08	&	-0.31	&	0.0293	&	101.03	&	9.22	&	3.1 	&	44	&	1.67	&	8.55	&	3	&	S+S	&	N	\\
8656-3703	&	7.75250 	&	0.43575 	&	0.01917	&	9.36	&	-0.55	&	0.0232	&	101.21	&	9.22	&	3.3 	&	54	&	1.63	&	8.11	&	1	&	S+S	&	N	\\
8657-12704	&	10.40833	&	0.25764 	&	0.01807	&	9.08	&	-0.50	&	0.0998	&	47.61	&	9.41	&	3.1 	&	35	&	1.74	&	8.10	&	2	&	S+S	&	N	\\
8728-12701	&	57.72958	&	-7.06065	&	0.02841	&	10.50	&	-0.03	&	0.0716	&	66.78	&	9.78	&	4.8 	&	20	&	2.19	&	9.36	&	3	&	S+E	&	N	\\
8981-6101	&	185.94406	&	36.15281	&	0.03332	&	10.82	&	0.09	&	0.0320	&	81.77	&	9.89	&	5.0 	&	49	&	1.94	&	9.68	&	3	&	S+S	&	Y	\\
8981-6101B	&	186.00444	&	36.15589	&	0.03336	&	10.82	&	-0.22	&	0.0320	&	81.77	&	9.59	&	2.9 	&	24	&	1.75	&	9.10	&	3	&	S+S	&	Y	\\
9027-12702	&	243.98368	&	31.40677	&	0.02256	&	10.00	&	0.06	&	0.0210	&	129.06	&	9.80	&	5.7 	&	65	&	1.83	&	9.02	&	1	&	S+S	&	N	\\
9027-9101	&	243.90740	&	31.32130	&	0.02220	&	10.15	&	-0.11	&	0.0352	&	126.97	&	9.59	&	3.8 	&	66	&	1.59	&	8.76	&	3	&	S+S	&	Y	\\
9030-3702	&	241.23416	&	30.52573	&	0.05540	&	10.67	&	0.86	&	0.0219	&	48.27	&	10.49	&	2.5 	&	30	&	1.66	&	9.79	&	1	&	S+S	&	Y	\\
9032-3704	&	241.21561	&	30.52601	&	0.05537	&	10.49	&	0.73	&	0.0152	&	48.27	&	10.49	&	2.3 	&	34	&	1.56	&	9.99	&	1	&	S+S	&	Y	\\
9050-9101	&	245.99844	&	21.79503	&	0.03209	&	10.24	&	0.52	&	0.0310	&	41.92	&	9.20	&	3.9 	&	43	&	1.82	&	9.71	&	2	&	S+S	&	Y	\\
9050-9101B	&	245.99284	&	21.82049	&	0.03209	&	10.24	&	0.28	&	0.0310	&	41.92	&	9.20	&	2.8 	&	35	&	1.67	&	9.45	&	2	&	S+S	&	Y	\\
9093-12701	&	239.75272	&	27.98539	&	0.05140	&	10.61	&	0.30	&	0.1533	&	24.43	&	...  	&	4.0 	&	68	&	1.58	&	9.90	&	2	&	S+S	&	Y	\\
9094-12703	&	239.74204	&	27.98753	&	0.05210	&	10.63	&	0.78	&	0.0411	&	24.43	&	... 	&	4.6 	&	73	&	1.60	&	10.00	&	2	&	S+S	&	Y	\\
9094-12705	&	240.46430	&	26.31944	&	0.04393	&	9.85	&	-0.11	&	0.0445	&	176.34	&	10.05	&	4.1 	&	74	&	1.54	&	9.24	&	3	&	S+E	&	N	\\
9185-9101	&	256.21228	&	34.81733	&	0.05683	&	10.81	&	1.39	&	0.193	&	7.25	&	10.09	&	2.3 	&	21	&	1.62	&	10.49	&	2	&	S+S	&	Y	\\
9488-9102	&	126.70413	&	20.36485	&	0.02510	&	9.75	&	0.27	&	0.0804	&	190.82	&	10.31	&	4.4 	&	23	&	2.14	&	8.48	&	3	&	S+S	&	N	\\
9499-12703	&	118.42323	&	26.49270	&	0.03742	&	10.82	&	0.07	&	0.0281	&	78.04	&	10.17	&	4.4 	&	40	&	1.98	&	9.60	&	1	&	S+E	&	N	\\
9507-12701	&	128.25074	&	26.01405	&	0.01763	&	9.87	&	-0.47	&	0.0668	&	21.83	&	9.62	&	12.8	&	82	&	2.14	&	9.09	&	2	&	S+E	&	N	\\
9509-6103	&	123.06447	&	26.20257	&	0.02508	&	10.04	&	-0.06	&	0.0147	&	87.35	&	9.71	&	3.1 	&	33	&	1.75	&	9.30	&	1	&	S+S	&	N	\\
9881-6102	&	205.21316	&	24.47331	&	0.02705	&	10.77	&	-0.05	&	0.0195	&	152.23	&	9.99	&	5.4 	&	19	&	2.32	&	9.81	&	1	&	S+E	&	N	\\
9889-1902	&	234.85860	&	24.94357	&	0.02286	&	9.98	&	0.63	&	0.0240	&	7.92	&	9.13	&	3.0 	&	53	&	1.58	&	8.91	&	1	&	S+S	&	Y	\\
9889-1902B	&	234.86460	&	24.94762	&	0.02291	&	10.41	&	0.99	&	0.0240	&	7.92	&	9.13	&	4.7 	&	46	&	1.93	&	9.65	&	1	&	S+S	&	Y	\\
8252-9101	&	144.69238	&	48.56287	&	0.02478	&	9.45	&	0.13	&	0.0383	&	82.20	&	9.27	&	0.7 	&	25	&	1.09	&	9.04	&	3	&	S+S	&	Y	\\
8624-12703	&	264.23954	&	59.20029	&	0.03070	&	10.46	&	0.62	&	0.0501	&	143.76	&	9.51	&	5.0 	&	66	&	1.34	&	9.54	&	3	&	S+S	&	Y	\\
8945-12701	&	171.89808	&	47.37939	&	0.03273	&	9.94	&	0.45	&	0.0578	&	100.36	&	10.32	&	3.9 	&	29	&	1.43	&	8.89	&	3	&	S+S	&	Y	\\
8979-3704	&	244.42297	&	40.93382	&	0.06209	&	10.80	&	0.96	&	0.0376	&	188.99	&	... 	&	4.3 	&	51	&	1.37	&	9.57	&	3	&	S+S	&	N	\\
8987-3701	&	136.24989	&	28.34772	&	0.04864	&	10.04	&	0.85	&	0.0157	&	44.75	&	10.08	&	2.2 	&	38	&	1.24	&	9.12	&	1	&	S+S	&	Y	\\
9034-3702	&	226.28033	&	46.96366	&	0.03771	&	9.83	&	0.39	&	0.0174	&	63.88	&	... 	&	3.2 	&	51	&	1.29	&	9.10	&	1	&	S+S	&	Y	\\
9508-6104	&	127.55302	&	26.62732	&	0.05299	&	10.82	&	1.12	&	0.0294	&	105.66	&	10.03	&	3.7 	&	66	&	1.26	&	9.77	&	3	&	S+S	&	Y	\\
\enddata
\tablecomments{A detailed description of this table's contents is given in Section \ref{subsec:sample}. }
\end{deluxetable*}
% \end{rotatetable*}

\begin{figure*}[t]
	\includegraphics[width=1\textwidth]{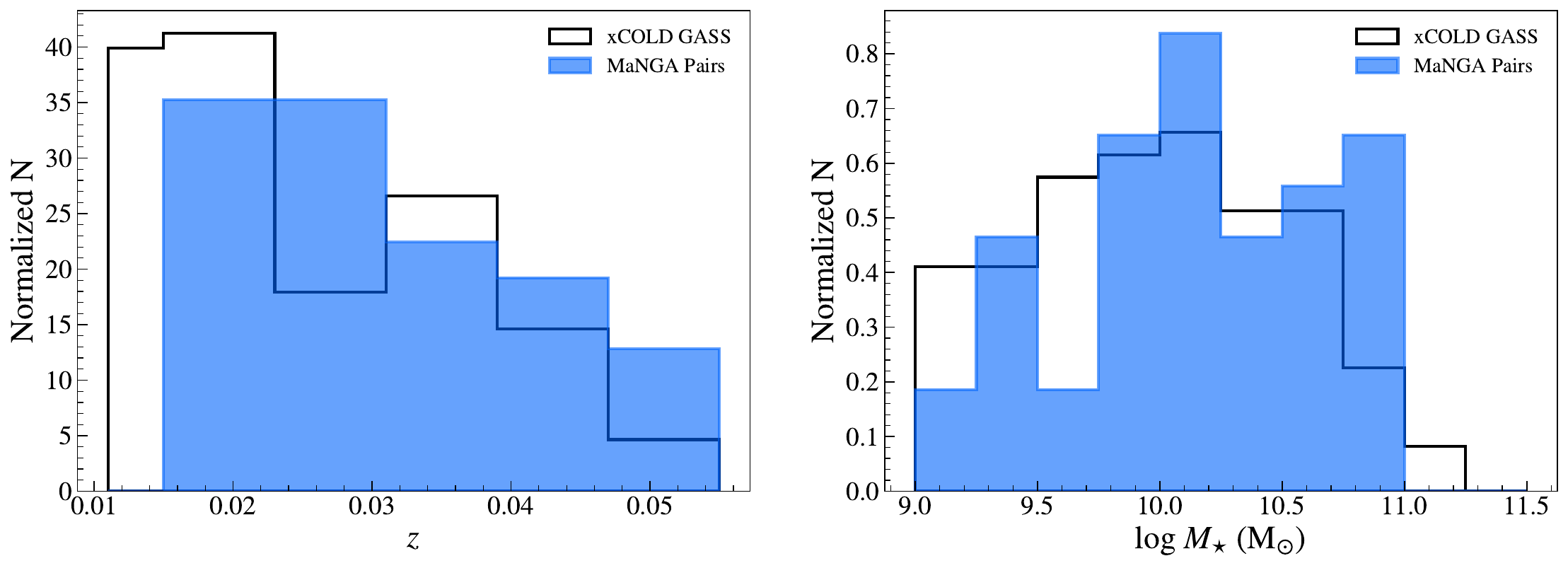}
	\caption{The left panel shows the distributions of redshift for the galaxy pair and the control sample. The right panel shows the distributions of stellar mass for the galaxy pair and the control sample. 
		\label{fig:co_pair_cs_z}
	}
\end{figure*}

\begin{figure}[t]
	\includegraphics[width=0.48\textwidth]{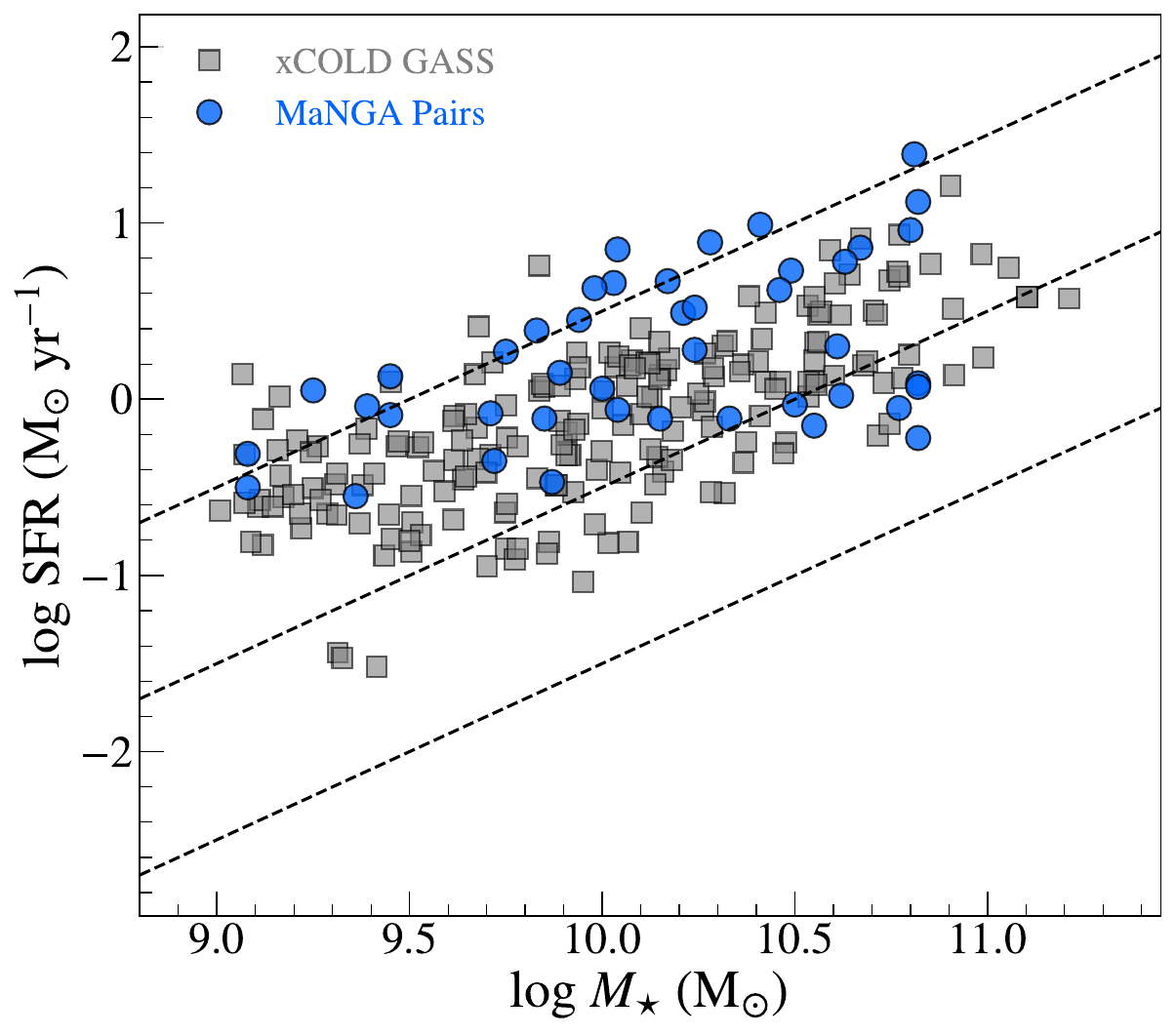}
	\caption{Global SFR as a function of stellar mass. The blue circles represent galaxy pairs selected from the MaNGA survey, while the grey squares are isolated galaxies from the control sample. From top to bottom, the three dashed lines correspond to ${\rm log(sSFR/yr^{-1})}=-9.5$, $-10.5$, and $-11.5$.
		\label{fig:co_pair_cs_sfr_m}
	}
\end{figure}

\subsection{Subsamples and Control Sample}\label{subsec:subsample}

The definition of the merger stage follows previous \hi\ observations \citep{Yu2022} and the statistical studies of SFR enhancement in MaNGA galaxy pairs \citep{Feng2020}, which relies on the combination of kinematic asymmetry ($\overline{v}_{\rm asym}$) and projected separation ($d_{\rm p}$). Adopting the statistical results of \cite{Feng2020}, we divide the pair sample into two subsamples based on their $\overline{v}_{\rm asym}$ values: low asymmetry ($0.007<\overline{v}_{\rm asym}<0.029$) and high asymmetry ($0.029<\overline{v}_{\rm asym}<0.316$). We adopt the boundary of $\overline{v}_{\rm asym} =0.029$ based on previous \hi\ observations \citep{Yu2022}, which has also been confirmed to be effective in determining merger stages by simulations of 24 galaxy mergers using the FIRE-2 model \citep{McElroy2022}. In particular, paired galaxies after the pericentric passage exhibit a more disturbed velocity field compared to pairs before the passage \citep{Hung2016,McElroy2022}. Hence, we consider the galaxy pairs with low $\overline{v}_{\rm asym}$ values as pairs to be in the pre-passage stage, indicating no significant interactions between member galaxies. During the sequence of galaxy interactions in pair-phase, simulations suggest a constant decrease in the physical separation before the pericentric passage, and then the separation will increase until the paired galaxies are approaching apocenter \citep{Torrey2012,Moreno2015,Moreno2019}. In this work, we use the projected separation ($d_{\rm p}$) as a reference to distinguish the pericenter and apocenter stages of galaxy pairs. Galaxy pairs with high $\overline{v}_{\rm asym}$ value and $d_{\rm p}<$ 50 kpc are defined to be at the stage of pericenter passage, and those with high $\overline{v}_{\rm asym}$ value and $d_{\rm p}>$ 50 kpc are regarded as being in the apocenter stage. The number of pairs in each stage is as follows: (1) pre-passage: 13 ; (2) pericenter passage: 12; (3) apocenter passage: 18.

To systematically investigate the impacts of galaxy interactions on star formation and molecular gas properties, we compiled a control sample of isolated galaxies. The isolated galaxies are selected from the Extended CO Legacy Database for GASS \citep[xCOLD GASS;][]{Saintonge2017}, with the requirement of having no bright neighbors ($r < 17.77$) within $d_{\rm p} \leqslant$  200 $h^{-1}$ kpc and $|\Delta v|\leqslant$ 500 km $\rm s^{-1}$ \citep{Feng2020}. To match with the star-forming galaxies in our pair sample, we require that the isolated galaxies have log$(\rm{sSFR/yr}^{-1})>-11$. In total, we selected 195 isolated galaxies with both \hi\ and CO data from xCOLD GASS as our control sample pool. To fairly compare with the paired galaxies, we also adopt the SFR and stellar mass ($M_{\star}$) from the public catalog of MPA-JHU DR7 for the isolated galaxies. As shown in Figure \ref{fig:co_pair_cs_z}, we compared the distributions of redshift and stellar mass between the galaxy pair sample and the control sample. The control sample tends to include more nearby galaxies at lower redshift ($z<0.015$) and more massive galaxies ($\text{log}(M_{\star}/M_{\odot})> 11$) at the high-mass end. To compare the galaxy properties in pairs and controls fairly, we conducted a galaxy-by-galaxy matching between the paired galaxies and isolated galaxies from the control sample pool for further analysis (see details in Section \ref{subsec:offset}). Regarding kinematic asymmetry in isolated galaxies, a systematic study suggests that $\overline{v}_{\rm asym}$ shows a significant anticorrelation with stellar mass only in low-mass galaxies at $\text{log}(M_{\star}/M_{\odot})< 9.7$, and it is independent of \hi\ gas content \citep{Feng2022}. Thus, after matching for stellar mass, we minimized potential bias due to the dependence of kinematic asymmetry on stellar mass at the low-mass end. In Figure \ref{fig:co_pair_cs_sfr_m}, we present the distributions of paired galaxies and isolated galaxies in the SFR--$M_{\star}$ plane. Galaxies in our galaxy pair sample are marked by blue circles. For the control sample, isolated galaxies selected from the xCOLD GASS survey are represented by gray squares. At a given stellar mass, the mean SFR of paired galaxies is higher than that of the isolated galaxies in the control sample.

\subsection{Observations and Data}
\subsubsection{James Clerk Maxwell Telescope Observations}
For part of the sources in our galaxy pair sample, we performed CO(2$-$1) observations through our PI program (M21BP051, PI: Q. Yu) with the James Clerk Maxwell Telescope (JCMT). $^{12}$CO(2$-$1) observations for 23 paired galaxies were carried out between 2021 September and 2021 December, using the N$\rm \overline{a}$makanui receiver at 230 GHz on JCMT. The beam size is $\sim$21$\arcsec$ at the observed frequency. The observed sample consists of nine major-merger galaxy pairs and 14 minor-merger pairs. The observations were conducted in beam-switching mode with a throw of 180$\arcsec$ in the azimuthal direction. The integration time for each source ranged between 10 and 60 minutes. The data were reduced using the Starlink software \citep{Currie2014}. We followed the standard procedure to calibrate individual exposures of each galaxy, and subsequently, we co-added the spectra for each source. After calibration, we rebinned the spectra to a velocity resolution of 10 $\rm km\ s^{-1}$. In some cases, the spectra were rebinned to a velocity of $\sim$20 $\rm km\ s^{-1}$ to achieve tentative detections. By utilizing line-free channels as the fitting region, we subtracted a linear or second-order polynomial baseline from the spectrum. We converted the spectra from antenna temperature in Kelvin to Jansky, adopting a conversion factor of 15.6/$\eta_a$, where the aperture efficiency $\eta_a$ is 0.57. In total, we detected 17 CO(2$-$1) lines out of 23 sources, including nine major-merger galaxies and eight minor-merger galaxies. For each detected spectrum, we visually determined the line widths at zero level ($\Delta V$) and derived the velocity-integrated flux. For nondetections, we calculated the upper limits of velocity-integrated flux as
\begin{equation}\label{eq:uplim_Ico}
I_{\rm CO(2-1)} = 3\times rms\times\sqrt{\delta V\ \Delta V},
\end{equation} 
where $\delta V$ is the channel width in kilometers per second, $\Delta V$ is the zero-level line width in kilometers per second, and rms is the root mean square noise in Kelvin. The non-detections are all minor mergers and are not included in the final pair sample. All the observing results are presented in Section \ref{subsec:obs_res} and Table \ref{tab:emission}.

\subsubsection{Institut de Radioastronomie Milimetrique 30\,m Observations}

Through our PI programs (E01-21, 029-22, PI: Q. Yu) with the Institut de Radioastronomie Milimetrique (IRAM) 30\,m telescope on Pico Veleta, we performed CO observations for 29 galaxies in the galaxy pair sample between 2022 March and 2022 July. For each target, the $^{12}$CO(1$-$0) and $^{12}$CO(2$-$1) lines were observed simultaneously, using the dual polarization receiver EMIR in combination with the autocorrelator FTS at a frequency resolution of 0.195 MHz ($\sim$0.5 $\rm km\ s^{-1}$ at CO(1$-$0)) and with the autocorrelator WILMA with a frequency resolution of 2 MHz ($\sim$5 $\rm km\ s^{-1}$ at CO(1$-$0)). The beam sizes are $\sim$22$\arcsec$ and $\sim$11$\arcsec$ for the CO(1$-$0) and CO(2$-$1) observations, respectively. We used the wobbler-switching mode to observe each source with a wobbler throw ranging between 90$\arcsec$ and 110$\arcsec$ in the azimuthal direction. The wobbler throw was individually chosen to mitigate potential contamination from the companion galaxy. We divided the observations of paired galaxies into different groups based on their redshifts, ensuring that both the CO(1$-$0) and CO(2$-$1) lines can be covered within the bandwidth of the receiver for each source. In our observing strategy, we continued observing each source until it achieved a detection with S/N $\geqslant$ 5 or until an rms of $<$2 mK ($T_{\rm A}^{*}$) was achieved for a velocity resolution of $\sim$20 $\rm km\ s^{-1}$.  

The data were reduced using the CLASS software in the GILDAS package\footnote{\url{http://www.iram.fr/IRAMFR/GILDAS}} \citep{Gildas2013}. We initially inspected the spectra in individual scans and discarded low-quality scans with severe baseline issues before standard calibration. The baseline was subtracted with a constant or a linear function for each scan. Subsequently, we averaged and rebinned the spectra to achieve a velocity resolution of $\sim$20 $\rm km\ s^{-1}$. The final spectra were converted to the main beam temperature scale using the equation $T_{\rm mb}=(F_{\rm eff}/B_{\rm eff})\times T_{\rm A}^{*}$, where $F_{\rm eff}$ and $B_{\rm eff}$ represent the IRAM forward efficiency and the beam efficiency, respectively. For 115 GHz (230 GHz), the IRAM 30\,m telescope has the $F_{\rm eff}$ of 0.95 (0.91) and the $B_{\rm eff}$ of 0.77 (0.58). In the subsequent analysis, we obtained the flux of the spectra by applying the $T_{\rm mb}$-to-flux conversion factor of the IRAM 30\,m telescope (5 $\rm Jy\ K^{-1}$). We visually determined the line widths at zero level ($\Delta V$) and derived the velocity-integrated flux ($S_{\rm CO}$) for each spectrum. Given that the CO(1$-$0) and CO(2$-$1) lines were observed simultaneously, we constrained the line widths between the two transitions based on detected lines with higher S/N.

\subsubsection{MASCOT Data}
MASCOT \citep{Wylezalek2022} is a CO(1$-$0) follow-up survey of MaNGA galaxies using the 12\,m millimeter single-dish telescope at the Arizona Radio Observatory (ARO). Aiming to measure the molecular gas content in MaNGA star-forming galaxies with $\text{log}(M_{\star}/M_{\odot})> 9.5$, the MASCOT survey presents CO(1$-$0) observations for 187 MaNGA galaxies in its first data release \citep{Wylezalek2022}, with a beam size of $\sim$55\arcsec. After crossmatching with the parent sample described in Section \ref{subsec:sample}, we included seven major-merger paired galaxies in our final galaxy pair sample, with CO data from the MASCOT survey.

\subsection{Aperture Correction and Molecular Gas Mass}\label{subsec:aperture_correct}
Considering the optical sizes of some galaxies are more extended than the telescope beams, single-pointing observations at the center of a galaxy need an aperture correction to derive the CO flux for the entire galaxy. Thus, we applied an aperture correction to all of the measured CO line fluxes following the method of \cite{Lisenfeld2011,Lisenfeld2019}. We assume an exponential molecular gas disk with a scale length of $r_e$, the CO flux can be described as
\begin{equation}\label{eq:exp_sco}
S_{\rm CO}(r) = S_{\rm CO,center}\ \text{exp}(-r/r_e),
\end{equation}
where $S_{\rm CO,center}$ represents the CO flux in the central position and derived from the measured $I_{\rm CO}$, $r_e$ is the CO scale length. \cite{Lisenfeld2011} have found the CO scale length well correlated to the optical radius at the 25 mag isophote, $r_{25}$, and it can be derived as $r_e = 0.2\times r_{25}$ based on different CO observations of local spiral galaxies \citep{Nishiyama2001,Regan2001,Leroy2008}. Therefore, we collected the $r_{25}$ (see Table \ref{tab:major_pair}) measured in the $r$ band from the NASA/IPAC Extragalactic Database (NED) and calculated the $r_e$ for each source in the pair sample. Adopting the derived $r_e$ and inclination angle ($i$) from NED, we performed a 2D integration of the exponential disk following Equation (\ref{eq:exp_sco}) to compute the total CO flux from the entire disk \citep[for details, see][]{Lisenfeld2011}. After deriving the central CO flux ($S_{\rm CO,center}$) and the total CO flux ($S_{\rm CO,tot}$), the aperture correction factor ($f_{\rm aper}$) is defined as
\begin{equation}
\begin{aligned}\label{eq:f_aper}
f_{\rm aper} & = S_{\rm CO,tot} / S_{\rm CO,center}\\
 & =  M_{\rm H_2}/M_{\rm H_2,center} ,
\end{aligned}
\end{equation}
where $M_{\rm H_2}$ and $M_{\rm H_2,center}$ are the molecular masses of the entire galaxy and the central pointing, respectively. The molecular gas mass is derived from the CO(1$-$0) luminosity as:
\begin{equation}\label{eq:h2_mass}
M_{\rm H_2}[M_{\odot}]=\alpha_{\rm CO} L_{\rm CO}^{\prime},
\end{equation}
where $\alpha_{\rm CO}$ is the CO-to-\hh\ conversion factor. Considering the paired galaxies in our sample are mostly massive ($\text{log}(M_{\star}/M_{\odot})> 9.7$) and do not exhibit extreme starbursts, we adopt the Galactic value with $\alpha_{\rm CO} = 3.2\ M_{\sun}/{\rm(K\ km\ s^{-1}\ pc^{-2}})$ \citep{Bolatto2013}. For the CO flux from the MASCOT survey, we performed the same aperture correction and adopted the Galactic value of $\alpha_{\rm CO}$. For isolated galaxies from xCOLD GASS, we adopted the same $\alpha_{\rm CO}$ and recalculated the estimated \hh\ mass. The $L_{\rm CO}^{\prime}$ is calculated following \cite{Solomon1997} as
\begin{equation}\label{eq:h2_lco}
L_{\rm CO}^{\prime}[{\rm K\, km\, s^{\!-\!1}\, pc^{2}}]\!=\!3.25\!\times\! 10^{7}\!S_{\rm {CO,tot}}\nu_{\rm {rest}}^{\!-\!2}D_{L}^{2}(1\!+\!z)^{\!-\!1},
\end{equation}
where $S_{\rm{CO,tot}}$ is the total flux of CO(1$-$0) line in $\text{Jy km s}^{-1}$, $\nu_{\rm {rest}}$ represents the rest frequency of the spectral line in GHz, $D_{L}$ is the luminosity distance in Mpc, and $z$ is the redshift of the galaxy.

\subsection{Atomic Gas}

We use the \hi\ data for 39 galaxy pairs from the pair sample taken from \cite{Yu2022}. For the \hi\ data used in this work, six sources were observed by our PI programs with the Five-hundred-meter Aperture Spherical radio Telescope (FAST). The remaining data are extracted from the HI-MaNGA survey. The HI-MaNGA survey is designed as an \hi\ follow-up project for the SDSS-VI MaNGA survey, which mainly uses the Robert C. Byrd Green Bank Telescope to perform \hi\ observations in combination with data crossmatched from the Arecibo Legacy Fast ALFA survey \citep{Haynes2018}. The HI-MaNGA survey focuses on galaxies with a stellar mass in the range of $8.5 <\text{log}(M_{\star}/M_{\odot})< 11.2$ and a redshift of $z<0.05$ \citep{Masters2019,Stark2021}. Due to the relatively large beam sizes, the \hi\ observations of paired galaxies often suffered from the source blending problem \citep{Zuo2018,Yu2022}. Therefore, we adopt the same method of \cite{Yu2022} and consider a galaxy pair as a single system when calculating \hi-related properties. For the S+S pairs, the $M_{\star}$ and SFR are sums of two subcomponents, and then we divide these values and the \hi\ mass by two to obtain the typical value for one of the member galaxies. For the S+E pairs, we follow the approach of previous studies \citep{Zuo2018,Lisenfeld2019,Yu2022} and assume the \hi\ content is mostly contributed by the spiral component, so only the spiral component is considered for $M_{\star}$ and SFR.

\begin{deluxetable*}{lrrrrrrc}\setlength\tabcolsep{15pt}
\tablenum{2}
\tablecaption{CO(2-1) emission line results from JCMT observations\label{tab:emission}}
\tabletypesize{\small}
\tablewidth{0pt}
\tablehead{
\colhead{Source name} & \colhead{R.A.} & \colhead{Decl.} & \colhead{rms} & \colhead{S/N} & \colhead{$I_{\text{CO(2-1)}}$} & \colhead{$\Delta V_{\text{CO(2-1)}}$} & \colhead{$f_{\rm aper}$}\\
\colhead{} & \colhead{(deg)} & \colhead{(deg)} & \colhead{(mK)} & \colhead{} & \colhead{($\text{K km s}^{-1}$)} & \colhead{($\text{km s}^{-1}$)} & \colhead{}
}
\decimalcolnumbers
\startdata
8078-6103 & 42.41654 & -0.06985 & 2.10 & 26.4 & 3.47 $\pm$ 0.37 & 415 $\pm$ 10 & 1.93 \\
8250-6101 & 138.75315 & 42.02439 & 2.11 & 39.7 & 4.94 $\pm$ 0.51 & 368 $\pm$ 10 & 1.79 \\
8260-6101 & 182.40876 & 42.00967 & 2.35 & 6.1 & 1.02 $\pm$ 0.19 & 320 $\pm$ 10 & 1.69 \\
8450-6102 & 171.74883 & 21.14168 & 2.02 & 11.2 & 0.93 $\pm$ 0.12 & 174 $\pm$ 10 & 1.77 \\
8456-12702 & 149.96826 & 45.28310 & 1.78 & 5.4 & 0.43 $\pm$ 0.09 & 215 $\pm$ 10 & 1.76 \\
8567-6102 & 119.31545 & 47.80304 & 0.75 & -- & $<$ 0.28 & -- & -- \\
8606-9102 & 255.70905 & 36.70675 & 1.84 & 10.4 & 1.07 $\pm$ 0.15 & 324 $\pm$ 10 & 1.69 \\
8715-12704 & 121.17245 & 50.71853 & 0.98 & -- & $<$ 0.37 & -- & -- \\
8952-6103 & 204.67418 & 27.74241 & 1.95 & 19.9 & 2.5 $\pm$ 0.28 & 431 $\pm$ 10 & 1.85 \\
8982-9102 & 202.68920 & 26.52161 & 1.02 & -- & $<$ 0.39 & -- & -- \\
9025-3701 & 246.41772 & 29.14761 & 1.18 & -- & $<$ 0.46 & -- & -- \\
9030-3702 & 241.23417 & 30.52578 & 3.08 & 9.3 & 1.51 $\pm$ 0.22 & 278 $\pm$ 10 & 1.66\\
9093-12703 & 239.75272 & 27.98539 & 1.94 & 10.2 & 1.24 $\pm$ 0.17 & 395 $\pm$ 10 & 1.60 \\
9185-9101 & 256.21228 & 34.81733 & 2.41 & 27.3 & 4.96 $\pm$ 0.53 & 569 $\pm$ 10 & 1.65 \\
9488-9102 & 126.70413 & 20.36485 & 1.52 & 6.3 & 0.28 $\pm$ 0.05 & 94 $\pm$ 10 & 2.14 \\
9498-12704 & 118.18742 & 24.29509 & 0.81 & -- & $<$ 0.32 & -- & -- \\
9499-12703 & 118.42323 & 26.49270 & 1.33 & 21.0 & 1.80 $\pm$ 0.20 & 432 $\pm$ 10 & 1.98 \\
9501-12704 & 129.37073 & 26.01438 & 1.62 & 7.1 & 0.68 $\pm$ 0.12 & 350 $\pm$ 10 & 1.56 \\
9503-6104 & 120.40348 & 24.43890 & 1.07 & 4.6 & 0.34 $\pm$ 0.08 & 229 $\pm$ 20 & 1.46 \\
9510-6101 & 127.85323 & 27.58001 & 0.82 & -- & $<$ 0.31 & -- & -- \\
9866-12702 & 243.20452 & 31.99319 & 4.02 & 10.1 & 2.43 $\pm$ 0.34 & 380 $\pm$ 10 & 1.99 \\
9882-12705 & 207.78113 & 24.01823 & 1.47 & 25.7 & 2.42 $\pm$ 0.26 & 434 $\pm$ 10 & 1.88 \\
10218-3704 & 119.16881 & 16.92011 & 2.26 & 34.9 & 3.94 $\pm$ 0.41 & 264 $\pm$ 10 & 1.86 \\
\enddata
\tablecomments{The columns show (1) name of our targets. (2) R.A. in degrees. (3) Decl. in degrees. (4) rms noise of the spectra at $\sim$10--40 $\text{km s}^{-1}$, only non-detected spectra were rebinned to a velocity resolution of $\sim$40 $\text{km s}^{-1}$. (5) S/N. (5) Integrated flux in $\text{K km s}^{-1}$. (6) The CO line width in $\text{km s}^{-1}$ measured at zero level. (7) Aperture correction.}
\end{deluxetable*}

\section{Results} \label{sec:results}

\begin{deluxetable*}{lrrrcrrrrc}\setlength\tabcolsep{8pt}
\tablenum{3}
\centering
\caption{CO(1$-$0) and CO(2$-$1) emission line results from IRAM 30m observations \label{tab:iram_co_line}}
\tablewidth{0pt}
\tablehead{
\colhead{MaNGA ID} & \colhead{$\rm rms_{10}$} & \colhead{${\rm S/N_{10}}$} & \colhead{$I_{\rm{CO(1-0)}}$} & \colhead{$\Delta V_{\rm{CO(1-0)}}$} & \colhead{$f_{\rm aper}$} & \colhead{$\rm rms_{21}$} & \colhead{${\rm S/N_{21}}$} & \colhead{$I_{\rm{CO(2-1)}}$} & \colhead{$\Delta V_{\rm{CO(2-1)}}$} \\
\colhead{plate-ifu}  & \colhead{(mK)} & \colhead{(mK)} & \colhead{($\text{K km s}^{-1}$)} & \colhead{($\text{km s}^{-1}$)} & \colhead{} & \colhead{(mK)} & \colhead{(mK)} & \colhead{($\text{K km s}^{-1}$)} & \colhead{($\text{km s}^{-1}$)} 
}
\decimalcolnumbers
\startdata
8078-6104	&	1.13	&	14.3	&	1.14 $\pm$ 0.39	&	223 $\pm$ 21	&	1.52	&	3.99	&	11.2	&	2.05  $\pm$ 0.27	&	189 $\pm$ 10  \\	
8078-6104B	&	2.01	&	9.6 	&	1.67 $\pm$ 0.24	&	335 $\pm$ 21	&	1.67	&	4.29	&	21.7	&	5.68  $\pm$ 0.62	&	334 $\pm$ 10	\\
8082-12703	&	3.76	&	21.4	&	8.07 $\pm$ 0.89	&	471 $\pm$ 21	&	1.73	&	9.98	&	27.0	&	16.73 $\pm$ 1.78	&	363 $\pm$ 10	\\
8085-12704	&	1.83	&	4.8 	&	1.43 $\pm$ 0.34	&	611 $\pm$ 42	&	1.69	&	3.43	&	8.1 	&	2.40  $\pm$ 0.38	&	347 $\pm$ 21	\\
8153-12701	&	0.72	&	4.2 	&	0.33 $\pm$ 0.08	&	265 $\pm$ 42	&	1.65	&	1.74	&	6.2 	&	0.91  $\pm$ 0.17	&	330 $\pm$ 21	\\
8338-6102	&	1.10	&	5.8 	&	0.36 $\pm$ 0.07	&	149 $\pm$ 21	&	1.68	&	2.11	&	8.2 	&	1.29  $\pm$ 0.20	&	257 $\pm$ 21	\\
8547-12702	&	1.54	&	5.5 	&	0.80 $\pm$ 0.17	&	403 $\pm$ 21	&	1.64	&	3.43	&	9.2 	&	3.72  $\pm$ 0.55	&	626 $\pm$ 21	\\
8552-12702	&	1.23	&	5.1 	&	0.97 $\pm$ 0.21	&	561 $\pm$ 42	&	1.55	&	2.63	&	7.0 	&	1.43  $\pm$ 0.25	&	280 $\pm$ 21	\\
8588-12702	&	1.26	&	8.5 	&	0.65 $\pm$ 0.10	&	174 $\pm$ 21	&	1.91	&	6.06	&	4.1 	&	1.69  $\pm$ 0.45	&	217 $\pm$ 21	\\
8656-1901	&	0.80	&	5.1 	&	0.81 $\pm$ 0.18	&	937 $\pm$ 42	&	1.67	&	2.18	&	7.7 	&	1.68  $\pm$ 0.27	&	467 $\pm$ 21	\\
8656-3703	&	0.61	&	5.1 	&	0.30 $\pm$ 0.07	&	212 $\pm$ 42	&	1.63	&	2.27	&	8.4 	&	1.52  $\pm$ 0.24	&	298 $\pm$ 21	\\
8657-12704	&	0.65	&	5.6 	&	0.31 $\pm$ 0.06	&	169 $\pm$ 42	&	1.74	&	3.02	&	3.3 	&	1.02  $\pm$ 0.33	&	254 $\pm$ 42	\\
8728-12701	&	3.92	&	11.6	&	1.77 $\pm$ 0.23	&	141 $\pm$ 10	&	2.19	&	7.03	&	6.7 	&	3.24  $\pm$ 0.58	&	216 $\pm$ 21	\\
8981-6101	&	1.75	&	12.7	&	3.02 $\pm$ 0.38	&	836 $\pm$ 21	&	1.94	&	3.93	&	9.9 	&	4.66  $\pm$ 0.66	&	656 $\pm$ 21	\\
8981-6101B	&	1.62	&	10.0	&	0.89 $\pm$ 0.13	&	274 $\pm$ 10	&	1.75	&	3.27	&	13.1	&	1.98  $\pm$ 0.25	&	196 $\pm$ 10	\\
9027-12702	&	2.55	&	11.2	&	1.56 $\pm$ 0.21	&	279 $\pm$ 10	&	1.83	&	4.70	&	7.4 	&	2.77  $\pm$ 0.47	&	299 $\pm$ 21	\\
9027-9101	&	2.09	&	6.3 	&	1.01 $\pm$ 0.19	&	278 $\pm$ 21	&	1.59	&	3.41	&	15.8	&	4.60  $\pm$ 0.54	&	341 $\pm$ 21	\\
9032-3704	&	3.42	&	12.6	&	2.79 $\pm$ 0.36	&	366 $\pm$ 10	&	1.56	&	13.5	&	10.2	&	8.47  $\pm$ 1.19	&	331 $\pm$ 10	\\
9050-9101	&	3.43	&	19.3	&	3.74 $\pm$ 0.42	&	294 $\pm$ 10	&	1.82	&	6.00	&	32.9	&	12.34 $\pm$ 1.29	&	359 $\pm$ 10	\\
9050-9101B	&	2.95	&	16.1	&	2.26 $\pm$ 0.27	&	207 $\pm$ 10	&	1.67	&	5.86	&	22.5	&	6.57  $\pm$ 0.72	&	228 $\pm$ 10	\\
9093-12701	&	2.63	&	13.9	&	2.58 $\pm$ 0.32	&	442 $\pm$ 10	&	1.58	&	5.39	&	5.4 	&	2.99  $\pm$ 0.63	&	476 $\pm$ 21	\\
9094-12703	&	2.33	&	13.1	&	3.16 $\pm$ 0.40	&	475 $\pm$ 21	&	1.60	&	8.85	&	10.5	&	10.07  $\pm$ 1.39	&	522 $\pm$ 21	\\
9094-12705	&	1.01	&	6.1 	&	0.79 $\pm$ 0.15	&	359 $\pm$ 42	&	1.54	&	2.62	&	6.5 	&	1.37  $\pm$ 0.25	&	289 $\pm$ 21	\\
9185-9101	&	6.00	&	16.5	&	8.00 $\pm$ 0.94	&	572 $\pm$ 10	&	1.62	&	10.8	&	34.3	&	26.7  $\pm$ 2.78	&	456 $\pm$ 10	\\
9507-12701	&	2.48	&	12.2	&	2.57 $\pm$ 0.33	&	341 $\pm$ 21	&	2.14	&	5.19	&	8.3 	&	3.40  $\pm$ 0.53	&	297 $\pm$ 21	\\
9509-6103	&	2.35	&	19.6	&	2.49 $\pm$ 0.28	&	270 $\pm$ 10	&	1.75	&	5.13	&	19.5	&	4.56  $\pm$ 0.51	&	193 $\pm$ 10	\\
9881-6102	&	4.27	&	23.5	&	5.21 $\pm$ 0.57	&	249 $\pm$ 10	&	2.32	&	11.75	&	16.5	&	10.02 $\pm$ 1.17	&	248 $\pm$ 10	\\
9889-1902	&	1.69	&	8.1 	&	1.34 $\pm$ 0.21	&	452 $\pm$ 21	&	1.58	&	2.91	&	17.8	&	5.30  $\pm$ 0.61	&	492 $\pm$ 21	\\
9889-1902B	&	4.32	&	15.3	&	6.05 $\pm$ 0.72	&	388 $\pm$ 21	&	1.93	&	9.52	&	8.0 	&	6.72  $\pm$ 1.08	&	364 $\pm$ 21	\\
\enddata
\tablecomments{The columns are described as follows: (1) Source name. (2) rms noise of CO(1$-$0) line at the velocity resolution of $\sim$10$-$42 $\text{km s}^{-1}$. (3) S/N achieved in the CO(1$-$0) line. (4) Intergated CO(1$-$0) line intensity measured after rebinning the spectra to velocity resolution of $\sim$10$-$42 $\text{km s}^{-1}$. (5) Zero-level line width of CO(1$-$0). (6) Aperture correction. (7) rms noise of CO(2$-$1) line at the velocity resolution of $\sim$10$-$42 $\text{km s}^{-1}$. (8) S/N achieved in the CO(2$-$1) line. (9) Intergated CO(2$-$1) line intensity measured after rebinning the spectra to velocity resolution of $\sim$10$-$42 $\text{km s}^{-1}$. (10) Zero-level line width of CO(2$-$1).}
\end{deluxetable*}

\subsection{Molecular Gas Properties and SFR}\label{subsec:obs_res}

We detected 17 CO(2$-$1) emission lines out of 23 paired galaxies from JCMT observations. The detailed results of JCMT observations are listed in Table \ref{tab:emission}, and the profiles of the emission line are shown in Figure \ref{fig:jcmt_co_2-1} in the appendix. In our IRAM 30\,m observations, we observed 29 paired galaxies and all sources have detections of CO(1$-$0) and  CO(2$-$1) lines, including a couple of tentative detections (S/N $<5$). The results of IRAM observations are listed in Table \ref{tab:iram_co_line}, and the CO spectra are presented in appendix (Figures \ref{fig:iram_co_1-0} and \ref{fig:iram_co_2-1_b}). The molecular gas mass is estimated from the CO luminosity using Equations (\ref{eq:h2_mass}) and (\ref{eq:h2_lco}). Given that JCMT observations only cover the CO(2$-$1) line, we adopt a CO(2$-$1)-to-CO(1$-$0) ratio of 0.8 when calculating the CO luminosity \citep{Leroy2009}. Considering JCMT observations contain minor-merger galaxy pairs that are beyond the scope of this paper, we performed the following analysis using major-merger galaxy pairs as described in Section \ref{subsec:sample}. Therefore, we only consider 43 major-merger paired galaxies that have CO data from JCMT observations, IRAM 30\,m observations, and the MASCOT survey. For the tentative detections with S/N $<5$, we treat these measurements as upper limits in the statistical analysis, adopting the Kaplan--Meier estimator \citep{KM1958,Feigelson1985} to calculate the mean values and errors for the corresponding physical properties. 

After deriving the molecular mass, we compared the molecular gas fraction (\fhh) and star formation efficiency (SFE) between paired galaxies and isolated galaxies. The molecular gas fraction \fhh\ is defined as
\begin{equation}
f_{\rm H_{2}} = \frac{M_{\rm H_2}}{M_{\star}}  . 
\label{eq:fhh}
\end{equation}
Despite the definition of $f_{\rm H_{2}} = M_{\rm H_2}/(M_{\rm H_2}+M_{\star})$ used in some studies, here we adopt the definition frequently adopted by previous observational studies \citep{Violino2018,Lisenfeld2019} for comparison. 

\begin{figure*}[t]
	\includegraphics[width=1\textwidth]{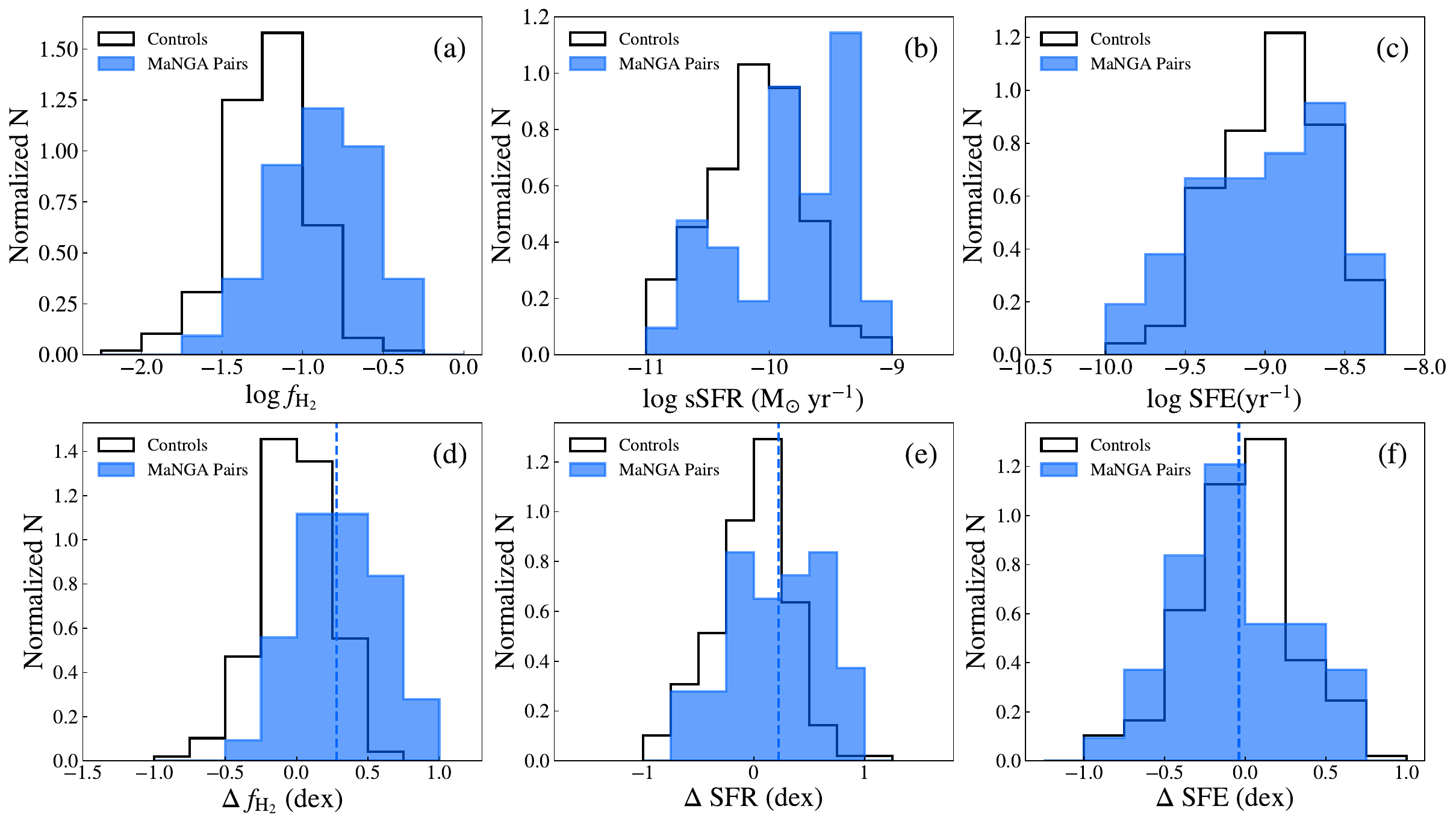}
	\caption{The upper panels show distributions of galaxy properties of the galaxy pair sample and the control sample, while the lower panels present the offset of these properties relative to the controls. The paired galaxies and control galaxies are plotted as filled and open histograms, respectively. The dashed lines represents the mean value of distributions for paired galaxies. (a) Distributions of \fhh. (b) Distributions of sSFR. (c) Distributions of SFE. (d) Distributions of $\Delta f_{\text{\hh}}$. (e) Distributions of $\Delta \text{SFR}$. (f) Distributions of $\Delta$SFE.  
	\label{fig:h2_full_pair_cs_dlt}
	}
\end{figure*}

In Figure \ref{fig:h2_full_pair_cs_dlt}, we compare the distributions of galaxy properties in paired galaxies and control galaxies. The blue histogram represents the distribution of the pair sample, while the open histogram represents the control sample. As shown in Figure \ref{fig:h2_full_pair_cs_dlt}(a), we compare the distributions of \fhh\ for galaxies in pairs and controls. The mean molecular gas fraction of the pair sample in logarithm is log \fhh$=-0.89\pm 0.05$. In contrast, the mean molecular gas fraction of the control sample (log \fhh$=-1.22\pm 0.02$) is lower by a factor of $\sim$2 compared to the pair sample. We conduct the Anderson--Darling (AD) two-sample test to check any discrepancy between distributions. The AD test is more sensitive than the classic Kolmogorov--Smirnov test to test whether the two distributions are drawn from the same parent distribution \citep{Anderson1952,Feigelson2012}. We use the {\tt scipy.stats.anderson\_ksamp} to perform the AD two-sample test, which calculates the truncated $p\text{-value}$ in the range between 0.001 and 0.25. In the following analysis, we adopt $p\text{-values}$ of $<$0.001 (significant) and $<$0.01 (moderate) as significance levels to reject the null hypothesis that the two samples are drawn from the same parent population. Therefore, the AD test suggests the distributions of \fhh\ for galaxies in pairs and controls are different ($p\text{-value}<0.001$). However, multiple factors may contribute to the difference in molecular gas fraction. For example, more galaxies with lower stellar mass in the sample can bias the average \fhh\ toward a higher value, since \fhh\ is tightly anticorrelated with $M_{\star}$ for star-forming galaxies \citep{Boselli2014,Cicone2017,Saintonge2017}. Therefore, in Section \ref{subsec:offset}, we conduct a galaxy-by-galaxy matching between the galaxy pair sample and the control sample to minimize bias.

As shown in Figure \ref{fig:h2_full_pair_cs_dlt}(b), the sSFR of the pair sample is distributed at the higher region relative to the control sample. The average sSFR of the pair sample (log$(\text{sSFR}/\text{yr}^{-1})=-9.88\pm0.08$) is significantly higher than that of the control sample (log$(\text{sSFR}/\text{yr}^{-1})=-10.11\pm0.03$). The AD test suggests the distributions of sSFR in the pair sample and the control sample are distinct at the significance level of $p\text{-value}<0.001$.

The SFE is defined as
\begin{equation}
{\rm SFE\ [yr^{-1}]}= \frac{\rm{SFR}}{M_{\rm H_2}}  . 
\label{eq:sfe}
\end{equation}
In Figure \ref{fig:h2_full_pair_cs_dlt}(c), we present the distributions of SFE in the pair sample and the control sample. The SFE of paired galaxies distributes from $\rm -9.86$ to $\rm -8.21\ yr^{-1}$, with the mean $\rm log\ (SFE/yr^{-1})=-8.99 \pm 0.06$. In contrast, the SFE of isolated galaxies distributes from $\rm -9.90$ to $\rm -7.90\ yr^{-1}$, with the mean $\rm log\ (SFE/yr^{-1})=-8.90 \pm 0.03$. The AD test returns a $p\text{-value}=0.24$, suggesting the distributions of SFE in paired galaxies and isolated galaxies are likely drawn from the same distribution.

\subsection{Molecular-to-atomic Gas Mass Ratio and Total Gas Mass}
\begin{figure*}[t]
	\includegraphics[width=1\textwidth]{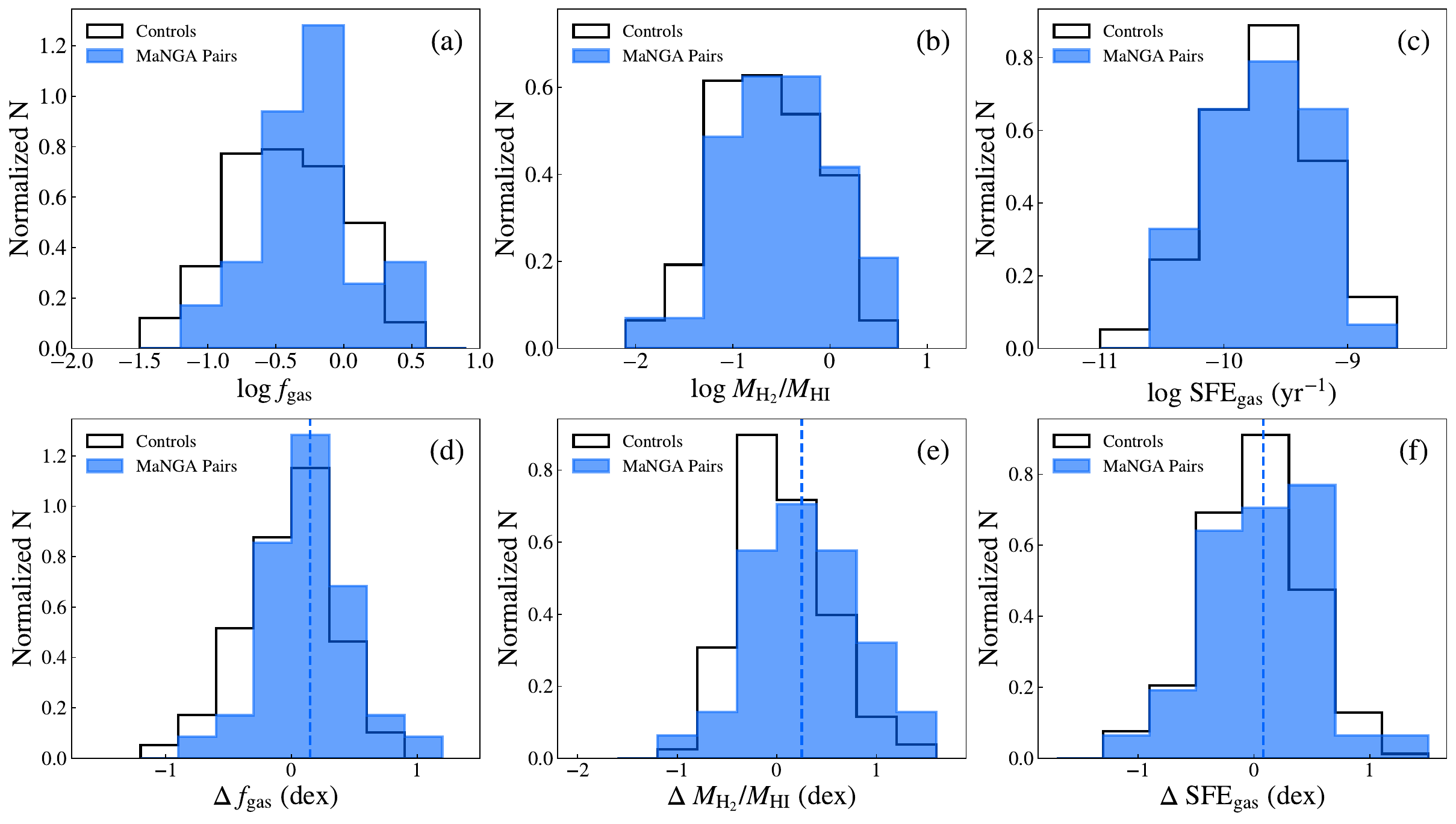}
	\caption{The upper panels show distributions of galaxy properties of the galaxy pair sample and the control sample, while the lower panels present the offset of these properties relative to the controls. The paired galaxies and control galaxies are plotted as filled and open histograms, respectively. The dashed lines represent the mean value of distributions for paired galaxies. (a) Distributions of the total gas fraction $f_{\rm gas}$. (b) Distributions of molecular-to atomic gas mass ratio \mhh$/\mhi$. (c) Distributions of the star formation efficiency of the total gas ${\rm SFE}_{\rm gas}$. (d) Distributions of $\Delta f_{\rm gas}$. (e) Distributions of $\Delta M_{\rm H_{2}}/\mhi$. (f) Distributions of $\Delta {\rm SFE}_{\rm gas}$.
	\label{fig:h2_full_pair_hih2_dlt}
	}
\end{figure*}

Utilizing the combination of molecular gas properties and \hi\ data from previous observations \citep{Stark2021,Yu2022}, we further derived and analyzed the molecular-to-atomic gas mass ratio (\mhh$/\mhi$), the total gas mass fraction ($f_{\rm gas}$), and the SFE of the total gas (${\rm SFE}_{\rm gas}$). The total gas mass is the sum of $\mhi$ and \mhh, and the $f_{\rm gas}$ is defined as
\begin{equation}
f_{\rm gas} = \frac{\mhi+M_{\rm H_2}}{M_{\star}}  . 
\label{eq:fgas}
\end{equation}
As shown in Figure \ref{fig:h2_full_pair_hih2_dlt}(a), we compare the distributions of $f_{\rm gas}$ for galaxies in pairs and controls. The $f_{\rm gas}$ of paired galaxies tends to be distributed at higher values, with a mean log $f_{\rm gas}$ of $-0.28\pm0.06$. In contrast, the average total gas mass fraction is $f_{\rm gas}=-0.44\pm0.03$. Based on the AD test result ($p\text{-value}=0.03$), the two distributions of $f_{\rm gas}$ could potentially be drawn from the same parent distribution. Figure \ref{fig:h2_full_pair_hih2_dlt}(b) presents the distributions of \mhh$/\mhi$ for the pair sample and the control sample. The pair sample has an average log(\mhh$/\mhi$) of $-0.38\pm0.10$, while the mean log(\mhh$/\mhi$) for the control sample is $-0.64\pm0.04$. An AD test suggests the two distributions may be drawn from the same distribution ($p\text{-value}=0.01$). The ${\rm SFE}_{\rm gas}$ is defined as
\begin{equation}
{\rm SFE_{ gas}\ [yr^{-1}]} = \frac{\rm SFR}{\mhi+M_{\rm H_2}}  . 
\label{eq:sfe_gas}
\end{equation}
As shown in Figure \ref{fig:h2_full_pair_hih2_dlt}(c), we present the distributions of ${\rm SFE}_{\rm gas}$ for the paired galaxies and isolated galaxies. The mean values of log ${\rm SFE}_{\rm gas}$ are $-9.62\pm0.07$ and $-9.67\pm0.03$ for the pair sample and the control sample, respectively. The AD test returned a $p\text{-value}$ of $>$0.25, indicating that the distributions of ${\rm SFE}_{\rm gas}$ in the galaxy pair sample and the control sample are probably drawn from the same parent population.

\subsection{Offset of Galaxy Properties}\label{subsec:offset}

\begin{figure*}[t]
	\includegraphics[width=1\textwidth]{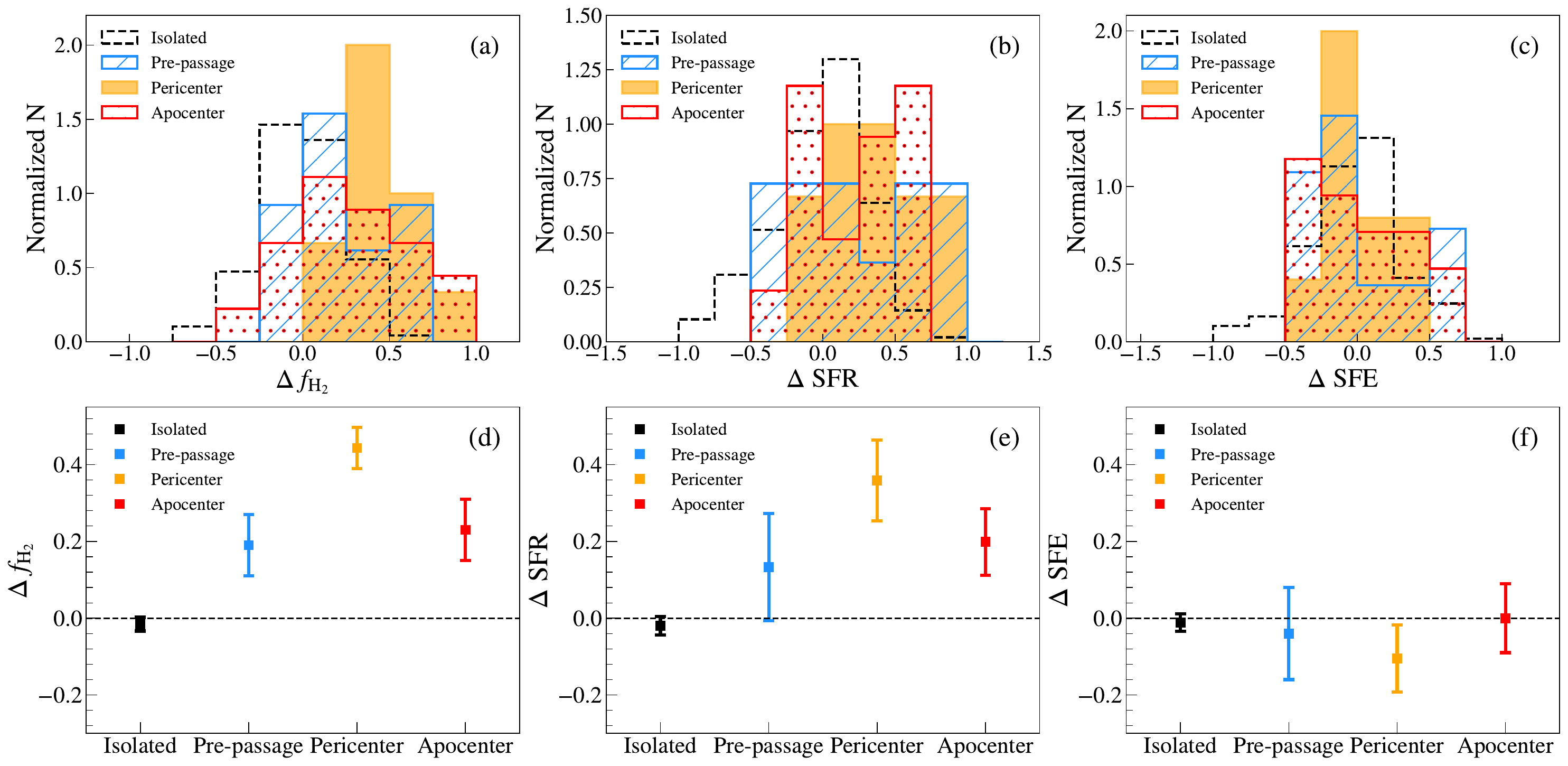}
	\caption{The upper panels show histograms of different galaxy property offsets. The lower panels show the corresponding mean values for the offset distributions. In each plot, paired galaxies at different merger stages and the control sample are marked with blue, orange, red, and black, respectively. In the lower panels, the mean value of each distribution is indicated by a square point, and the error bar represents the standard error of the mean for each distribution. (a) The distribution of $\Delta f_{\text{\hh}}$. (b) The distribution of $\Delta \text{SFR}$. (c) The distribution of $\Delta$SFE. (d) The mean value of $\Delta f_{\text{\hh}}$. (e) The mean value of $\Delta \text{SFR}$. (f) The mean value of $\Delta$SFE. 
	\label{fig:h2_offset}
	}
\end{figure*}

To reduce bias and fairly compare the SFR and gas properties of galaxies in pairs and controls, we follow the method of previous studies \citep{Ellison2018,Pan2018} to calculate the {\it offset} quantities based on a galaxy-by-galaxy matching process. For the galaxy pair sample, each galaxy is matched in stellar mass and redshift with at least five isolated galaxies in the control pool. We require the matched galaxies to meet the criteria of $|\Delta \text{log}\ (M_{\star}/M_{\odot})| < 0.2$ and $|\Delta z|< 0.01$ \citep{Yu2022}. If the minimum number of matched controls is not attained, the tolerances for stellar mass and redshift are increased by 0.1 dex and 0.005, respectively. Except for three sources requiring 2--3 iterations of matching, most paired galaxies were matched with at least five control galaxies in the first round of matching.

After matching stellar mass and redshift, the offset of molecular gas fraction, for example, is calculated as 
\begin{equation}
    \Delta f_{\text{\hh}} = \text{log}\ f_{\text{\hh,\ pair}} - \text{log}\ \text{median}(f_{\text{\hh,\ control}}) ,
\end{equation}
where the $\text{log}\ f_{\text{\hh,\ pair}}$ represents the molecular gas fraction of paired galaxy, and the $\text{log}\ \text{median}(f_{\text{\hh,\ control}})$ is the median \hh\ gas fraction of its matched control galaxies in the logarithmic scale. We apply the same procedure to compute the offsets of $\Delta \text{SFR}$, $\Delta$SFE, $\Delta f_{\text{gas}}$, $\Delta M_{\rm H_{2}}/\mhi$, and $\Delta \rm SFE_{\text{gas}}$ (Table \ref{tab:offsets}). Notice that offsets of these galaxy properties are calculated in the logarithmic scale. 

As shown in Figure \ref{fig:h2_full_pair_cs_dlt}(d), the mean \hh\ gas fraction offset ($\Delta f_{\text{\hh}}$) for paired galaxies is $0.28\pm0.05$ dex, indicating that the \fhh\ of the galaxy pair sample is significantly higher than that of the control sample. The AD test returns a $p\text{-value}$ of $<$0.001, suggesting the distributions of $\Delta f_{\text{\hh}}$ in paired galaxies and control galaxies are distinct distributions. The distributions of the SFR offset ($\Delta \text{SFR}$) are presented in Figure \ref{fig:h2_full_pair_cs_dlt}(e). Paired galaxies have a mean $\Delta \text{SFR} = 0.22\pm0.06$ dex, suggesting that the SFR of galaxies in pairs is enhanced by a factor of $\sim$1.7 compared to isolated galaxies at the significance level of $\sim$3.7$\sigma$. The AD test suggests a significant discrepancy between the distributions of $\Delta \text{SFR}$ in the pair sample and control sample ($p\text{-value}<0.001$). Figure \ref{fig:h2_full_pair_cs_dlt}(f) presents the distributions of $\Delta$SFE for the pair sample and the control sample. For the galaxy pair sample, the average value of $\Delta$SFE is $-0.04\pm0.06$, indicating that the SFE of paired galaxies is comparable to isolated galaxies. Based on the AD test ($p\text{-value}=0.21$), the distribution of $\Delta$SFE in paired galaxies is consistent with that in isolated galaxies.

As shown in Figure \ref{fig:h2_full_pair_hih2_dlt}(d), the total gas fraction offset ($\Delta f_{\rm gas}$) of the pair sample has a mean value of $0.15\pm0.06$, implying consistent $f_{\rm gas}$ in paired galaxies and isolated galaxies ($<$ 3$\sigma$). The AD test indicates that $\Delta f_{\rm gas}$) in two samples may be drawn from the same parent sample with a $p\text{-value}=0.01$. In Figure \ref{fig:h2_full_pair_hih2_dlt}(e), we present the distributions of $\Delta M_{\rm H_{2}}/\mhi$ for galaxies in pairs and controls. Paired galaxies have a mean $\Delta M_{\rm H_{2}}/\mhi$ of $0.25\pm 0.09$, suggesting the \mhh$/\mhi$ of the pair sample is not significantly different from that of the control sample within the error bar ($<$ 3$\sigma$). The $p\text{-value}$ of the AD test is 0.01, indicating the distributions of $\Delta M_{\rm H_{2}}/\mhi$ in two samples may be drawn from the same parent population. For $\Delta {\rm SFE}_{\rm gas}$ (Figure \ref{fig:h2_full_pair_hih2_dlt}(f)), both the mean values ($\Delta {\rm SFE}_{\rm gas}=0.06\pm0.07$) and AD test ($p\text{-value}>0.25$) suggest consistent distributions between the pair sample and the control sample. 

\subsection{Galaxy Properties along the Merger Sequence}\label{subsec:offset_stage}

\begin{figure*}[t]
	\includegraphics[width=1\textwidth]{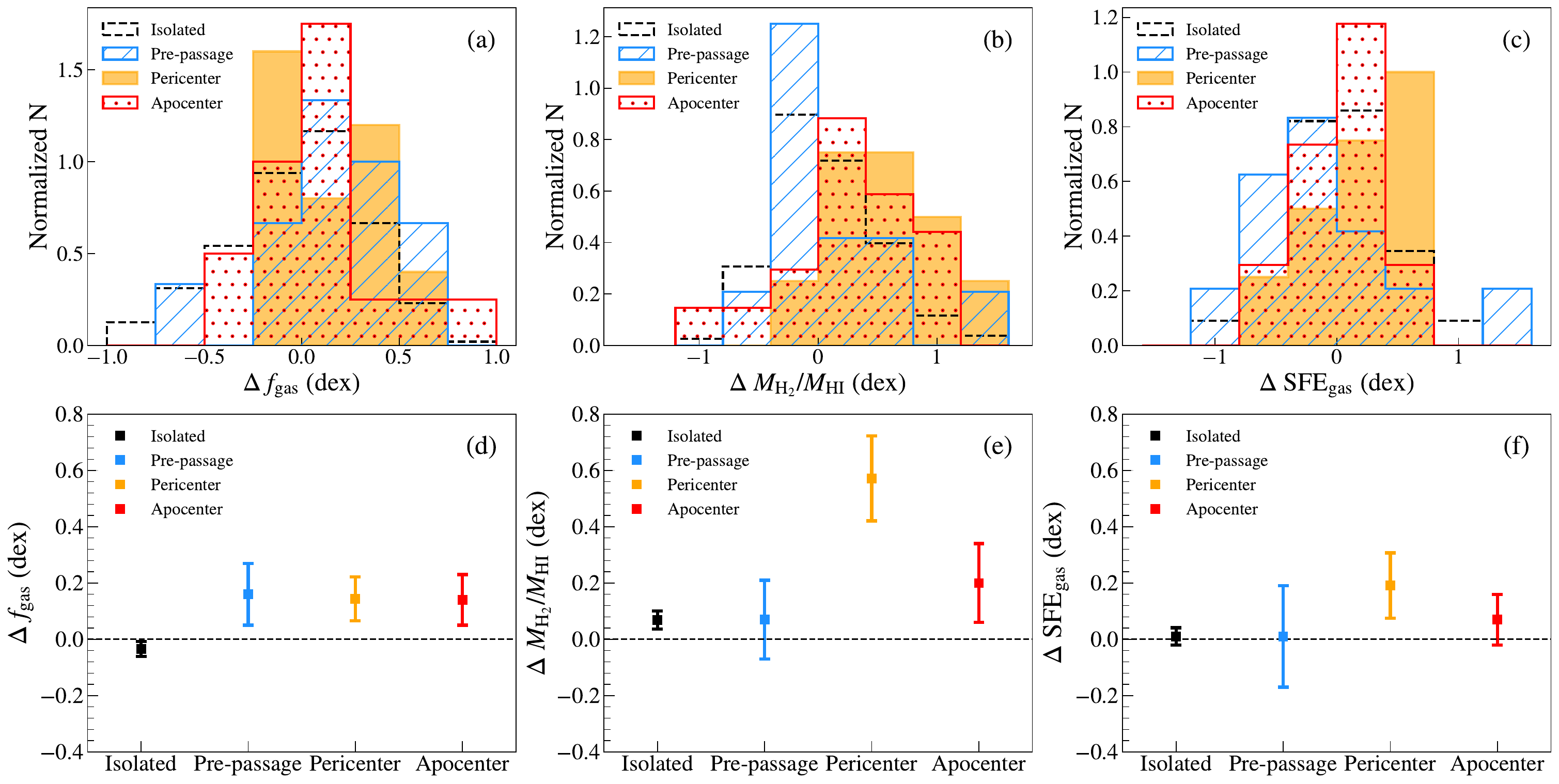}
	\caption{The upper panels show histograms of different galaxy property offsets. The lower panels show the corresponding mean values for the offset distributions. In each plot, paired galaxies at different merger stages and the control sample are marked with blue, orange, red, and black, respectively. In the lower panels, the mean value of each distribution is indicated by a square point, and the error bar represents the standard error of the mean for each distribution. (a) The distribution of $\Delta f_{\rm gas}$. (b) The distribution of $\Delta M_{\rm H_{2}}/\mhi$. (c) The distribution of $\Delta {\rm SFE}_{\rm gas}$. (d) The mean value of $\Delta f_{\rm gas}$. (e) The mean value of $\Delta M_{\rm H_{2}}/\mhi$. (f) The mean value of $\Delta {\rm SFE}_{\rm gas}$.
	\label{fig:hih2_offset}
	}
\end{figure*}

Based on the kinematic asymmetry and projected distance, we divided the galaxy pair sample into three subsamples at different merger stages. To further investigate the interplay between cold gas and galaxy interaction along the merger sequence, we compare the aforementioned gas properties and SFR between subsamples (Table \ref{tab:offsets}). Our main results are presented in Figures \ref{fig:h2_offset} and \ref{fig:hih2_offset}, in which the upper panels show histograms of each galaxy property offset for different samples, and the lower panels represent the corresponding mean values with errors for each distribution. The histograms of isolated galaxies and paired galaxies at the pre-passage, pericentric, and apocenter stages are plotted with black, blue, orange, and red bars, respectively. In the lower panels, the mean values of each distribution are indicated using the same color scheme as the histograms, and the error bar represents the error of the mean value for each distribution.

In Figure \ref{fig:h2_offset}(a) and (d), we plot the distributions of \hh\ gas fraction offsets for isolated galaxies and paired galaxies at different merger stages. Compared to the controls, the mean values of each distribution indicate a significant elevation in the \hh\ gas fraction of pairs at the pericenter stage by 0.44 $\pm$ 0.05 dex. The AD test also returns a significant discrepancy ($p\text{-value}<0.001$) of $\Delta f_{\text{\hh}}$ between pericenter-stage pairs and isolated galaxies. At the pre-passage stage, the \hh\ gas fraction of paired galaxies ($\Delta f_{\text{\hh}} = 0.19\pm0.08$ dex) is consistent with that of isolated galaxies considering the uncertainty ($<$3$\sigma$). The $p\text{-value}$ of AD test indicates the distributions of $\Delta f_{\text{\hh}}$ in two samples may be drawn from the same parent population ($p\text{-value}$=0.01). In contrast, the paired galaxies at the apocenter stage have $\Delta f_{\text{\hh}} = 0.23\pm0.08$ dex), suggesting the mean \fhh\ is not significantly different from isolated galaxies ($<$ 3$\sigma$). Based on the AD test, the distribution of $\Delta f_{\text{\hh}}$ in paired galaxies at the apocenter stage is different from that in isolated galaxies with $p\text{-value}<0.001$.

The offset of SFR reveals enhanced star formation of galaxy pairs during galaxy interactions, especially when pairs are at the pericenter stage. In Figure \ref{fig:h2_offset}(b) and (e), we plot the distributions of global SFR offsets for isolated galaxies and paired galaxies at different merger stages. Figure \ref{fig:h2_offset}(b) shows that the $\Delta \text{SFR}$ of galaxy pairs at all stages tends to be distributed with positive values, indicating enhanced SFR during galaxy interactions. As shown in Figure \ref{fig:h2_offset}(e), the paired galaxies during pericenter passage have an SFR enhancement by a factor of 2.3 with $\Delta \text{SFR} = 0.36\pm0.11$ dex ($\sim$3.3$\sigma$). The AD test suggests the $\Delta \text{SFR}$ in two samples are distinct distributions at the significance of $p\text{-value}<0.001$. In contrast, the mean $\Delta \text{SFR}$ value of paired galaxies during pre-passage is $0.13 \pm 0.14$ dex, consistent with isolated galaxies. The $p\text{-value}$ of AD test also supports the consistency of $\Delta \text{SFR}$ between two samples ($p\text{-value}=0.07$). Paired galaxies at the apocenter stage present consistent SFR ($0.20 \pm 0.09$ dex) compared to isolated galaxies. The AD test indicates the distributions of $\Delta \text{SFR}$ in the two samples are moderately different at the significance level of $p\text{-value}=0.003$.

In terms of SFE, our data suggest the SFE of paired galaxies at different stages is comparable to that of isolated galaxies. In Figure \ref{fig:h2_offset}(c) and (f), we plot the distribution of SFE offsets for isolated galaxies and paired galaxies along the merger sequence. As shown in Figure \ref{fig:h2_offset}(c), the distributions of $\Delta$SFE for pairs at different stages are mostly consistent with those of isolated galaxies within error bars. Our results in Figure \ref{fig:h2_offset}(f) indicate that the mean $\Delta \rm SFE$ for paired galaxies at the pre-passage stage is $-0.04 \pm 0.12$ dex, and the paired galaxies at pericenter passage have mean $\Delta \rm SFE= -0.10 \pm 0.09$ dex. The mean SFE offset for pairs at the apocenter stage is $\Delta \rm SFE= 0.00 \pm 0.09$ dex. For each subsample of paired galaxies at different stages, the AD test indicates consistent distributions of $\Delta$SFE compared to isolated galaxies ($p\text{-value}>0.25$). 

Combining with the \hi\ data, we find the total gas fraction of paired galaxies is comparable to that of isolated galaxies. In Figures \ref{fig:hih2_offset}(a) and (d), we present the distributions of $\Delta f_{\rm gas}$ for galaxies in controls and pairs at different merger stages. As shown in Figure \ref{fig:hih2_offset}(d), our results ($\Delta f_{\rm gas}=0.16\pm0.11$) suggest the $f_{\rm gas}$ of paired galaxies at the pre-passage stage is consistent with that of isolated galaxies with the significance of $<$ 3$\sigma$. The AD test returns a $p\text{-value}$ of 0.06, indicating the two distributions may be drawn from the same parent population. Galaxy pairs ($\Delta f_{\rm gas}=0.14\pm0.08$) during pericentric passage also present consistent $f_{\rm gas}$ with isolated galaxies. Based on the AD test ($p\text{-value}=0.22$), the distributions of $\Delta f_{\rm gas}$ in pericenter-stage pairs and isolated galaxies are consistent. Similarly, paired galaxies at the apocenter stage have the average $\Delta f_{\rm gas}=0.14\pm0.09$, suggesting the $f_{\rm gas}$ is comparable to that of isolated galaxies within the error bar. The AD test indicates the distribution of $\Delta f_{\rm gas}$ in galaxies approaching the apocenter is consistent with that in isolated galaxies with $p\text{-value}=0.09$.

The offset of \mhh$/\mhi$ suggests paired galaxies have various molecular-to-atomic gas ratios along the merger sequence, with pairs at the pericenter stage exhibiting increased \mhh$/\mhi$. We present the distributions of $\Delta M_{\rm H_{2}}/\mhi$ in Figures \ref{fig:hih2_offset}(b) and (e). As shown in Figure \ref{fig:hih2_offset}(e), the pre-passage galaxy pairs ($\Delta M_{\rm H_{2}}/\mhi=0.07\pm0.14$) show comparable \mhh$/\mhi$ to those of isolated galaxies, with a $p\text{-value}$ of $>0.25$ from the AD test. Paired galaxies during the pericentric passage have a mean $\Delta M_{\rm H_{2}}/\mhi=0.57\pm0.15$, suggesting an enhancement of $M_{\rm H_{2}}/\mhi$ by a factor of 3.7 with respect to isolated galaxies at the significance of $\sim$3.8$\sigma$. The AD test indicates a moderate difference ($p\text{-value}$=0.002) of $\Delta M_{\rm H_{2}}/\mhi$ distributions in pericenter-stage pairs and isolated galaxies. In contrast, galaxy pairs at the apocenter stage present a mean $\Delta M_{\rm H_{2}}/\mhi$ of $0.20\pm0.14$, indicating that the $M_{\rm H_{2}}/\mhi$ is consistent with that of isolated galaxies. The AD test also suggests the two distributions are not significantly different ($p\text{-value}=0.08$).

In Figures \ref{fig:hih2_offset}(c) and (f), we compare the distributions of the total gas fraction for galaxies in controls and pairs at different merger stages. As shown in Figure \ref{fig:hih2_offset}(f), paired galaxies during the pericentric passage present an average $\Delta {\rm SFE}_{\rm gas}=0.19\pm0.12$, indicating a consistent ${\rm SFE}_{\rm gas}$ with isolated galaxies. The AD test suggests the two distributions are likely drawn from the same parent population ($p\text{-value}=0.20$). In contrast, galaxy pairs at the pre-passage and the apocenter stages have comparable ${\rm SFE}_{\rm gas}$ to that of isolated galaxies, with the mean $\Delta {\rm SFE}_{\rm gas}$ of $0.01\pm0.18$ and $0.07\pm0.09$, respectively. For these two subsamples of paired galaxies, the AD test indicates consistent distributions of $\Delta {\rm SFE}_{\rm gas}$ compared to isolated galaxies ($p\text{-value}>0.25$).

\section{Discussions} \label{sec:discussions}

\begin{deluxetable*}{lrrcr}\setlength\tabcolsep{25pt}
\tablenum{4}
\centering
\tablewidth{0pt}
\caption{Mean offsets of galaxy properties for different subsamples \label{tab:offsets}}
\tablehead{
\colhead{Property offset} & \colhead{Pre-passage} & \colhead{Pericenter} & \colhead{Apocenter} & \colhead{Total}	}
%\decimalcolnumbers
\startdata
$\Delta f_{\rm H_{2}}$ & 0.19$\pm$0.09 & {\bf 0.44$\pm$0.05} & 0.23$\pm$0.08 & {\bf 0.28$\pm$0.05} \\
$\Delta \text{SFR}$ & 0.13$\pm$0.14 & {\bf 0.36$\pm$0.11} & 0.20$\pm$0.09 & {\bf 0.22$\pm$0.06} \\
$\Delta \rm SFE$ & $-$0.04$\pm$0.12 & $-$0.10$\pm$0.09 & 0.00$\pm$0.09 & $-$0.04$\pm$0.06 \\
$\Delta f_{\rm gas}$ & 0.16$\pm$0.11 & 0.14$\pm$0.08 & 0.14$\pm$0.09 & 0.15$\pm$0.06 \\
$\Delta M_{\rm H_{2}}/\mhi$ & 0.07$\pm$0.14 & {\bf 0.57$\pm$0.15} & 0.20$\pm$0.14 & 0.25$\pm$0.09\\
${\rm SFE}_{\rm gas}$ & 0.01$\pm$0.18 & 0.19$\pm$0.12 & 0.07$\pm$0.09 & 0.08$\pm$0.07\\
\enddata
\tablecomments{ Mean offsets of galaxy properties for which the difference $>$3$\sigma$ are shown in boldface. 
}
\end{deluxetable*}

\subsection{Gas Fractions and SFR at Different Merger Stages}

In this section, we further discuss the interplay between properties of the molecular gas and star formation along the merger sequence. By performing galaxy-by-galaxy matching between the paired galaxies and isolated galaxies, we derived the offset of \fhh, SFR, $f_{\rm gas}$, and \mhh$/\mhi$ for each subsample in Section \ref{subsec:offset}. When paired galaxies are at the pre-passage stage, the gravitational influence between member galaxies is not significant, so the gas properties and star formation of paired galaxies should be similar to those of isolated galaxies. Indeed, our observations of paired galaxies are consistent with this expectation, suggesting that the defined merger sequence is effective in the identification of pre-passage galaxy pairs.  

At the pericenter stage, our results reveal that the average SFR of paired galaxies is significantly enhanced by a factor of $\sim$2.3 with respect to isolated galaxies. In contrast, the enhancement of SFR in apocenter stage pairs is weaker and less significant than in paired galaxies at the pericenter stage. The strongest enhancement of SFR occurs only during the pericentric passage, consistent with previous studies of interacting galaxy pairs \citep[e.g.,][]{Barton2000,Ellison2008,Scudder2012,Patton2013}. In the meantime, the CO data suggest pairs at this stage exhibit a significantly elevated \fhh. The increased \fhh\ has also been found by previous CO observations of close galaxy pairs \citep{Combes1994,Casasola2004,Pan2018,Violino2018,Lisenfeld2019}, as well as recent FIRE-2 simulations \citep{Moreno2019}. The increase of \fhh\ in paired galaxies is in good agreement with our previous \hi\ observations \citep{Yu2022}, indicating that the marginal depletion of \hi\ gas in paired galaxies may originate from the accelerated formation of the molecular gas. During the pericentric passage, merger-induced external pressure may lead to the accelerated transition from atomic to molecular gas \citep{Braine1993,Elmegreen1993,Kaneko2017}, which consumes the \hi\ gas, thereby increasing the \fhh. In this scenario, the molecular-to-atomic gas ratio is theoretically predicted to be elevated, which is confirmed by our results at the significance of 3.8$\sigma$ (Figure \ref{fig:hih2_offset}(e)). The enhancement of \mhh$/\mhi$ has also been detected in previous observations, including the infrared luminous galaxies at the mid- and late-merger stages \citep{Mirabel1989} and galaxy pairs encountering close interactions \citep{Lisenfeld2019}. 

On the other hand, our data suggest SFR of pairs approaching the apocenter is consistent with that of isolated galaxies. The marginal difference in SFR aligns with previous observations indicating the enhancement of SFR declines as the separations become larger \citep{Scudder2012,Patton2013,Feng2020}. Recent simulations also indicate that the merger-induced star formation will be gradually alleviated when galaxy pairs are heading toward the apocenter after the close encounter \citep{Moreno2015,Moreno2019}. Similarly, the increase in \fhh\ becomes weaker with the significance $<$3$\sigma$ when paired galaxies are at the apocenter stage (Figure \ref{fig:h2_offset}(d)). Correspondingly, the mean \mhh$/\mhi$ of the apocenter-stage pairs is comparable to that of isolated galaxies, indicating that the accelerated transition from atomic to molecular gas may only occur when paired galaxies undergo strong interactions. Thus, the decrease of \fhh\ relative to pericenter-stage pairs could be explained by the consumption due to star formation and a decelerated transition from \hi\ to \hh\ gas.

The enhancement of \mhh$/\mhi$ at the pericenter stage may play a crucial role in the rise of \fhh. While the enhanced \mhh$/\mhi$ is in agreement with the findings of \cite{Lisenfeld2019}, early observations of interacting galaxies yielded a contrary result that the \mhh$/\mhi$ is not significantly different from that of the control sample \citep{Casasola2004}. The inconsistent results may be contributed by the mixed pair types and merger stages in the sample of \cite{Casasola2004}, as the significant enhancement of \mhh$/\mhi$ is detected in close S$+$S pairs with interaction signs \cite{Lisenfeld2019}, corresponding to the pairs at the pericenter stage in our sample.  

In terms of the total gas fraction, paired galaxies at all stages show comparable $f_{\rm gas}$ with respect to isolated galaxies. The consistency of $f_{\rm gas}$ between paired galaxies and isolated galaxies has also been revealed by previous observations \citep{Lisenfeld2019}.

\subsection{SFE at Different Merger Stages}

As demonstrated in Section \ref{subsec:offset}, we calculated the offset of SFE and ${\rm SFE}_{\rm gas}$ and investigated the variations of $\Delta$SFE and $\Delta {\rm SFE}_{\rm gas}$ along the merger sequence. Similar to gas fractions, paired galaxies at the pre-passage stage also exhibit comparable SFE and ${\rm SFE}_{\rm gas}$ compared to the controls. Previous investigations on SFE of the \hi\ gas (\sfe) have also indicated the consistency of the pre-passage pairs and isolated galaxies \citep{Yu2022}.

\begin{figure*}[t]
\centering
	\includegraphics[width=1\textwidth]{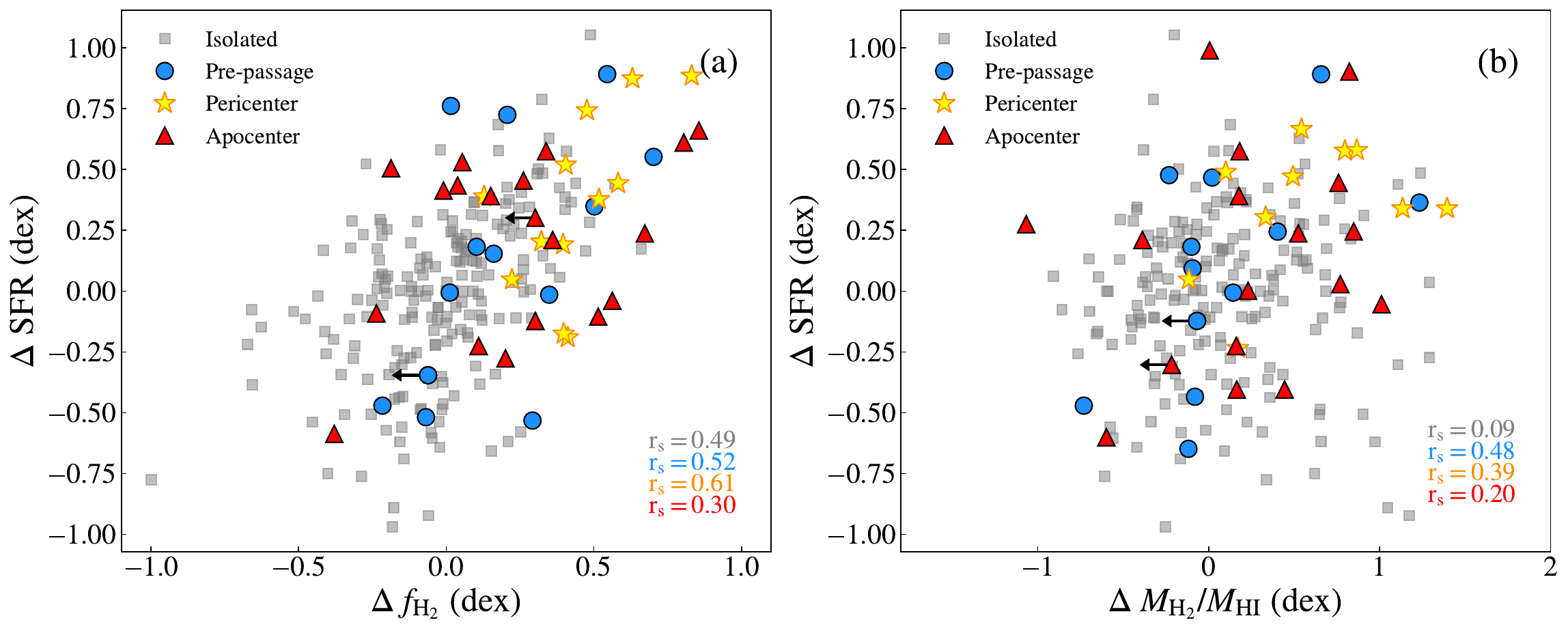}
	\caption{(a) $\Delta \text{SFR}$ as a function of $\Delta f_{\text{\hh}}$. (b) $\Delta \text{SFR}$ as a function of $\Delta M_{\rm H_{2}}/\mhi$. Paired galaxies at different merger stages and the control sample are marked with blue, orange, red, and black, respectively. In each plot, the Spearman coefficients of the relations in each subsample are labeled in the southeast corner.
		\label{fig:hih2_corre}
	}
\end{figure*}

In pericenter-stage pairs, the SFE is comparable to that of isolated galaxies (Figure \ref{fig:h2_offset}(f)), aligning with findings from previous observational studies \citep{Combes1994,Casasola2004,Lisenfeld2019,Li2023}. Similarly, galaxies in apocenter-stage pairs also exhibit consistent SFE. The observed consistency of integrated SFE between local interacting pairs and isolated systems is in line with recent simulations. \cite{Renaud2014} performed a suite of parsec-resolution simulations and found that the tidal interactions and inflows at the early stage are not strong enough to significantly increase the gas density, leading to a similar SFE as the isolated star-forming galaxies. However, the enhancement of SFE in close galaxy pairs has also been suggested by other previous CO observations \citep{Sofue1993,Michiyama2016,Pan2018,Violino2018}. Various factors may contribute to this discrepancy in integrated SFE. For example, as discussed by \cite{Pan2018}, the selection and comparison of galaxy pairs and control galaxies could introduce biased results if the global properties are not carefully controlled between samples. Furthermore, various types of interacting galaxy pairs may also contribute to the discrepancy. Taking into account the mass ratio, major-merger pairs tend to show a weakly enhanced SFE, while minor-merger pairs do not \citep{Pan2018}. \cite{Lisenfeld2019} suggest that S$+$S pairs present a higher SFE relative to S$+$E pairs. In addition, the enhancement of SFE at specific regions (e.g., the nuclear region) could be diluted in single-dish observations. A recent spatially resolved CO study of galaxy mergers has revealed the complexity and variations of \fhh\ and SFE among different individual mergers \citep{Thorp2022}, which demonstrates the importance of kpc-scale observations to study various SFR, \fhh\ and SFE in different regions of mergers. Therefore, spatially resolved CO observations for a large sample of interacting galaxies are required for understanding the fraction and specific regions of the enhanced SFE along the evolutionary sequence of interacting galaxies.

The ${\rm SFE}_{\rm gas}$ of paired galaxies at the pericenter stage is weakly enhanced compared to isolated galaxies, but the difference is marginal. In previous \hi\ observations, the \sfe\ is also found to be enhanced at the pericenter stage \citep{Yu2022}. The slight enhancement in ${\rm SFE}_{\rm gas}$ could be attributed to the \hi--\hh\ transition during the pericentric passage, consistent with the depletion of \hi\ gas detected in previous work \citep{Yu2022}. In contrast, paired galaxies at the apocenter stage pairs show consistent ${\rm SFE}_{\rm gas}$ compared to isolated galaxies.

\subsection{What Drives the Enhancement of SFR in Paired Galaxies?}

To investigate what galaxy property drives the enhancement of SFR during the merger, we analyzed the correlation between the enhanced SFR and gas properties in the full pair sample and subsamples. Using the derived offset of galaxy properties in Section \ref{subsec:offset}, we calculated the Spearman's rank order coefficients among $\Delta f_{\rm H_{2}}$, $\Delta M_{\rm H_{2}}/\mhi$, $\Delta f_{\rm gas}$, and $\Delta \text{SFR}$. During the analysis, we evaluate the strength of correlation using Spearman's rank order coefficient ($r_s$; $-1 \leqslant r_s \leqslant  1$), with $|r_s| \geqslant 0.6$ or $0.3 < |r_s| < 0.6$ regarded as a tight or weak correlation, and  $|r_s| \leqslant 0.3$ suggests no correlation. The $p\text{-value}$ of Spearman's rank order describes the significance of each correlation. In the following analysis, we adopt a $p\text{-value}<0.01$ as the significance level to reject the null hypothesis that there is no correlation between parameters.

As shown in Table \ref{tab:correlation}, we list Spearman's rank order coefficient $r_s$ with the p-value for each correlation in the control sample and subsamples of pairs. Our results show that both $\Delta f_{\rm H_{2}}$ and $\Delta M_{\rm H_{2}}/\mhi$ have positive correlations with $\Delta \text{SFR}$. In the full pair sample, $\Delta f_{\rm H_{2}}$ presents a weak correlation with $\Delta \text{SFR}$ $ (r_s=0.44,\ p\text{-value}=3.52\times10^{-3})$. $\Delta M_{\rm H_{2}}/\mhi$ is also weakly correlated to $\Delta \text{SFR}$ $ (r_s=0.43,\ p\text{-value}=5.96\times10^{-3})$. Contrastingly, there is no correlation between $\Delta f_{\rm gas}$ and $\Delta \text{SFR}$ $ (r_s=-0.17,\ p\text{-value}=0.31)$. For isolated galaxies, we find there is a weak correlation $ (r_s=0.49)$ between $\Delta f_{\rm H_{2}}$ and $\Delta \text{SFR}$ at the significance level of $p\text{-value}=6.58\times10^{-13}$. In terms of $\Delta M_{\rm H_{2}}/\mhi$ versus $\Delta \text{SFR}$, isolated galaxies present no correlation. In contrast, the $\Delta f_{\rm gas}$ of isolated galaxies exhibits no correlation with  $\Delta \text{SFR}$ ($r_s=0.24,\ p\text{-value}=1.94\times10^{-3}$).

\begin{deluxetable*}{lccr}\setlength\tabcolsep{27pt}
\tablenum{5}
\centering
\tablewidth{0pt}
\caption{Spearman's Rank Order Coefficient and the Significance \label{tab:correlation}}
\tablehead{
\colhead{Sample} & \colhead{$\Delta f_{\rm H_{2}}$ vs. $\Delta \text{SFR}$} & \colhead{$\Delta M_{\rm H_{2}}/\mhi$ vs. $\Delta \text{SFR}$} & \colhead{$\Delta f_{\rm gas}$ vs. $\Delta \text{SFR}$}	}
\decimalcolnumbers
\startdata
Pre-passage & 0.52 (0.07) & 0.48 (0.11) & $-$0.17 (0.60) \\
Pericenter & 0.61 (0.04) & 0.39 (0.27) & 0.13 (0.73) \\
Apocenter & 0.30 (0.23) & 0.20 (0.44) & $-$0.13 (0.62) \\
All pairs & {\bf 0.44} (3.52$\times10^{-3}$) & {\bf 0.43} (5.96$\times10^{-3}$) & $-$0.17 (0.31) \\
Isolated galaxies & {\bf 0.49} (6.58$\times10^{-13}$) & 0.09 (0.19) & 0.24 (1.94$\times10^{-3}$)\\
\enddata
\tablecomments{Spearman's rank order coefficient $r_s$ with the $p\text{-value}$ as significance in the parentheses. The coefficients of strong or weak correlations with $p\text{-value}<0.01$ are printed in boldface. The columns are (1) subsamples of paired galaxies at different stages, the full pair sample, and isolated galaxies as the control sample. (2) Spearman's rank order coefficient and significance of $\Delta f_{\rm H_{2}}$ versus $\Delta \text{SFR}$. (3) Spearman's rank order coefficient and significance of $\Delta M_{\rm H_{2}}/\mhi$ versus $\Delta \text{SFR}$. (4) Spearman's rank order coefficient and significance of $\Delta f_{\rm gas}$ versus $\Delta \text{SFR}$.
}
\end{deluxetable*}

To compare correlations in different merger stages, we plot $\Delta \text{SFR}$ as a function of $\Delta f_{\rm H_{2}}$ and a function of $\Delta M_{\rm H_{2}}/\mhi$ in Figure \ref{fig:hih2_corre}. In each panel, isolated galaxies and paired galaxies at different stages are represented by gray squares, blue circles, golden stars, and red triangles, respectively. Regarding $\Delta f_{\rm H_{2}}$ versus $\Delta \text{SFR}$, paired galaxies at the pericenter stage exhibit a hint of a strong correlation $(r_s=0.61)$, but the significance (see Table \ref{tab:correlation}, $p\text{-value}=0.04$) cannot reject the null hypothesis of no correlation. As shown in Figure \ref{fig:hih2_corre} and Table \ref{tab:correlation}, although the $r_s$ values show a hint of a weak correlation between $\Delta M_{\rm H_{2}}/\mhi$ and $\Delta \text{SFR}$ in paired galaxies at the pre-passage and pericenter stages, the $p\text{-values}$ suggest $\sim$11$-$27\% chance of no correlation. The $\Delta f_{\rm gas}$ of paired galaxies at all stages show no correlation with $\Delta \text{SFR}$. Given that the $p\text{-value}$ could be affected by the low numbers in the subsamples, a further correlation test with a larger sample is required to study any potential correlations.

Although the correlation tests between subsamples suggest marginal significance, our results in Section \ref{subsec:offset_stage} indicate that the enhancement of SFR is mainly driven by the elevation of \fhh\ in the pericenter stage, which is consistent with previous CO observations of galaxy pairs \citep{Pan2018,Lisenfeld2019}. Based on our results, the enhanced \fhh\ and SFR can be explained by the merger-induced external pressure that enhanced $M_{\rm H_{2}}/\mhi$ \citep{Braine1993,Elmegreen1993,Kaneko2017}, which has been detected in the pericenter stage (Figure \ref{fig:hih2_offset}(e)). When paired galaxies are approaching the pericenter, the shock-triggered external pressure accelerates the transition from atomic gas to molecular gas, resulting in the increment of \fhh\ and the further enhancement of SFR. However, we cannot rule out the contribution of enhanced SFE in specific regions during these stages, since the enhancement could be diluted by the unresolved single-dish observations. 

A recent study performed spatially resolved CO observations of galaxy mergers and found that \fhh\ and SFE have comparable contributions in the enhanced star formation, with the driving mechanisms varying among individual galaxies \citep{Thorp2022}. Global star formation of individual mergers can be driven by either the molecular gas fuel or the efficiency of forming stars, and some mergers are dominated by the fuel- and efficiency-driven star formation equally \citep{Thorp2022}. These authors claimed to find a trend that paired galaxies are more likely to have SFE enhancement than mergers at coalescence, and they proposed a scenario where efficiency-driven star formation becomes less important toward coalescence. This contradicts our results and those of \cite{Lisenfeld2019}, both found no SFE enhancement for paired galaxies. Also, ULIRGs, which are mostly mergers at coalescence, have very strong SFE enhancement \citep{Solomon1988,Mirabel1989}, probably because in a ULIRG (e.g., Arp 220), the molecular gas may be compressed into a central gas disk with a much higher dense gas ($n >10^4\,{\rm cm}^{-3}$) fraction than in normal giant molecular clouds, resulting in very high SFE \citep{Scoville1997,Scoville2017}. On the other hand, for paired galaxies, the effect of interaction on SFE can be very complex. \cite{Xu2021} resolved the CO(1$-$0) emission in two close major-merger pairs, Arp 238 (S+S) and Arp 142 (S+E). They found significant differences in local SFE among the two systems. The SFR in Arp 238 is concentrated in the two nuclei, with an SFE enhancement of  $\sim$0.7 dex. On the other hand, the disk of the star-forming galaxy of Arp 142 has an SFE $\sim$0.4 dex below the mean SFE of control galaxies. These authors attribute this difference to the very different interaction orbits: while the low-speed coplanar orbit of Arp 238 results in strong tidal torque, which triggers nuclear starbursts, the high-speed head-on collision in Arp 142 produces a density wave shock that suppresses the SFR in the system.

In summary, our results indicate that the elevation of \fhh\ can drive the enhancement of global SFR in paired galaxies at the pericenter stages, while the contribution of enhanced SFE in specific regions cannot be ruled out through current unresolved observations. Although our results suggest the contribution of \fhh, \mhh$/\mhi$, and SFE in the SFR enhancement may vary along the merger sequence, spatially resolved CO observations of a large sample are required to pinpoint the driving mechanisms and their variations during the evolution of the merger-induced star formation. A caveat should be noted that all these correlation analyses for the subsamples suffer from low number statistics and require further examination with larger samples. 

\section{Summary}\label{sec:Summary}

We investigate the interplay between molecular gas, atomic gas, and star formation of galaxies along the merger sequence, using a sample of 43 paired galaxies selected from the MaNGA survey. The merger stage is defined with the combination of the projected separation and kinematic asymmetry derived from the velocity field of \ha\ gas. We obtained the molecular gas properties from new CO observations with JCMT and IRAM 30\,m telescope, as well as data from the MASCOT survey. By combining these data with ancillary \hi\ data from FAST and HI-MaNGA, we study the molecular gas fraction (\fhh), molecular-to-atomic gas ratio (\mhh$/\mhi$), SFE, the total gas fraction ($f_{\rm gas}$), and the SFE of total gas (${\rm SFE}_{\rm gas}$) in paired galaxies at different merger stages, with isolated galaxies selected from the xCOLD GASS survey as the control sample. The main conclusions are as follows:
\begin{enumerate}
\setlength{\itemsep}{0pt}
\setlength{\parsep}{0pt}
\setlength{\parskip}{0pt}
	\item For the full pair sample, our results suggest the \fhh\ of paired galaxies is significantly increased compared to that of isolated galaxies, while the SFE is comparable. In contrast, We do not find statistically significant differences in $f_{\rm gas}$, \mhh$/\mhi$, and ${\rm SFE}_{\rm gas}$ between paired galaxies and isolated galaxies. Additionally, the correlation analyses suggest that the enhancement of SFR is weakly correlated to the elevation of \fhh\ and \mhh$/\mhi$ in paired galaxies, implying galaxy interactions may impact the transition from \hi\ to \hh\ gas and alter star formation.
	\item We find the molecular and atomic gas fractions in galaxy pairs vary along the merger sequence. When paired galaxies are at the pre-passage stage, the \fhh, $f_{\rm gas}$, and \mhh$/\mhi$ are consistent with the isolated galaxies. For galaxy pairs during the pericentric passage, the \fhh\ and \mhh$/\mhi$ are significantly elevated, while the $f_{\rm gas}$ has no difference with that of isolated galaxies. At the apocenter stage, the \fhh, \mhh$/\mhi$, and $f_{\rm gas}$ of paired galaxies are comparable to those of isolated galaxies.
    \item We find the SFE and ${\rm SFE}_{\rm gas}$ of paired galaxies have no significant variations along the merger sequence. For paired galaxies at different merger stages, the SFE and ${\rm SFE}_{\rm gas}$ are consistent with those of isolated galaxies.
    \item Our results indicate that the elevation of \fhh\ can drive the enhancement of global SFR in paired galaxies at the pericenter stage, while the contribution of enhanced SFE in specific regions cannot be ruled out through current unresolved observations. As for paired galaxies in different merger stages, further investigations with larger samples are required to test any correlations between the enhancements of SFR and gas properties.
\end{enumerate}

Our results indicate the enhancement of star formation in major-merger galaxy pairs is mainly driven by the elevation of the molecular gas fraction. The significant increase in molecular gas fraction occurs at the pericenter stage, indicating the accelerated transition from atomic gas to molecular gas due to external pressure induced by strong tidal forces after close interactions. However, the contribution of enhanced SFE in specific regions cannot be ruled out through current unresolved observations. Furthermore, the effects due to different orbital parameters of the interacting galaxies require further investigation. In future work, it is necessary to conduct spatially resolved observations of both \hi\ and CO lines for a large galaxy merger sample that spans a wide range of merger stages, allowing us to extend our understanding and constraints on where and how the SFR, \fhh, \mhh$/\mhi$, SFE vary along the evolutionary sequence. 
%% The "ht!" tells LaTeX to put the figure "here" first, at the "top" next
%% and to override the normal way of calculating a float position

\begin{acknowledgments}{\small
This work is supported by the National Natural Science Foundation of China (NSFC) under Nos. 11890692, 12133008, and 12221003. We acknowledge the science research grants from the China Manned Space Project with No. CMS-CSST-2021-A04. S.F. acknowledges support from the NSFC under No. 12103017 and the Natural Science Foundation of Hebei Province (No. A2021205001). X.-J.J. acknowledges support from the NSFC under No. 12373026. U.L. acknowledges support from the research project PID2020-114414GB-I00 financed by MCIN/AEI/10.13039/501100011033, and the Junta de Andaluc\'ia (Spain) grant FQM108.

The James Clerk Maxwell Telescope is operated by the East Asian Observatory on behalf of The National Astronomical Observatory of Japan; Academia Sinica Institute of Astronomy and Astrophysics; the Korea Astronomy and Space Science Institute; the National Astronomical Research Institute of Thailand; Center for Astronomical Mega-Science (as well as the National Key R\&D Program of China with No. 2017YFA0402700). Additional funding support is provided by the Science and Technology Facilities Council of the United Kingdom and participating universities and organizations in the United Kingdom and Canada. The JCMT data used in this work is obtained under the program M21BP051. N$\rm \overline{a}$makanui was constructed and funded by ASIAA in Taiwan, with funding for the mixers provided by ASIAA and at 230GHz by EAO. The N$\rm \overline{a}$makanui instrument is a backup receiver for the GLT. The authors wish to recognize and acknowledge the very significant cultural role and reverence that the summit of Maunakea has always had within the indigenous Hawaiian community.  We are most fortunate to have the opportunity to conduct observations from this mountain.

This work is based on observations carried out under project numbers E01-21 and 029-22 with the IRAM 30\,m telescope. IRAM is supported by INSU/CNRS (France), MPG (Germany), and IGN (Spain). We acknowledge the staff at IRAM for their help during the observations.

SDSS-IV is managed by the Astrophysical Research Consortium for the Participating Institutions of the SDSS Collaboration including the Brazilian Participation Group, the Carnegie Institution for Science, Carnegie Mellon University, the Chilean Participation Group, the French Participation Group, Harvard-Smithsonian Center for Astrophysics, Instituto de Astrof{\'i}sica de Canarias, The Johns Hopkins University, Kavli Institute for the Physics and Mathematics of the Universe (IPMU)/University of Tokyo, the Korean Participation Group, Lawrence Berkeley National Laboratory, Leibniz Institut f{\"u}r Astrophysik Potsdam (AIP), Max-Planck-Institut f{\"u}r Astronomie (MPIA Heidelberg), Max-Planck-Institut f{\"u}r Astrophysik (MPA Garching), Max-Planck-Institut f{\"u}r Extraterrestrische Physik (MPE), National Astronomical Observatories of China, New Mexico State University, New York University, University of Notre Dame, Observat{\'a}rio Nacional/MCTI, The Ohio State University, Pennsylvania State University, Shanghai Astronomical Observatory, United Kingdom Participation Group, Universidad Nacional Aut{\'o}noma de M{\'e}xico, University of Arizona, University of Colorado Boulder, University of Oxford, University of Portsmouth, University of Utah, University of Virginia, University of Washington, University of Wisconsin, Vanderbilt University, and Yale University.

}
\end{acknowledgments}


\vspace{5mm}
\facilities{JCMT; IRAM 30m telescope}


\software{ Starlink \citep{Currie2014}, GILDAS \citep{Gildas2013}, Astropy \citep{Astropy2013}, SciPy \citep{Virtanen2020}.
          }


%\newpage
\appendix
\section{CO Spectra}
We present the calibrated and rebinned CO spectra from JCMT observations and IRAM 30\,m observations as follows. The data can be obtained from the corresponding author upon reasonable request.

\bibliography{ref}{}

@ARTICLE{DiMatteo2007,
       author = {{Di Matteo}, P. and {Combes}, F. and {Melchior}, A. -L. and {Semelin}, B.},
        title = "{Star formation efficiency in galaxy interactions and mergers: a statistical study}",
      journal = {\aap},
         year = 2007,
        month = jun,
       volume = {468},
       number = {1},
        pages = {61-81},
          doi = {10.1051/0004-6361:20066959},
archivePrefix = {arXiv},
       eprint = {astro-ph/0703212},
 primaryClass = {astro-ph},
       adsurl = {https://ui.adsabs.harvard.edu/abs/2007A&A...468...61D}
}

@ARTICLE{Cox2008,
       author = {{Cox}, T.~J. and {Jonsson}, Patrik and {Somerville}, Rachel S. and {Primack}, Joel R. and {Dekel}, Avishai},
        title = "{The effect of galaxy mass ratio on merger-driven starbursts}",
      journal = {\mnras},
         year = 2008,
        month = feb,
       volume = {384},
       number = {1},
        pages = {386-409},
          doi = {10.1111/j.1365-2966.2007.12730.x},
archivePrefix = {arXiv},
       eprint = {0709.3511},
 primaryClass = {astro-ph},
       adsurl = {https://ui.adsabs.harvard.edu/abs/2008MNRAS.384..386C}
}

@ARTICLE{Ellison2008,
       author = {{Ellison}, Sara L. and {Patton}, David R. and {Simard}, Luc and {McConnachie}, Alan W.},
        title = "{Galaxy Pairs in the Sloan Digital Sky Survey. I. Star Formation, Active Galactic Nucleus Fraction, and the Mass-Metallicity Relation}",
      journal = {\aj},
         year = 2008,
        month = may,
       volume = {135},
       number = {5},
        pages = {1877-1899},
          doi = {10.1088/0004-6256/135/5/1877},
archivePrefix = {arXiv},
       eprint = {0803.0161},
 primaryClass = {astro-ph},
       adsurl = {https://ui.adsabs.harvard.edu/abs/2008AJ....135.1877E}
}

@ARTICLE{Satyapal2014,
       author = {{Satyapal}, Shobita and {Ellison}, Sara L. and {McAlpine}, William and {Hickox}, Ryan C. and {Patton}, David R. and {Mendel}, J. Trevor},
        title = "{Galaxy pairs in the Sloan Digital Sky Survey - IX. Merger-induced AGN activity as traced by the Wide-field Infrared Survey Explorer}",
      journal = {\mnras},
         year = 2014,
        month = jun,
       volume = {441},
       number = {2},
        pages = {1297-1304},
          doi = {10.1093/mnras/stu650},
archivePrefix = {arXiv},
       eprint = {1403.7531},
 primaryClass = {astro-ph.GA},
       adsurl = {https://ui.adsabs.harvard.edu/abs/2014MNRAS.441.1297S}
}

@ARTICLE{Moreno2015,
       author = {{Moreno}, Jorge and {Torrey}, Paul and {Ellison}, Sara L. and {Patton}, David R. and {Bluck}, Asa F.~L. and {Bansal}, Gunjan and {Hernquist}, Lars},
        title = "{Mapping galaxy encounters in numerical simulations: the spatial extent of induced star formation}",
      journal = {\mnras},
         year = 2015,
        month = apr,
       volume = {448},
       number = {2},
        pages = {1107-1117},
          doi = {10.1093/mnras/stv094},
archivePrefix = {arXiv},
       eprint = {1501.03573},
 primaryClass = {astro-ph.GA},
       adsurl = {https://ui.adsabs.harvard.edu/abs/2015MNRAS.448.1107M}
}

@ARTICLE{Hani2018,
       author = {{Hani}, Maan H. and {Sparre}, Martin and {Ellison}, Sara L. and {Torrey}, Paul and {Vogelsberger}, Mark},
        title = "{Galaxy mergers moulding the circum-galactic medium - I. The impact of a major merger}",
      journal = {\mnras},
         year = 2018,
        month = mar,
       volume = {475},
       number = {1},
        pages = {1160-1176},
          doi = {10.1093/mnras/stx3252},
archivePrefix = {arXiv},
       eprint = {1801.06183},
 primaryClass = {astro-ph.GA},
       adsurl = {https://ui.adsabs.harvard.edu/abs/2018MNRAS.475.1160H}
}

@ARTICLE{Combes1994,
       author = {{Combes}, F. and {Prugniel}, P. and {Rampazzo}, R. and {Sulentic}, J.~W.},
        title = "{CO in paired galaxies : star formation induced by gas flow.}",
      journal = {\aap},
         year = 1994,
        month = jan,
       volume = {281},
        pages = {725-736},
       adsurl = {https://ui.adsabs.harvard.edu/abs/1994A&A...281..725C}
}

@ARTICLE{Casasola2004,
       author = {{Casasola}, V. and {Bettoni}, D. and {Galletta}, G.},
        title = "{The gas content of peculiar galaxies: Strongly interacting systems}",
      journal = {\aap},
         year = 2004,
        month = aug,
       volume = {422},
        pages = {941-950},
          doi = {10.1051/0004-6361:20040283},
archivePrefix = {arXiv},
       eprint = {astro-ph/0405112},
 primaryClass = {astro-ph},
       adsurl = {https://ui.adsabs.harvard.edu/abs/2004A&A...422..941C}
}

@ARTICLE{Solomon1988,
       author = {{Solomon}, P.~M. and {Sage}, L.~J.},
        title = "{Star-Formation Rates, Molecular Clouds, and the Origin of the Far-Infrared Luminosity of Isolated and Interacting Galaxies}",
      journal = {\apj},
         year = 1988,
        month = nov,
       volume = {334},
        pages = {613},
          doi = {10.1086/166865},
       adsurl = {https://ui.adsabs.harvard.edu/abs/1988ApJ...334..613S}
}

@ARTICLE{Sofue1993,
       author = {{Sofue}, Yoshiaki and {Wakamatsu}, Ken-Ichi and {Taniguchi}, Yoshiaki and {Nakai}, Naomasa},
        title = "{CO Observations of Arp's Interacting Galaxies}",
      journal = {\pasj},
         year = 1993,
        month = feb,
       volume = {45},
        pages = {43-55},
       adsurl = {https://ui.adsabs.harvard.edu/abs/1993PASJ...45...43S}
}

@ARTICLE{Michiyama2016,
       author = {{Michiyama}, Tomonari and {Iono}, Daisuke and {Nakanishi}, Kouichiro and {Ueda}, Junko and {Saito}, Toshiki and {Ando}, Misaki and {Kaneko}, Hiroyuki and {Yamashita}, Takuji and {Matsuda}, Yuichi and {Hatsukade}, Bunyo and {Kikuchi}, Kenichi and {Komugi}, Shinya and {Muto}, Takayuki},
        title = "{Investigating the relation between CO (3-2) and far-infrared luminosities for nearby merging galaxies using ASTE}",
      journal = {\pasj},
         year = 2016,
        month = dec,
       volume = {68},
       number = {6},
          eid = {96},
        pages = {96},
          doi = {10.1093/pasj/psw087},
archivePrefix = {arXiv},
       eprint = {1608.05075},
 primaryClass = {astro-ph.GA},
       adsurl = {https://ui.adsabs.harvard.edu/abs/2016PASJ...68...96M}
}

@ARTICLE{Pan2018,
       author = {{Pan}, Hsi-An and {Lin}, Lihwai and {Hsieh}, Bau-Ching and {Xiao}, Ting and {Gao}, Yang and {Ellison}, Sara L. and {Scudder}, Jillian M. and {Barrera-Ballesteros}, Jorge and {Yuan}, Fangting and {Saintonge}, Am{\'e}lie and {Wilson}, Christine D. and {Hwang}, Ho Seong and {De Looze}, Ilse and {Gao}, Yu and {Ho}, Luis C. and {Brinks}, Elias and {Mok}, Angus and {Brown}, Toby and {Davis}, Timothy A. and {Williams}, Thomas G. and {Chung}, Aeree and {Parsons}, Harriet and {Bureau}, Martin and {Sargent}, Mark T. and {Chung}, Eun Jung and {Kim}, Eunbin and {Liu}, Tie and {Micha{\l}owski}, Micha{\l} J. and {Tosaki}, Tomoka},
        title = "{The Effect of Galaxy Interactions on Molecular Gas Properties}",
      journal = {\apj},
         year = 2018,
        month = dec,
       volume = {868},
       number = {2},
          eid = {132},
        pages = {132},
          doi = {10.3847/1538-4357/aaeb92},
archivePrefix = {arXiv},
       eprint = {1810.10162},
 primaryClass = {astro-ph.GA},
       adsurl = {https://ui.adsabs.harvard.edu/abs/2018ApJ...868..132P}
}

@ARTICLE{Violino2018,
       author = {{Violino}, Giulio and {Ellison}, Sara L. and {Sargent}, Mark and {Coppin}, Kristen E.~K. and {Scudder}, Jillian M. and {Mendel}, Trevor J. and {Saintonge}, Amelie},
        title = "{Galaxy pairs in the SDSS - XIII. The connection between enhanced star formation and molecular gas properties in galaxy mergers}",
      journal = {\mnras},
         year = 2018,
        month = may,
       volume = {476},
       number = {2},
        pages = {2591-2604},
          doi = {10.1093/mnras/sty345},
archivePrefix = {arXiv},
       eprint = {1802.02166},
 primaryClass = {astro-ph.GA},
       adsurl = {https://ui.adsabs.harvard.edu/abs/2018MNRAS.476.2591V}
}

@ARTICLE{Soares2007,
       author = {{Soares}, D.~S.~L.},
        title = "{The Identification of Physical Close Galaxy Pairs}",
      journal = {\aj},
         year = 2007,
        month = jul,
       volume = {134},
       number = {1},
        pages = {71-76},
          doi = {10.1086/518240},
archivePrefix = {arXiv},
       eprint = {astro-ph/0703552},
 primaryClass = {astro-ph},
       adsurl = {https://ui.adsabs.harvard.edu/abs/2007AJ....134...71S}
}

@ARTICLE{Torrey2012,
       author = {{Torrey}, Paul and {Cox}, T.~J. and {Kewley}, Lisa and {Hernquist}, Lars},
        title = "{The Metallicity Evolution of Interacting Galaxies}",
      journal = {\apj},
         year = 2012,
        month = feb,
       volume = {746},
       number = {1},
          eid = {108},
        pages = {108},
          doi = {10.1088/0004-637X/746/1/108},
archivePrefix = {arXiv},
       eprint = {1107.0001},
 primaryClass = {astro-ph.GA},
       adsurl = {https://ui.adsabs.harvard.edu/abs/2012ApJ...746..108T}
}

@ARTICLE{Bolatto2013,
       author = {{Bolatto}, Alberto D. and {Wolfire}, Mark and {Leroy}, Adam K.},
        title = "{The CO-to-H$_{2}$ Conversion Factor}",
      journal = {\araa},
         year = 2013,
        month = aug,
       volume = {51},
       number = {1},
        pages = {207-268},
          doi = {10.1146/annurev-astro-082812-140944},
archivePrefix = {arXiv},
       eprint = {1301.3498},
 primaryClass = {astro-ph.GA},
       adsurl = {https://ui.adsabs.harvard.edu/abs/2013ARA&A..51..207B}
}

@ARTICLE{Bundy2015,
       author = {{Bundy}, Kevin and {Bershady}, Matthew A. and {Law}, David R. and
         {Yan}, Renbin and {Drory}, Niv and {MacDonald}, Nicholas and
         {Wake}, David A. and {Cherinka}, Brian and
         {S{\'a}nchez-Gallego}, Jos{\'e} R. and {Weijmans}, Anne-Marie and
         {Thomas}, Daniel and {Tremonti}, Christy and {Masters}, Karen and
         {Coccato}, Lodovico and {Diamond-Stanic}, Aleksandar M. and
         {Arag{\'o}n-Salamanca}, Alfonso and {Avila-Reese}, Vladimir and
         {Badenes}, Carles and {Falc{\'o}n-Barroso}, J{\'e}sus and
         {Belfiore}, Francesco and {Bizyaev}, Dmitry and {Blanc}, Guillermo A. and
         {Bland-Hawthorn}, Joss and {Blanton}, Michael R. and
         {Brownstein}, Joel R. and {Byler}, Nell and {Cappellari}, Michele and
         {Conroy}, Charlie and {Dutton}, Aaron A. and {Emsellem}, Eric and
         {Etherington}, James and {Frinchaboy}, Peter M. and {Fu}, Hai and
         {Gunn}, James E. and {Harding}, Paul and {Johnston}, Evelyn J. and
         {Kauffmann}, Guinevere and {Kinemuchi}, Karen and {Klaene}, Mark A. and
         {Knapen}, Johan H. and {Leauthaud}, Alexie and {Li}, Cheng and
         {Lin}, Lihwai and {Maiolino}, Roberto and {Malanushenko}, Viktor and
         {Malanushenko}, Elena and {Mao}, Shude and {Maraston}, Claudia and
         {McDermid}, Richard M. and {Merrifield}, Michael R. and
         {Nichol}, Robert C. and {Oravetz}, Daniel and {Pan}, Kaike and
         {Parejko}, John K. and {Sanchez}, Sebastian F. and {Schlegel}, David and
         {Simmons}, Audrey and {Steele}, Oliver and {Steinmetz}, Matthias and
         {Thanjavur}, Karun and {Thompson}, Benjamin A. and {Tinker}, Jeremy L. and
         {van den Bosch}, Remco C.~E. and {Westfall}, Kyle B. and
         {Wilkinson}, David and {Wright}, Shelley and {Xiao}, Ting and
         {Zhang}, Kai},
        title = "{Overview of the SDSS-IV MaNGA Survey: Mapping nearby Galaxies at Apache Point Observatory}",
      journal = {\apj},
         year = 2015,
        month = jan,
       volume = {798},
       number = {1},
          eid = {7},
        pages = {7},
          doi = {10.1088/0004-637X/798/1/7},
archivePrefix = {arXiv},
       eprint = {1412.1482},
 primaryClass = {astro-ph.GA},
       adsurl = {https://ui.adsabs.harvard.edu/abs/2015ApJ...798....7B}
}

@ARTICLE{Feng2020,
	author = {{Feng}, Shuai and {Shen\vspace{0mm}}, Shi-Yin and {Yuan}, Fang-Ting and
	{Riffel}, Rogemar A. and {Pan}, Kaike},
	title = "{SDSS-IV MaNGA: Kinematic Asymmetry as an Indicator of Galaxy Interaction in Paired Galaxies}",
	journal = {\apjl},
	year = 2020,
	month = apr,
	volume = {892},
	number = {2},
	eid = {L20},
	pages = {L20},
	doi = {10.3847/2041-8213/ab7dba},
	archivePrefix = {arXiv},
	eprint = {2003.02493},
	primaryClass = {astro-ph.GA},
	adsurl = {https://ui.adsabs.harvard.edu/abs/2020ApJ...892L..20F}
}

@ARTICLE{Feng2019,
	author = {{Feng}, Shuai and {Shen}, Shi-Yin and {Yuan\vspace{0mm}}, Fang-Ting and
	{Luo}, A. -Li and {Zhang}, Jian-Nan and {Wang}, Meng-Xin and
	{Wang}, Xia and {Li}, Yin-Bi and {Hou}, Wen and {Kong}, Xiao and
	{Guo}, Yan-Xin and {Zuo}, Fang},
	title = "{Bivariate Luminosity Function of Galaxy Pairs}",
	journal = {\apj},
	year = 2019,
	month = aug,
	volume = {880},
	number = {2},
	eid = {114},
	pages = {114},
	doi = {10.3847/1538-4357/ab24da},
	archivePrefix = {arXiv},
	eprint = {1905.07276},
	primaryClass = {astro-ph.GA},
	adsurl = {https://ui.adsabs.harvard.edu/abs/2019ApJ...880..114F}
}

@ARTICLE{Stark2021,
       author = {{Stark}, David V. and {Masters}, Karen L. and {Avila-Reese}, Vladimir and {Riffel}, Rogemar and {Riffel}, Rogerio and {Boardman}, Nicholas Fraser and {Zheng}, Zheng and {Weijmans}, Anne-Marie and {Dillon}, Sean and {Fielder}, Catherine and {Finnegan}, Daniel and {Fofie}, Patricia and {Goddy}, Julian and {Harrington}, Emily and {Pace}, Zachary and {Rujopakarn}, Wiphu and {Samanso}, Nattida and {Shamsi}, Shoaib and {Sharma}, Anubhav and {Warrick}, Elizabeth and {Witherspoon}, Catherine and {Wolthuis}, Nathan},
        title = "{H I-MaNGA: tracing the physics of the neutral and ionized ISM with the second data release}",
      journal = {\mnras},
         year = 2021,
        month = may,
       volume = {503},
       number = {1},
        pages = {1345-1366},
          doi = {10.1093/mnras/stab566},
archivePrefix = {arXiv},
       eprint = {2101.12680},
 primaryClass = {astro-ph.GA},
       adsurl = {https://ui.adsabs.harvard.edu/abs/2021MNRAS.503.1345S}
}

@ARTICLE{Lisenfeld2019,
       author = {{Lisenfeld}, Ute and {Xu}, Cong Kevin and {Gao}, Yu and {Domingue}, Donovan L. and {Cao}, Chen and {Yun}, Min S. and {Zuo}, Pei},
        title = "{CO observations of major merger pairs at z = 0: molecular gas mass and star formation}",
      journal = {\aap},
         year = 2019,
        month = jul,
       volume = {627},
          eid = {A107},
        pages = {A107},
          doi = {10.1051/0004-6361/201935536},
archivePrefix = {arXiv},
       eprint = {1906.00682},
 primaryClass = {astro-ph.GA},
       adsurl = {https://ui.adsabs.harvard.edu/abs/2019A&A...627A.107L}
}

@ARTICLE{Moreno2019,
       author = {{Moreno}, Jorge and {Torrey}, Paul and {Ellison\vspace{0mm}}, Sara L. and
         {Patton}, David R. and {Hopkins}, Philip F. and {Bueno}, Michael and
         {Hayward}, Christopher C. and {Narayanan}, Desika and
         {Kere{\v{s}}}, Du{\v{s}}an and {Bluck}, Asa F.~L. and {Hernquist}, Lars},
        title = "{Interacting galaxies on FIRE-2: the connection between enhanced star formation and interstellar gas content}",
      journal = {\mnras},
         year = 2019,
        month = may,
       volume = {485},
       number = {1},
        pages = {1320-1338},
          doi = {10.1093/mnras/stz417},
archivePrefix = {arXiv},
       eprint = {1902.02305},
 primaryClass = {astro-ph.GA},
       adsurl = {https://ui.adsabs.harvard.edu/abs/2019MNRAS.485.1320M}
}

@ARTICLE{Saintonge2017,
       author = {{Saintonge}, Am{\'e}lie and {Catinella}, Barbara and {Tacconi}, Linda J. and {Kauffmann}, Guinevere and {Genzel}, Reinhard and {Cortese}, Luca and {Dav{\'e}}, Romeel and {Fletcher}, Thomas J. and {Graci{\'a}-Carpio}, Javier and {Kramer}, Carsten and {Heckman}, Timothy M. and {Janowiecki}, Steven and {Lutz}, Katharina and {Rosario}, David and {Schiminovich}, David and {Schuster}, Karl and {Wang}, Jing and {Wuyts}, Stijn and {Borthakur}, Sanchayeeta and {Lamperti}, Isabella and {Roberts-Borsani}, Guido W.},
        title = "{xCOLD GASS: The Complete IRAM 30 m Legacy Survey of Molecular Gas for Galaxy Evolution Studies}",
      journal = {\apjs},
         year = 2017,
        month = dec,
       volume = {233},
       number = {2},
          eid = {22},
        pages = {22},
          doi = {10.3847/1538-4365/aa97e0},
archivePrefix = {arXiv},
       eprint = {1710.02157},
 primaryClass = {astro-ph.GA},
       adsurl = {https://ui.adsabs.harvard.edu/abs/2017ApJS..233...22S}
}

@ARTICLE{Haynes2018,
	author = {{Haynes}, Martha P. and {Giovanelli}, Riccardo and {Kent}, Brian R. and
	{Adams}, Elizabeth A.~K. and {Balonek}, Thomas J. and
	{Craig}, David W. and {Fertig}, Derek and {Finn}, Rose and
	{Giovanardi}, Carlo and {Hallenbeck}, Gregory and {Hess}, Kelley M. and
	{Hoffman}, G. Lyle and {Huang}, Shan and {Jones}, Michael G. and
	{Koopmann}, Rebecca A. and {Kornreich}, David A. and {Leisman}, Lukas and
	{Miller}, Jeffrey and {Moorman}, Crystal and {O'Connor}, Jessica and
	{O'Donoghue}, Aileen and {Papastergis}, Emmanouil and
	{Troischt}, Parker and {Stark}, David and {Xiao}, Li},
	title = "{The Arecibo Legacy Fast ALFA Survey: The ALFALFA Extragalactic H I Source Catalog}",
	journal = {\apj},
	year = 2018,
	month = jul,
	volume = {861},
	number = {1},
	eid = {49},
	pages = {49},
	doi = {10.3847/1538-4357/aac956},
	archivePrefix = {arXiv},
	eprint = {1805.11499},
	primaryClass = {astro-ph.GA},
	adsurl = {https://ui.adsabs.harvard.edu/abs/2018ApJ...861...49H}
}

@ARTICLE{Ellison2018,
       author = {{Ellison}, Sara L. and {Catinella}, Barbara and {Cortese}, Luca},
        title = "{Enhanced atomic gas fractions in recently merged galaxies: quenching is not a result of post-merger gas exhaustion}",
      journal = {\mnras},
         year = 2018,
        month = aug,
       volume = {478},
       number = {3},
        pages = {3447-3466},
          doi = {10.1093/mnras/sty1247},
archivePrefix = {arXiv},
       eprint = {1805.03604},
 primaryClass = {astro-ph.GA},
       adsurl = {https://ui.adsabs.harvard.edu/abs/2018MNRAS.478.3447E}
}

@ARTICLE{Hibbard1996,
	author = {{Hibbard}, J.~E. and {van Gorkom}, J.~H.},
	title = "{HI, HII, and R-Band Observations of a Galactic Merger Sequence}",
	journal = {\aj},
	year = 1996,
	month = feb,
	volume = {111},
	pages = {655},
	doi = {10.1086/117815},
	archivePrefix = {arXiv},
	eprint = {astro-ph/9512035},
	primaryClass = {astro-ph},
	adsurl = {https://ui.adsabs.harvard.edu/abs/1996AJ....111..655H}
}

@ARTICLE{Zuo2018,
	author = {{Zuo}, Pei and {Xu}, Cong K. and {Yun}, Min S. and {Lisenfeld}, Ute and
	{Li}, Di and {Cao}, Chen},
	title = "{H I Observations of Major-merger Pairs at z = 0: Atomic Gas and Star Formation}",
	journal = {\apjs},
	year = 2018,
	month = jul,
	volume = {237},
	number = {1},
	eid = {2},
	pages = {2},
	doi = {10.3847/1538-4365/aabd30},
	archivePrefix = {arXiv},
	eprint = {1804.03500},
	primaryClass = {astro-ph.GA},
	adsurl = {https://ui.adsabs.harvard.edu/abs/2018ApJS..237....2Z}
}

@ARTICLE{Ellison2011,
       author = {{Ellison}, Sara L. and {Patton}, David R. and {Mendel}, J. Trevor and
         {Scudder}, Jillian M.},
        title = "{Galaxy pairs in the Sloan Digital Sky Survey - IV. Interactions trigger active galactic nuclei}",
      journal = {\mnras},
         year = 2011,
        month = dec,
       volume = {418},
       number = {3},
        pages = {2043-2053},
          doi = {10.1111/j.1365-2966.2011.19624.x},
archivePrefix = {arXiv},
       eprint = {1108.2711},
 primaryClass = {astro-ph.CO},
       adsurl = {https://ui.adsabs.harvard.edu/abs/2011MNRAS.418.2043E}
}

@ARTICLE{Mihos1996,
       author = {{Mihos}, J. Christopher and {Hernquist}, Lars},
        title = "{Gasdynamics and Starbursts in Major Mergers}",
      journal = {\apj},
         year = 1996,
        month = jun,
       volume = {464},
        pages = {641},
          doi = {10.1086/177353},
archivePrefix = {arXiv},
       eprint = {astro-ph/9512099},
 primaryClass = {astro-ph},
       adsurl = {https://ui.adsabs.harvard.edu/abs/1996ApJ...464..641M}
}

@ARTICLE{Zhang2020,
       author = {{Zhang}, Huanian and {Fang}, Taotao and {Zaritsky}, Dennis and {Behroozi}, Peter and {Werk}, Jessica and {Yang}, Xiaohu},
        title = "{Observing the Effects of Galaxy Interactions on the Circumgalactic Medium}",
      journal = {\apjl},
         year = 2020,
        month = apr,
       volume = {893},
       number = {1},
          eid = {L3},
        pages = {L3},
          doi = {10.3847/2041-8213/ab8068},
archivePrefix = {arXiv},
       eprint = {2003.08393},
 primaryClass = {astro-ph.GA},
       adsurl = {https://ui.adsabs.harvard.edu/abs/2020ApJ...893L...3Z}
}

@ARTICLE{Smith2018,
       author = {{Smith}, Beverly J. and {Campbell}, Kristen and {Struck}, Curtis and {Soria}, Roberto and {Swartz}, Douglas and {Magno}, Macon and {Dunn}, Brianne and {Giroux}, Mark L.},
        title = "{Diffuse X-Ray-emitting Gas in Major Mergers}",
      journal = {\aj},
         year = 2018,
        month = feb,
       volume = {155},
       number = {2},
          eid = {81},
        pages = {81},
          doi = {10.3847/1538-3881/aaa1a6},
archivePrefix = {arXiv},
       eprint = {1712.04049},
 primaryClass = {astro-ph.GA},
       adsurl = {https://ui.adsabs.harvard.edu/abs/2018AJ....155...81S}
}

@ARTICLE{Pan2019,
       author = {{Pan}, Hsi-An and {Lin}, Lihwai and {Hsieh\vspace{0mm}}, Bau-Ching and {Barrera-Ballesteros}, Jorge K. and {S{\'a}nchez}, Sebasti{\'a}n F. and {Hsu}, Chin-Hao and {Keenan}, Ryan and {Tissera}, Patricia B. and {Boquien}, M{\'e}d{\'e}ric and {Dai}, Y. Sophia and {Knapen}, Johan H. and {Riffel}, Rog{\'e}rio and {Argudo-Fern{\'a}ndez}, Maria and {Xiao}, Ting and {Yuan}, Fang-Ting},
        title = "{SDSS-IV MaNGA: Spatial Evolution of Star Formation Triggered by Galaxy Interactions}",
      journal = {\apj},
         year = 2019,
        month = aug,
       volume = {881},
       number = {2},
          eid = {119},
        pages = {119},
          doi = {10.3847/1538-4357/ab311c},
archivePrefix = {arXiv},
       eprint = {1907.04491},
 primaryClass = {astro-ph.GA},
       adsurl = {https://ui.adsabs.harvard.edu/abs/2019ApJ...881..119P}
}

@ARTICLE{Sanchez2012,
       author = {{S{\'a}nchez}, S.~F. and {Kennicutt}, R.~C. and {Gil de Paz}, A. and {van de Ven}, G. and {V{\'\i}lchez}, J.~M. and {Wisotzki}, L. and {Walcher}, C.~J. and {Mast}, D. and {Aguerri}, J.~A.~L. and {Albiol-P{\'e}rez}, S. and {Alonso-Herrero}, A. and {Alves}, J. and {Bakos}, J. and {Bart{\'a}kov{\'a}}, T. and {Bland-Hawthorn}, J. and {Boselli}, A. and {Bomans}, D.~J. and {Castillo-Morales}, A. and {Cortijo-Ferrero}, C. and {de Lorenzo-C{\'a}ceres}, A. and {Del Olmo}, A. and {Dettmar}, R. -J. and {D{\'\i}az}, A. and {Ellis}, S. and {Falc{\'o}n-Barroso}, J. and {Flores}, H. and {Gallazzi}, A. and {Garc{\'\i}a-Lorenzo}, B. and {Gonz{\'a}lez Delgado}, R. and {Gruel}, N. and {Haines}, T. and {Hao}, C. and {Husemann}, B. and {Igl{\'e}sias-P{\'a}ramo}, J. and {Jahnke}, K. and {Johnson}, B. and {Jungwiert}, B. and {Kalinova}, V. and {Kehrig}, C. and {Kupko}, D. and {L{\'o}pez-S{\'a}nchez}, {\'A}. R. and {Lyubenova}, M. and {Marino}, R.~A. and {M{\'a}rmol-Queralt{\'o}}, E. and {M{\'a}rquez}, I. and {Masegosa}, J. and {Meidt}, S. and {Mendez-Abreu}, J. and {Monreal-Ibero}, A. and {Montijo}, C. and {Mour{\~a}o}, A.~M. and {Palacios-Navarro}, G. and {Papaderos}, P. and {Pasquali}, A. and {Peletier}, R. and {P{\'e}rez}, E. and {P{\'e}rez}, I. and {Quirrenbach}, A. and {Rela{\~n}o}, M. and {Rosales-Ortega}, F.~F. and {Roth}, M.~M. and {Ruiz-Lara}, T. and {S{\'a}nchez-Bl{\'a}zquez}, P. and {Sengupta}, C. and {Singh}, R. and {Stanishev}, V. and {Trager}, S.~C. and {Vazdekis}, A. and {Viironen}, K. and {Wild}, V. and {Zibetti}, S. and {Ziegler}, B.},
        title = "{CALIFA, the Calar Alto Legacy Integral Field Area survey. I. Survey presentation}",
      journal = {\aap},
         year = 2012,
        month = feb,
       volume = {538},
          eid = {A8},
        pages = {A8},
          doi = {10.1051/0004-6361/201117353},
archivePrefix = {arXiv},
       eprint = {1111.0962},
 primaryClass = {astro-ph.CO},
       adsurl = {https://ui.adsabs.harvard.edu/abs/2012A&A...538A...8S}
}

@ARTICLE{Croom2012,
       author = {{Croom}, Scott M. and {Lawrence}, Jon S. and {Bland-Hawthorn}, Joss and {Bryant}, Julia J. and {Fogarty}, Lisa and {Richards}, Samuel and {Goodwin}, Michael and {Farrell}, Tony and {Miziarski}, Stan and {Heald}, Ron and {Jones}, D. Heath and {Lee}, Steve and {Colless}, Matthew and {Brough}, Sarah and {Hopkins}, Andrew M. and {Bauer}, Amanda E. and {Birchall}, Michael N. and {Ellis}, Simon and {Horton}, Anthony and {Leon-Saval}, Sergio and {Lewis}, Geraint and {L{\'o}pez-S{\'a}nchez}, {\'A}. R. and {Min}, Seong-Sik and {Trinh}, Christopher and {Trowland}, Holly},
        title = "{The Sydney-AAO Multi-object Integral field spectrograph}",
      journal = {\mnras},
         year = 2012,
        month = mar,
       volume = {421},
       number = {1},
        pages = {872-893},
          doi = {10.1111/j.1365-2966.2011.20365.x},
archivePrefix = {arXiv},
       eprint = {1112.3367},
 primaryClass = {astro-ph.CO},
       adsurl = {https://ui.adsabs.harvard.edu/abs/2012MNRAS.421..872C}
}

@ARTICLE{Barrera2015,
       author = {{Barrera-Ballesteros}, J.~K. and {Garc{\'\i}a-Lorenzo}, B. and {Falc{\'o}n-Barroso}, J. and {van de Ven}, G. and {Lyubenova}, M. and {Wild}, V. and {M{\'e}ndez-Abreu}, J. and {S{\'a}nchez}, S.~F. and {Marquez}, I. and {Masegosa}, J. and {Monreal-Ibero}, A. and {Ziegler}, B. and {del Olmo}, A. and {Verdes-Montenegro}, L. and {Garc{\'\i}a-Benito}, R. and {Husemann}, B. and {Mast}, D. and {Kehrig}, C. and {Iglesias-Paramo}, J. and {Marino}, R.~A. and {Aguerri}, J.~A.~L. and {Walcher}, C.~J. and {V{\'\i}lchez}, J.~M. and {Bomans}, D.~J. and {Cortijo-Ferrero}, C. and {Gonz{\'a}lez Delgado}, R.~M. and {Bland-Hawthorn}, J. and {McIntosh}, D.~H. and {Bekerait{\.{e}}}, S.},
        title = "{Tracing kinematic (mis)alignments in CALIFA merging galaxies. Stellar and ionized gas kinematic orientations at every merger stage}",
      journal = {\aap},
         year = 2015,
        month = oct,
       volume = {582},
          eid = {A21},
        pages = {A21},
          doi = {10.1051/0004-6361/201424935},
archivePrefix = {arXiv},
       eprint = {1506.03819},
 primaryClass = {astro-ph.GA},
       adsurl = {https://ui.adsabs.harvard.edu/abs/2015A&A...582A..21B}
}

@ARTICLE{Bloom2018,
       author = {{Bloom}, J.~V. and {Croom}, S.~M. and {Bryant}, J.~J. and {Schaefer}, A.~L. and {Bland-Hawthorn}, J. and {Brough}, S. and {Callingham}, J. and {Cortese}, L. and {Federrath}, C. and {Scott}, N. and {van de Sande}, J. and {D'Eugenio}, F. and {Sweet}, S. and {Tonini}, C. and {Allen}, J.~T. and {Goodwin}, M. and {Green}, A.~W. and {Konstantopoulos}, I.~S. and {Lawrence}, J. and {Lorente}, N. and {Medling}, A.~M. and {Owers}, M.~S. and {Richards}, S.~N. and {Sharp}, R.},
        title = "{The SAMI Galaxy Survey: gas content and interaction as the drivers of kinematic asymmetry}",
      journal = {\mnras},
         year = 2018,
        month = may,
       volume = {476},
       number = {2},
        pages = {2339-2351},
          doi = {10.1093/mnras/sty273},
archivePrefix = {arXiv},
       eprint = {1801.06628},
 primaryClass = {astro-ph.GA},
       adsurl = {https://ui.adsabs.harvard.edu/abs/2018MNRAS.476.2339B}
}

@ARTICLE{Scudder2012,
       author = {{Scudder}, Jillian M. and {Ellison}, Sara L. and {Torrey}, Paul and {Patton}, David R. and {Mendel}, J. Trevor},
        title = "{Galaxy pairs in the Sloan Digital Sky Survey - V. Tracing changes in star formation rate and metallicity out to separations of 80 kpc}",
      journal = {\mnras},
         year = 2012,
        month = oct,
       volume = {426},
       number = {1},
        pages = {549-565},
          doi = {10.1111/j.1365-2966.2012.21749.x},
archivePrefix = {arXiv},
       eprint = {1207.4791},
 primaryClass = {astro-ph.CO},
       adsurl = {https://ui.adsabs.harvard.edu/abs/2012MNRAS.426..549S}
}

@ARTICLE{Patton2013,
       author = {{Patton}, David R. and {Torrey}, Paul and {Ellison}, Sara L. and {Mendel}, J. Trevor and {Scudder}, Jillian M.},
        title = "{Galaxy pairs in the Sloan Digital Sky Survey - VI. The orbital extent of enhanced star formation in interacting galaxies}",
      journal = {\mnras},
         year = 2013,
        month = jun,
       volume = {433},
       number = {1},
        pages = {L59-L63},
          doi = {10.1093/mnrasl/slt058},
archivePrefix = {arXiv},
       eprint = {1305.1595},
 primaryClass = {astro-ph.CO},
       adsurl = {https://ui.adsabs.harvard.edu/abs/2013MNRAS.433L..59P}
}

@INPROCEEDINGS{Currie2014,
       author = {{Currie}, M.~J. and {Berry}, D.~S. and {Jenness}, T. and {Gibb}, A.~G. and {Bell}, G.~S. and {Draper}, P.~W.},
        title = "{Starlink Software in 2013}",
    booktitle = {Astronomical Data Analysis Software and Systems XXIII},
         year = 2014,
       editor = {{Manset}, N. and {Forshay}, P.},
       series = {Astronomical Society of the Pacific Conference Series},
       volume = {485},
        month = may,
        pages = {391},
       adsurl = {https://ui.adsabs.harvard.edu/abs/2014ASPC..485..391C}
}

@ARTICLE{Hung2016,
       author = {{Hung}, Chao-Ling and {Hayward}, Christopher C. and {Smith}, Howard A. and {Ashby}, Matthew L.~N. and {Lanz}, Lauranne and {Mart{\'\i}nez-Galarza}, Juan R. and {Sanders}, D.~B. and {Zezas}, Andreas},
        title = "{Merger Signatures in the Dynamics of Star-forming Gas}",
      journal = {\apj},
         year = 2016,
        month = jan,
       volume = {816},
       number = {2},
          eid = {99},
        pages = {99},
          doi = {10.3847/0004-637X/816/2/99},
archivePrefix = {arXiv},
       eprint = {1511.08481},
 primaryClass = {astro-ph.GA},
       adsurl = {https://ui.adsabs.harvard.edu/abs/2016ApJ...816...99H}
}

@ARTICLE{Masters2019,
       author = {{Masters}, Karen L. and {Stark}, David V. and {Pace}, Zachary J. and {Phipps}, Frederika and {Rujopakarn}, Wiphu and {Samanso}, Nattida and {Harrington}, Emily and {S{\'a}nchez-Gallego}, Jos{\'e} R. and {Avila-Reese}, Vladimir and {Bershady}, Matthew and {Cherinka}, Brian and {Fielder}, Catherine E. and {Finnegan}, Daniel and {Riffel}, Rogemar A. and {Rowlands}, Kate and {Shamsi}, Shoaib and {Newnham}, Lucy and {Weijmans}, Anne-Marie and {Witherspoon}, Catherine A.},
        title = "{H I-MaNGA: H I follow-up for the MaNGA survey}",
      journal = {\mnras},
         year = 2019,
        month = sep,
       volume = {488},
       number = {3},
        pages = {3396-3405},
          doi = {10.1093/mnras/stz1889},
archivePrefix = {arXiv},
       eprint = {1901.05579},
 primaryClass = {astro-ph.GA},
       adsurl = {https://ui.adsabs.harvard.edu/abs/2019MNRAS.488.3396M}
}

@ARTICLE{Solomon1997,
       author = {{Solomon}, P.~M. and {Downes}, D. and {Radford}, S.~J.~E. and {Barrett}, J.~W.},
        title = "{The Molecular Interstellar Medium in Ultraluminous Infrared Galaxies}",
      journal = {\apj},
         year = 1997,
        month = mar,
       volume = {478},
       number = {1},
        pages = {144-161},
          doi = {10.1086/303765},
archivePrefix = {arXiv},
       eprint = {astro-ph/9610166},
 primaryClass = {astro-ph},
       adsurl = {https://ui.adsabs.harvard.edu/abs/1997ApJ...478..144S}
}

@ARTICLE{Leroy2009,
       author = {{Leroy}, Adam K. and {Walter}, Fabian and {Bigiel}, Frank and {Usero}, Antonio and {Weiss}, Axel and {Brinks}, Elias and {de Blok}, W.~J.~G. and {Kennicutt}, Robert C. and {Schuster}, Karl-Friedrich and {Kramer}, Carsten and {Wiesemeyer}, H.~W. and {Roussel}, H{\'e}l{\`e}ne},
        title = "{Heracles: The HERA CO Line Extragalactic Survey}",
      journal = {\aj},
         year = 2009,
        month = jun,
       volume = {137},
       number = {6},
        pages = {4670-4696},
          doi = {10.1088/0004-6256/137/6/4670},
archivePrefix = {arXiv},
       eprint = {0905.4742},
 primaryClass = {astro-ph.CO},
       adsurl = {https://ui.adsabs.harvard.edu/abs/2009AJ....137.4670L}
}

@ARTICLE{Yu2022,
	author = {{Yu}, Qingzheng and {Fang}, Taotao and {Feng}, Shuai and {Zhang}, Bo and {Xu}, C. Kevin and {Wang}, Yunting and {Hao}, Lei},
	title = "{On the H I Content of MaNGA Major Merger Pairs}",
	journal = {\apj},
	year = 2022,
	month = aug,
	volume = {934},
	number = {2},
	eid = {114},
	pages = {114},
	doi = {10.3847/1538-4357/ac78e6},
	archivePrefix = {arXiv},
	eprint = {2206.06330},
	primaryClass = {astro-ph.GA},
	adsurl = {https://ui.adsabs.harvard.edu/abs/2022ApJ...934..114Y}
}

@ARTICLE{Lisenfeld2011,
	author = {{Lisenfeld}, U. and {Espada}, D. and {Verdes-Montenegro}, L. and {Kuno}, N. and {Leon}, S. and {Sabater}, J. and {Sato}, N. and {Sulentic}, J. and {Verley}, S. and {Yun}, M.~S.},
	title = "{The AMIGA sample of isolated galaxies. IX. Molecular gas properties}",
	journal = {\aap},
	year = 2011,
	month = oct,
	volume = {534},
	eid = {A102},
	pages = {A102},
	doi = {10.1051/0004-6361/201117056},
	archivePrefix = {arXiv},
	eprint = {1108.2130},
	primaryClass = {astro-ph.CO},
	adsurl = {https://ui.adsabs.harvard.edu/abs/2011A&A...534A.102L}
}

@ARTICLE{Wylezalek2022,
	author = {{Wylezalek}, D. and {Cicone}, C. and {Belfiore}, F. and {Bertemes}, C. and {Cazzoli}, S. and {Wagg}, J. and {Wang (王无忌)}, W. and {Aravena}, M. and {Maiolino}, R. and {Martin}, S. and {Bothwell}, M.~S. and {Brownstein}, J.~R. and {Bundy}, K. and {De Breuck}, C.},
	title = "{MASCOT: an ESO-ARO legacy survey of molecular gas in nearby SDSS-MaNGA galaxies - I. First data release, and global and resolved relations between H$_{2}$ and stellar content}",
	journal = {\mnras},
	year = 2022,
	month = mar,
	volume = {510},
	number = {3},
	pages = {3119-3131},
	doi = {10.1093/mnras/stab3356},
	archivePrefix = {arXiv},
	eprint = {2111.08719},
	primaryClass = {astro-ph.GA},
	adsurl = {https://ui.adsabs.harvard.edu/abs/2022MNRAS.510.3119W}
}

@ARTICLE{Kaneko2017,
	author = {{Kaneko}, Hiroyuki and {Kuno}, Nario and {Iono}, Daisuke and {Tamura}, Yoichi and {Tosaki}, Tomoka and {Nakanishi}, Kouichiro and {Sawada}, Tsuyoshi},
	title = "{Properties of molecular gas in galaxies in the early and mid stages of interaction. II. Molecular gas fraction}",
	journal = {\pasj},
	year = 2017,
	month = aug,
	volume = {69},
	number = {4},
	eid = {66},
	pages = {66},
	doi = {10.1093/pasj/psx041},
	archivePrefix = {arXiv},
	eprint = {1411.2660},
	primaryClass = {astro-ph.GA},
	adsurl = {https://ui.adsabs.harvard.edu/abs/2017PASJ...69...66K}
}

@ARTICLE{Astropy2013,
	author = {{Astropy Collaboration} and {Robitaille}, Thomas P. and {Tollerud}, Erik J. and {Greenfield}, Perry and {Droettboom}, Michael and {Bray}, Erik and {Aldcroft}, Tom and {Davis}, Matt and {Ginsburg}, Adam and {Price-Whelan}, Adrian M. and {Kerzendorf}, Wolfgang E. and {Conley}, Alexander and {Crighton}, Neil and {Barbary}, Kyle and {Muna}, Demitri and {Ferguson}, Henry and {Grollier}, Fr{\'e}d{\'e}ric and {Parikh}, Madhura M. and {Nair}, Prasanth H. and {Unther}, Hans M. and {Deil}, Christoph and {Woillez}, Julien and {Conseil}, Simon and {Kramer}, Roban and {Turner}, James E.~H. and {Singer}, Leo and {Fox}, Ryan and {Weaver}, Benjamin A. and {Zabalza}, Victor and {Edwards}, Zachary I. and {Azalee Bostroem}, K. and {Burke}, D.~J. and {Casey}, Andrew R. and {Crawford}, Steven M. and {Dencheva}, Nadia and {Ely}, Justin and {Jenness}, Tim and {Labrie}, Kathleen and {Lim}, Pey Lian and {Pierfederici}, Francesco and {Pontzen}, Andrew and {Ptak}, Andy and {Refsdal}, Brian and {Servillat}, Mathieu and {Streicher}, Ole},
	title = "{Astropy: A community Python package for astronomy}",
	journal = {\aap},
	year = 2013,
	month = oct,
	volume = {558},
	eid = {A33},
	pages = {A33},
	doi = {10.1051/0004-6361/201322068},
	archivePrefix = {arXiv},
	eprint = {1307.6212},
	primaryClass = {astro-ph.IM},
	adsurl = {https://ui.adsabs.harvard.edu/abs/2013A&A...558A..33A}
}

@ARTICLE{Virtanen2020,
	author = {{Virtanen}, Pauli and {Gommers}, Ralf and {Oliphant}, Travis E. and {Haberland}, Matt and {Reddy}, Tyler and {Cournapeau}, David and {Burovski}, Evgeni and {Peterson}, Pearu and {Weckesser}, Warren and {Bright}, Jonathan and {van der Walt}, St{\'e}fan J. and {Brett}, Matthew and {Wilson}, Joshua and {Millman}, K. Jarrod and {Mayorov}, Nikolay and {Nelson}, Andrew R.~J. and {Jones}, Eric and {Kern}, Robert and {Larson}, Eric and {Carey}, C.~J. and {Polat}, {\.I}lhan and {Feng}, Yu and {Moore}, Eric W. and {VanderPlas}, Jake and {Laxalde}, Denis and {Perktold}, Josef and {Cimrman}, Robert and {Henriksen}, Ian and {Quintero}, E.~A. and {Harris}, Charles R. and {Archibald}, Anne M. and {Ribeiro}, Ant{\^o}nio H. and {Pedregosa}, Fabian and {van Mulbregt}, Paul and {SciPy 1. 0 Contributors}},
	title = "{SciPy 1.0: fundamental algorithms for scientific computing in Python}",
	journal = {Nature Methods},
	year = 2020,
	month = feb,
	volume = {17},
	pages = {261-272},
	doi = {10.1038/s41592-019-0686-2},
	archivePrefix = {arXiv},
	eprint = {1907.10121},
	primaryClass = {cs.MS},
	adsurl = {https://ui.adsabs.harvard.edu/abs/2020NatMe..17..261V}
}

@MISC{Gildas2013,
	author = {{Gildas Team}},
	title = "{GILDAS: Grenoble Image and Line Data Analysis Software}",
	howpublished = {Astrophysics Source Code Library, record ascl:1305.010},
	year = 2013,
	month = may,
	eid = {ascl:1305.010},
	pages = {ascl:1305.010},
	archivePrefix = {ascl},
	eprint = {1305.010},
	adsurl = {https://ui.adsabs.harvard.edu/abs/2013ascl.soft05010G}
}

@ARTICLE{McElroy2022,
       author = {{McElroy}, Rebecca and {Bottrell}, Connor and {Hani}, Maan H. and {Moreno}, Jorge and {Croom}, Scott M. and {Hayward}, Christopher C. and {Twum}, Angela and {Feldmann}, Robert and {Hopkins}, Philip F. and {Hernquist}, Lars and {Husemann}, Bernd},
        title = "{The observability of galaxy merger signatures in nearby gas-rich spirals}",
      journal = {\mnras},
         year = 2022,
        month = sep,
       volume = {515},
       number = {3},
        pages = {3406-3419},
          doi = {10.1093/mnras/stac1715},
archivePrefix = {arXiv},
       eprint = {2206.07545},
 primaryClass = {astro-ph.GA},
       adsurl = {https://ui.adsabs.harvard.edu/abs/2022MNRAS.515.3406M}
}

@ARTICLE{Nishiyama2001,
       author = {{Nishiyama}, Kohta and {Nakai}, Naomasa and {Kuno}, Nario},
        title = "{CO Survey of Nearby Spiral Galaxies with the Nobeyama 45-m Telescope: II. Distribution and Dynamics of Molecular Gas}",
      journal = {\pasj},
         year = 2001,
        month = oct,
       volume = {53},
       number = {5},
        pages = {757-777},
          doi = {10.1093/pasj/53.5.757},
       adsurl = {https://ui.adsabs.harvard.edu/abs/2001PASJ...53..757N}
}

@ARTICLE{Regan2001,
       author = {{Regan}, Michael W. and {Thornley}, Michele D. and {Helfer}, Tamara T. and {Sheth}, Kartik and {Wong}, Tony and {Vogel}, Stuart N. and {Blitz}, Leo and {Bock}, Douglas C. -J.},
        title = "{The BIMA Survey of Nearby Galaxies. I. The Radial Distribution of CO Emission in Spiral Galaxies}",
      journal = {\apj},
         year = 2001,
        month = nov,
       volume = {561},
       number = {1},
        pages = {218-237},
          doi = {10.1086/323221},
archivePrefix = {arXiv},
       eprint = {astro-ph/0107211},
 primaryClass = {astro-ph},
       adsurl = {https://ui.adsabs.harvard.edu/abs/2001ApJ...561..218R}
}

@ARTICLE{Leroy2008,
       author = {{Leroy}, Adam K. and {Walter}, Fabian and {Brinks}, Elias and {Bigiel}, Frank and {de Blok}, W.~J.~G. and {Madore}, Barry and {Thornley}, M.~D.},
        title = "{The Star Formation Efficiency in Nearby Galaxies: Measuring Where Gas Forms Stars Effectively}",
      journal = {\aj},
         year = 2008,
        month = dec,
       volume = {136},
       number = {6},
        pages = {2782-2845},
          doi = {10.1088/0004-6256/136/6/2782},
archivePrefix = {arXiv},
       eprint = {0810.2556},
 primaryClass = {astro-ph},
       adsurl = {https://ui.adsabs.harvard.edu/abs/2008AJ....136.2782L}
}

@ARTICLE{Feigelson1985,
       author = {{Feigelson}, E.~D. and {Nelson}, P.~I.},
        title = "{Statistical methods for astronomical data with upper limits. I. Univariate distributions.}",
      journal = {\apj},
         year = 1985,
        month = jun,
       volume = {293},
        pages = {192-206},
          doi = {10.1086/163225},
       adsurl = {https://ui.adsabs.harvard.edu/abs/1985ApJ...293..192F}
}

@article{KM1958,
author = {E. L. Kaplan and Paul Meier},
title = {Nonparametric Estimation from Incomplete Observations},
journal = {Journal of the American Statistical Association},
volume = {53},
number = {282},
pages = {457-481},
year = {1958},
publisher = {Taylor & Francis},
doi = {10.1080/01621459.1958.10501452},
URL = { https://www.tandfonline.com/doi/abs/10.1080/01621459.1958.10501452},
eprint = {https://www.tandfonline.com/doi/pdf/10.1080/01621459.1958.10501452}
}

@ARTICLE{Barton2000,
       author = {{Barton}, Elizabeth J. and {Geller}, Margaret J. and {Kenyon}, Scott J.},
        title = "{Tidally Triggered Star Formation in Close Pairs of Galaxies}",
      journal = {\apj},
         year = 2000,
        month = feb,
       volume = {530},
       number = {2},
        pages = {660-679},
          doi = {10.1086/308392},
archivePrefix = {arXiv},
       eprint = {astro-ph/9909217},
 primaryClass = {astro-ph},
       adsurl = {https://ui.adsabs.harvard.edu/abs/2000ApJ...530..660B}
}

@ARTICLE{Thorp2022,
       author = {{Thorp}, Mallory D. and {Ellison}, Sara L. and {Pan}, Hsi-An and {Lin}, Lihwai and {Patton}, David R. and {Bluck}, Asa F.~L. and {Walters}, Dan and {Scudder}, Jillian M.},
        title = "{The ALMaQUEST Survey X: what powers merger induced star formation?}",
      journal = {\mnras},
         year = 2022,
        month = oct,
       volume = {516},
       number = {1},
        pages = {1462-1480},
          doi = {10.1093/mnras/stac2288},
archivePrefix = {arXiv},
       eprint = {2208.06426},
 primaryClass = {astro-ph.GA},
       adsurl = {https://ui.adsabs.harvard.edu/abs/2022MNRAS.516.1462T}
}

@ARTICLE{Elmegreen1993,
       author = {{Elmegreen}, B.~G.},
        title = "{The H to H 2 Transition in Galaxies: Totally Molecular Galaxies}",
      journal = {\apj},
         year = 1993,
        month = jul,
       volume = {411},
        pages = {170},
          doi = {10.1086/172816},
       adsurl = {https://ui.adsabs.harvard.edu/abs/1993ApJ...411..170E}
}

@ARTICLE{Braine1993,
       author = {{Braine}, J. and {Combes}, F.},
        title = "{A CO (1-0) and CO (2-1) survey of nearby spiral galaxies. III. More H2 gas in perturbed galaxies ?}",
      journal = {\aap},
         year = 1993,
        month = mar,
       volume = {269},
        pages = {7-14},
       adsurl = {https://ui.adsabs.harvard.edu/abs/1993A&A...269....7B}
}

@ARTICLE{Renaud2014,
       author = {{Renaud}, F. and {Bournaud}, F. and {Kraljic}, K. and {Duc}, P. -A.},
        title = "{Starbursts triggered by intergalactic tides andinterstellar compressive turbulence.}",
      journal = {\mnras},
         year = 2014,
        month = jul,
       volume = {442},
        pages = {L33-L37},
          doi = {10.1093/mnrasl/slu050},
archivePrefix = {arXiv},
       eprint = {1403.7316},
 primaryClass = {astro-ph.GA},
       adsurl = {https://ui.adsabs.harvard.edu/abs/2014MNRAS.442L..33R}
}

@ARTICLE{Mirabel1989,
       author = {{Mirabel}, I.~F. and {Sanders}, D.~B.},
        title = "{The Ratio of Molecular to Atomic Gas in Infrared Luminous Galaxies}",
      journal = {\apjl},
         year = 1989,
        month = may,
       volume = {340},
        pages = {L53},
          doi = {10.1086/185437},
       adsurl = {https://ui.adsabs.harvard.edu/abs/1989ApJ...340L..53M}
}

@ARTICLE{Li2008,
       author = {{Li}, Cheng and {Kauffmann}, Guinevere and {Heckman}, Timothy M. and {Jing}, Y.~P. and {White}, Simon D.~M.},
        title = "{Interaction-induced star formation in a complete sample of {}10$^{5}$ nearby star-forming galaxies}",
      journal = {\mnras},
         year = 2008,
        month = apr,
       volume = {385},
       number = {4},
        pages = {1903-1914},
          doi = {10.1111/j.1365-2966.2008.13000.x},
archivePrefix = {arXiv},
       eprint = {0711.3792},
 primaryClass = {astro-ph},
       adsurl = {https://ui.adsabs.harvard.edu/abs/2008MNRAS.385.1903L}
}

@ARTICLE{Kewley2006,
       author = {{Kewley}, Lisa J. and {Geller}, Margaret J. and {Barton}, Elizabeth J.},
        title = "{Metallicity and Nuclear Star Formation in Nearby Galaxy Pairs: Evidence for Tidally Induced Gas Flows}",
      journal = {\aj},
         year = 2006,
        month = apr,
       volume = {131},
       number = {4},
        pages = {2004-2017},
          doi = {10.1086/500295},
archivePrefix = {arXiv},
       eprint = {astro-ph/0511119},
 primaryClass = {astro-ph},
       adsurl = {https://ui.adsabs.harvard.edu/abs/2006AJ....131.2004K}
}

@ARTICLE{Rupke2010,
       author = {{Rupke}, David S.~N. and {Kewley}, Lisa J. and {Chien}, L. -H.},
        title = "{Gas-phase Oxygen Gradients in Strongly Interacting Galaxies. I. Early-stage Interactions}",
      journal = {\apj},
         year = 2010,
        month = nov,
       volume = {723},
       number = {2},
        pages = {1255-1271},
          doi = {10.1088/0004-637X/723/2/1255},
archivePrefix = {arXiv},
       eprint = {1009.0761},
 primaryClass = {astro-ph.GA},
       adsurl = {https://ui.adsabs.harvard.edu/abs/2010ApJ...723.1255R}
}

@ARTICLE{Kennicutt1984,
       author = {{Kennicutt}, R.~C., Jr. and {Keel}, W.~C.},
        title = "{Induced nuclear emission-line activity in interacting spiral galaxies.}",
      journal = {\apjl},
         year = 1984,
        month = apr,
       volume = {279},
        pages = {L5-L9},
          doi = {10.1086/184243},
       adsurl = {https://ui.adsabs.harvard.edu/abs/1984ApJ...279L...5K}
}

@ARTICLE{Xu2010,
       author = {{Xu}, C. Kevin and {Domingue}, Donovan and {Cheng}, Yi-Wen and {Lu}, Nanyao and {Huang}, Jiasheng and {Gao}, Yu and {Mazzarella}, Joseph M. and {Cutri}, Roc and {Sun}, Wei-Hsin and {Surace}, Jason},
        title = "{Local Benchmarks for the Evolution of Major-merger Galaxies{\textemdash}Spitzer Observations of a K-band Selected Sample}",
      journal = {\apj},
         year = 2010,
        month = apr,
       volume = {713},
       number = {1},
        pages = {330-355},
          doi = {10.1088/0004-637X/713/1/330},
archivePrefix = {arXiv},
       eprint = {1002.3648},
 primaryClass = {astro-ph.CO},
       adsurl = {https://ui.adsabs.harvard.edu/abs/2010ApJ...713..330X}
}

@ARTICLE{Goulding2018,
       author = {{Goulding}, Andy D. and {Greene}, Jenny E. and {Bezanson}, Rachel and {Greco}, Johnny and {Johnson}, Sean and {Leauthaud}, Alexie and {Matsuoka}, Yoshiki and {Medezinski}, Elinor and {Price-Whelan}, Adrian M.},
        title = "{Galaxy interactions trigger rapid black hole growth: An unprecedented view from the Hyper Suprime-Cam survey}",
      journal = {\pasj},
         year = 2018,
        month = jan,
       volume = {70},
          eid = {S37},
        pages = {S37},
          doi = {10.1093/pasj/psx135},
archivePrefix = {arXiv},
       eprint = {1706.07436},
 primaryClass = {astro-ph.GA},
       adsurl = {https://ui.adsabs.harvard.edu/abs/2018PASJ...70S..37G}
}

@ARTICLE{Kennicutt1987,
       author = {{Kennicutt}, Robert C., Jr. and {Keel}, William C. and {van der Hulst}, J.~M. and {Hummel}, E. and {Roettiger}, Kurt A.},
        title = "{The Effects of Interactions on Spiral Galaxies. II. Disk Star Formation Rates}",
      journal = {\aj},
         year = 1987,
        month = may,
       volume = {93},
        pages = {1011},
          doi = {10.1086/114384},
       adsurl = {https://ui.adsabs.harvard.edu/abs/1987AJ.....93.1011K}
}

@ARTICLE{Nikolic2004,
       author = {{Nikolic}, B. and {Cullen}, H. and {Alexander}, P.},
        title = "{Star formation in close pairs selected from the Sloan Digital Sky Survey}",
      journal = {\mnras},
         year = 2004,
        month = dec,
       volume = {355},
       number = {3},
        pages = {874-886},
          doi = {10.1111/j.1365-2966.2004.08366.x},
archivePrefix = {arXiv},
       eprint = {astro-ph/0407289},
 primaryClass = {astro-ph},
       adsurl = {https://ui.adsabs.harvard.edu/abs/2004MNRAS.355..874N}
}

@ARTICLE{Sparre2016,
       author = {{Sparre}, Martin and {Springel}, Volker},
        title = "{Zooming in on major mergers: dense, starbursting gas in cosmological simulations}",
      journal = {\mnras},
         year = 2016,
        month = nov,
       volume = {462},
       number = {3},
        pages = {2418-2430},
          doi = {10.1093/mnras/stw1793},
archivePrefix = {arXiv},
       eprint = {1604.08205},
 primaryClass = {astro-ph.GA},
       adsurl = {https://ui.adsabs.harvard.edu/abs/2016MNRAS.462.2418S}
}

@ARTICLE{Sanders1996,
       author = {{Sanders}, D.~B. and {Mirabel}, I.~F.},
        title = "{Luminous Infrared Galaxies}",
      journal = {\araa},
         year = 1996,
        month = jan,
       volume = {34},
        pages = {749},
          doi = {10.1146/annurev.astro.34.1.749},
       adsurl = {https://ui.adsabs.harvard.edu/abs/1996ARA&A..34..749S}
}

@ARTICLE{Veilleux2002,
       author = {{Veilleux}, S. and {Kim}, D. -C. and {Sanders}, D.~B.},
        title = "{Optical and Near-Infrared Imaging of the IRAS 1 Jy Sample of Ultraluminous Infrared Galaxies. II. The Analysis}",
      journal = {\apjs},
         year = 2002,
        month = dec,
       volume = {143},
       number = {2},
        pages = {315-376},
          doi = {10.1086/343844},
archivePrefix = {arXiv},
       eprint = {astro-ph/0207401},
 primaryClass = {astro-ph},
       adsurl = {https://ui.adsabs.harvard.edu/abs/2002ApJS..143..315V}
}

@ARTICLE{Haan2011,
       author = {{Haan}, S. and {Surace}, J.~A. and {Armus}, L. and {Evans}, A.~S. and {Howell}, J.~H. and {Mazzarella}, J.~M. and {Kim}, D.~C. and {Vavilkin}, T. and {Inami}, H. and {Sanders}, D.~B. and {Petric}, A. and {Bridge}, C.~R. and {Melbourne}, J.~L. and {Charmandaris}, V. and {Diaz-Santos}, T. and {Murphy}, E.~J. and {U}, V. and {Stierwalt}, S. and {Marshall}, J.~A.},
        title = "{The Nuclear Structure in Nearby Luminous Infrared Galaxies: Hubble Space Telescope NICMOS Imaging of the GOALS Sample}",
      journal = {\aj},
         year = 2011,
        month = mar,
       volume = {141},
       number = {3},
          eid = {100},
        pages = {100},
          doi = {10.1088/0004-6256/141/3/100},
archivePrefix = {arXiv},
       eprint = {1012.4012},
 primaryClass = {astro-ph.CO},
       adsurl = {https://ui.adsabs.harvard.edu/abs/2011AJ....141..100H}
}

@ARTICLE{Ellison2013,
       author = {{Ellison}, Sara L. and {Mendel}, J. Trevor and {Scudder}, Jillian M. and {Patton}, David R. and {Palmer}, Michael J.~D.},
        title = "{Galaxy pairs in the Sloan Digital Sky Survey - VII. The merger-luminous infrared galaxy connection}",
      journal = {\mnras},
         year = 2013,
        month = apr,
       volume = {430},
       number = {4},
        pages = {3128-3141},
          doi = {10.1093/mnras/sts546},
archivePrefix = {arXiv},
       eprint = {1301.5351},
 primaryClass = {astro-ph.CO},
       adsurl = {https://ui.adsabs.harvard.edu/abs/2013MNRAS.430.3128E}
}

@ARTICLE{Thorp2019,
       author = {{Thorp}, Mallory D. and {Ellison}, Sara L. and {Simard}, Luc and {S{\'a}nchez}, Sebastian F. and {Antonio}, Braulio},
        title = "{Spatially resolved star formation and metallicity profiles in post-merger galaxies from MaNGA}",
      journal = {\mnras},
         year = 2019,
        month = jan,
       volume = {482},
       number = {1},
        pages = {L55-L59},
          doi = {10.1093/mnrasl/sly185},
archivePrefix = {arXiv},
       eprint = {1810.00897},
 primaryClass = {astro-ph.GA},
       adsurl = {https://ui.adsabs.harvard.edu/abs/2019MNRAS.482L..55T}
}

@ARTICLE{Barnes1991,
       author = {{Barnes}, Joshua E. and {Hernquist}, Lars E.},
        title = "{Fueling Starburst Galaxies with Gas-rich Mergers}",
      journal = {\apjl},
         year = 1991,
        month = apr,
       volume = {370},
        pages = {L65},
          doi = {10.1086/185978},
       adsurl = {https://ui.adsabs.harvard.edu/abs/1991ApJ...370L..65B}
}

@ARTICLE{Barnes1996,
       author = {{Barnes}, Joshua E. and {Hernquist\vspace{0mm}}, Lars},
        title = "{Transformations of Galaxies. II. Gasdynamics in Merging Disk Galaxies}",
      journal = {\apj},
         year = 1996,
        month = nov,
       volume = {471},
        pages = {115},
          doi = {10.1086/177957},
       adsurl = {https://ui.adsabs.harvard.edu/abs/1996ApJ...471..115B}
}

@ARTICLE{Hopkins2009,
       author = {{Hopkins}, Philip F. and {Cox}, Thomas J. and {Younger}, Joshua D. and {Hernquist}, Lars},
        title = "{How do Disks Survive Mergers?}",
      journal = {\apj},
         year = 2009,
        month = feb,
       volume = {691},
       number = {2},
        pages = {1168-1201},
          doi = {10.1088/0004-637X/691/2/1168},
archivePrefix = {arXiv},
       eprint = {0806.1739},
 primaryClass = {astro-ph},
       adsurl = {https://ui.adsabs.harvard.edu/abs/2009ApJ...691.1168H}
}

@ARTICLE{Ueda2014,
       author = {{Ueda}, Junko and {Iono}, Daisuke and {Yun}, Min S. and {Crocker}, Alison F. and {Narayanan}, Desika and {Komugi}, Shinya and {Espada}, Daniel and {Hatsukade}, Bunyo and {Kaneko}, Hiroyuki and {Matsuda}, Yuichi and {Tamura}, Yoichi and {Wilner}, David J. and {Kawabe}, Ryohei and {Pan}, Hsi-An},
        title = "{Cold Molecular Gas in Merger Remnants. I. Formation of Molecular Gas Disks}",
      journal = {\apjs},
         year = 2014,
        month = sep,
       volume = {214},
       number = {1},
          eid = {1},
        pages = {1},
          doi = {10.1088/0067-0049/214/1/1},
archivePrefix = {arXiv},
       eprint = {1407.6873},
 primaryClass = {astro-ph.GA},
       adsurl = {https://ui.adsabs.harvard.edu/abs/2014ApJS..214....1U}
}

@ARTICLE{Iono2004,
       author = {{Iono}, Daisuke and {Yun}, Min S. and {Mihos}, J. Christopher},
        title = "{Radial Gas Flows in Colliding Galaxies: Connecting Simulations and Observations}",
      journal = {\apj},
         year = 2004,
        month = nov,
       volume = {616},
       number = {1},
        pages = {199-220},
          doi = {10.1086/424797},
archivePrefix = {arXiv},
       eprint = {astro-ph/0407609},
 primaryClass = {astro-ph},
       adsurl = {https://ui.adsabs.harvard.edu/abs/2004ApJ...616..199I}
}

@ARTICLE{Gao1999,
       author = {{Gao}, Yu and {Solomon}, Philip M.},
        title = "{Molecular Gas Depletion and Starbursts in Luminous Infrared Galaxy Mergers}",
      journal = {\apjl},
         year = 1999,
        month = feb,
       volume = {512},
       number = {2},
        pages = {L99-L103},
          doi = {10.1086/311878},
archivePrefix = {arXiv},
       eprint = {astro-ph/9812320},
 primaryClass = {astro-ph},
       adsurl = {https://ui.adsabs.harvard.edu/abs/1999ApJ...512L..99G}
}

@ARTICLE{Xu2021,
       author = {{Xu}, C.~K. and {Lisenfeld}, U. and {Gao}, Y. and {Renaud}, F.},
        title = "{NOEMA Observations of CO Emission in Arp 142 and Arp 238}",
      journal = {\apj},
         year = 2021,
        month = sep,
       volume = {918},
       number = {2},
          eid = {55},
        pages = {55},
          doi = {10.3847/1538-4357/ac0f77},
archivePrefix = {arXiv},
       eprint = {2106.15041},
 primaryClass = {astro-ph.GA},
       adsurl = {https://ui.adsabs.harvard.edu/abs/2021ApJ...918...55X}
}

@ARTICLE{Hou2023,
       author = {{Hou}, Meicun and {Li}, Zhiyuan and {Liu}, Xin and {Li}, Zongnan and {Li}, Ruancun and {Wang}, Ran and {Wang}, Jing and {Ho}, Luis C.},
        title = "{NOEMA Detection of Circumnuclear Molecular Gas in X-Ray Weak Dual Active Galactic Nuclei: No Evidence for Heavy Obscuration}",
      journal = {\apj},
         year = 2023,
        month = jan,
       volume = {943},
       number = {1},
          eid = {50},
        pages = {50},
          doi = {10.3847/1538-4357/acaaf9},
archivePrefix = {arXiv},
       eprint = {2212.06399},
 primaryClass = {astro-ph.GA},
       adsurl = {https://ui.adsabs.harvard.edu/abs/2023ApJ...943...50H}
}

@ARTICLE{Feng2022,
       author = {{Feng}, Shuai and {Shen}, Shi-Yin and {Yuan}, Fang-Ting and {Dai}, Y. Sophia and {Masters}, Karen L.},
        title = "{The Velocity Map Asymmetry of Ionized Gas in MaNGA. I. The Catalog and General Properties}",
      journal = {\apjs},
         year = 2022,
        month = sep,
       volume = {262},
       number = {1},
          eid = {6},
        pages = {6},
          doi = {10.3847/1538-4365/ac80f2},
archivePrefix = {arXiv},
       eprint = {2207.06050},
 primaryClass = {astro-ph.GA},
       adsurl = {https://ui.adsabs.harvard.edu/abs/2022ApJS..262....6F}
}

@ARTICLE{Yamashita2017,
       author = {{Yamashita}, Takuji and {Komugi}, Shinya and {Matsuhara}, Hideo and {Armus}, Lee and {Inami}, Hanae and {Ueda}, Junko and {Iono}, Daisuke and {Kohno}, Kotaro and {Evans}, Aaron S. and {Arimatsu}, Ko},
        title = "{Cold Molecular Gas Along the Merger Sequence in Local Luminous Infrared Galaxies}",
      journal = {\apj},
         year = 2017,
        month = aug,
       volume = {844},
       number = {2},
          eid = {96},
        pages = {96},
          doi = {10.3847/1538-4357/aa7af1},
archivePrefix = {arXiv},
       eprint = {1706.06271},
 primaryClass = {astro-ph.GA},
       adsurl = {https://ui.adsabs.harvard.edu/abs/2017ApJ...844...96Y}
}

@ARTICLE{Wilson2008,
       author = {{Wilson}, Christine D. and {Petitpas}, Glen R. and {Iono}, Daisuke and {Baker}, Andrew J. and {Peck}, Alison B. and {Krips}, Melanie and {Warren}, Bradley and {Golding}, Jennifer and {Atkinson}, Adam and {Armus}, Lee and {Cox}, T.~J. and {Ho}, Paul and {Juvela}, Mika and {Matsushita}, Satoki and {Mihos}, J. Christopher and {Pihlstrom}, Ylva and {Yun}, Min S.},
        title = "{Luminous Infrared Galaxies with the Submillimeter Array. I. Survey Overview and the Central Gas to Dust Ratio}",
      journal = {\apjs},
         year = 2008,
        month = oct,
       volume = {178},
       number = {2},
        pages = {189-224},
          doi = {10.1086/590910},
archivePrefix = {arXiv},
       eprint = {0806.3002},
 primaryClass = {astro-ph},
       adsurl = {https://ui.adsabs.harvard.edu/abs/2008ApJS..178..189W}
}

@ARTICLE{Shangguan2019,
       author = {{Shangguan}, Jinyi and {Ho}, Luis C. and {Li}, Ruancun and {Zhuang}, Ming-Yang and {Xie}, Yanxia and {Li}, Zhihui},
        title = "{Interstellar Medium and Star Formation of Starburst Galaxies on the Merger Sequence}",
      journal = {\apj},
         year = 2019,
        month = jan,
       volume = {870},
       number = {2},
          eid = {104},
        pages = {104},
          doi = {10.3847/1538-4357/aaf21a},
archivePrefix = {arXiv},
       eprint = {1811.05822},
 primaryClass = {astro-ph.GA},
       adsurl = {https://ui.adsabs.harvard.edu/abs/2019ApJ...870..104S}
}

@ARTICLE{Iono2009,
       author = {{Iono}, Daisuke and {Wilson}, Christine D. and {Yun}, Min S. and {Baker}, Andrew J. and {Petitpas}, Glen R. and {Peck}, Alison B. and {Krips}, Melanie and {Cox}, T.~J. and {Matsushita}, Satoki and {Mihos}, J. Christopher and {Pihlstrom}, Ylva},
        title = "{Luminous Infrared Galaxies with the Submillimeter Array. II. Comparing the CO (3-2) Sizes and Luminosities of Local and High-Redshift Luminous Infrared Galaxies}",
      journal = {\apj},
         year = 2009,
        month = apr,
       volume = {695},
       number = {2},
        pages = {1537-1549},
          doi = {10.1088/0004-637X/695/2/1537},
archivePrefix = {arXiv},
       eprint = {0902.0121},
 primaryClass = {astro-ph.CO},
       adsurl = {https://ui.adsabs.harvard.edu/abs/2009ApJ...695.1537I}
}

@ARTICLE{Li2023,
       author = {{Li}, Yang A. and {Ho}, Luis C. and {Shangguan}, Jinyi},
        title = "{The Subtle Effects of Mergers on Star Formation in Nearby Galaxies}",
      journal = {\apj},
         year = 2023,
        month = aug,
       volume = {953},
       number = {1},
          eid = {91},
        pages = {91},
          doi = {10.3847/1538-4357/acdddb},
archivePrefix = {arXiv},
       eprint = {2307.13462},
 primaryClass = {astro-ph.GA},
       adsurl = {https://ui.adsabs.harvard.edu/abs/2023ApJ...953...91L}
}

@ARTICLE{Barnes1992,
       author = {{Barnes}, Joshua E. and {Hernquist}, Lars},
        title = "{Dynamics of interacting galaxies.}",
      journal = {\araa},
         year = 1992,
        month = jan,
       volume = {30},
        pages = {705-742},
          doi = {10.1146/annurev.aa.30.090192.003421},
       adsurl = {https://ui.adsabs.harvard.edu/abs/1992ARA&A..30..705B}
}

@ARTICLE{Toomre1972,
       author = {{Toomre}, Alar and {Toomre}, Juri},
        title = "{Galactic Bridges and Tails}",
      journal = {\apj},
         year = 1972,
        month = dec,
       volume = {178},
        pages = {623-666},
          doi = {10.1086/151823},
       adsurl = {https://ui.adsabs.harvard.edu/abs/1972ApJ...178..623T}
}

@INPROCEEDINGS{Toomre1977,
       author = {{Toomre}, Alar},
        title = "{Mergers and Some Consequences}",
    booktitle = {Evolution of Galaxies and Stellar Populations},
         year = 1977,
       editor = {{Tinsley}, Beatrice M. and {Larson}, Richard B. Gehret, D. Campbell},
        month = jan,
        pages = {401},
       adsurl = {https://ui.adsabs.harvard.edu/abs/1977egsp.conf..401T}
}

@ARTICLE{Hopkins2013,
       author = {{Hopkins}, Philip F. and {Cox}, Thomas J. and {Hernquist}, Lars and {Narayanan}, Desika and {Hayward}, Christopher C. and {Murray}, Norman},
        title = "{Star formation in galaxy mergers with realistic models of stellar feedback and the interstellar medium}",
      journal = {\mnras},
         year = 2013,
        month = apr,
       volume = {430},
       number = {3},
        pages = {1901-1927},
          doi = {10.1093/mnras/stt017},
archivePrefix = {arXiv},
       eprint = {1206.0011},
 primaryClass = {astro-ph.CO},
       adsurl = {https://ui.adsabs.harvard.edu/abs/2013MNRAS.430.1901H}
}

@ARTICLE{Xu1991,
       author = {{Xu}, Cong and {Sulentic}, Jack W.},
        title = "{Infrared Emission in Paired Galaxies. II. Luminosity Functions and Far-Infrared Properties}",
      journal = {\apj},
         year = 1991,
        month = jun,
       volume = {374},
        pages = {407},
          doi = {10.1086/170132},
       adsurl = {https://ui.adsabs.harvard.edu/abs/1991ApJ...374..407X}
}

@ARTICLE{Boselli2014,
       author = {{Boselli}, A. and {Cortese}, L. and {Boquien}, M. and {Boissier}, S. and {Catinella}, B. and {Lagos}, C. and {Saintonge}, A.},
        title = "{Cold gas properties of the Herschel Reference Survey. II. Molecular and total gas scaling relations}",
      journal = {\aap},
         year = 2014,
        month = apr,
       volume = {564},
          eid = {A66},
        pages = {A66},
          doi = {10.1051/0004-6361/201322312},
archivePrefix = {arXiv},
       eprint = {1401.8101},
 primaryClass = {astro-ph.GA},
       adsurl = {https://ui.adsabs.harvard.edu/abs/2014A&A...564A..66B}
}

@ARTICLE{Cicone2017,
       author = {{Cicone}, C. and {Bothwell}, M. and {Wagg}, J. and {M{\o}ller}, P. and {De Breuck}, C. and {Zhang}, Z. and {Mart{\'\i}n}, S. and {Maiolino}, R. and {Severgnini}, P. and {Aravena}, M. and {Belfiore}, F. and {Espada}, D. and {Fl{\"u}tsch}, A. and {Impellizzeri}, V. and {Peng}, Y. and {Raj}, M.~A. and {Ram{\'\i}rez-Olivencia}, N. and {Riechers}, D. and {Schawinski}, K.},
        title = "{The final data release of ALLSMOG: a survey of CO in typical local low-M$_{{\ensuremath{*}}}$ star-forming galaxies}",
      journal = {\aap},
         year = 2017,
        month = aug,
       volume = {604},
          eid = {A53},
        pages = {A53},
          doi = {10.1051/0004-6361/201730605},
archivePrefix = {arXiv},
       eprint = {1705.05851},
 primaryClass = {astro-ph.GA},
       adsurl = {https://ui.adsabs.harvard.edu/abs/2017A&A...604A..53C}
}

@article{Anderson1952,
author = {T. W. Anderson and D. A. Darling},
title = {{Asymptotic Theory of Certain "Goodness of Fit" Criteria Based on Stochastic Processes}},
volume = {23},
journal = {The Annals of Mathematical Statistics},
number = {2},
publisher = {Institute of Mathematical Statistics},
pages = {193 -- 212},
year = {1952},
doi = {10.1214/aoms/1177729437},
URL = {https://doi.org/10.1214/aoms/1177729437}
}

@BOOK{Feigelson2012,
       author = {{Feigelson}, Eric D. and {Babu}, G. Jogesh},
        title = "{Modern Statistical Methods for Astronomy}",
         year = 2012,
          doi = {10.48550/arXiv.1205.2064},
    publisher = {Cambridge: Cambridge Univ. Press},
       adsurl = {https://ui.adsabs.harvard.edu/abs/2012msma.book.....F}
}

@ARTICLE{Keel1993,
       author = {{Keel}, William C.},
        title = "{Kinematic Regulation of Star Formation in Interacting Galaxies}",
      journal = {\aj},
         year = 1993,
        month = nov,
       volume = {106},
        pages = {1771},
          doi = {10.1086/116763},
       adsurl = {https://ui.adsabs.harvard.edu/abs/1993AJ....106.1771K}
}

@ARTICLE{Scoville1997,
       author = {{Scoville}, N.~Z. and {Yun}, M.~S. and {Bryant}, P.~M.},
        title = "{Arcsecond Imaging of CO Emission in the Nucleus of Arp 220}",
      journal = {\apj},
         year = 1997,
        month = jul,
       volume = {484},
       number = {2},
        pages = {702-719},
          doi = {10.1086/304368},
       adsurl = {https://ui.adsabs.harvard.edu/abs/1997ApJ...484..702S}
}

@ARTICLE{Scoville2017,
       author = {{Scoville}, Nick and {Murchikova}, Lena and {Walter}, Fabian and {Vlahakis}, Catherine and {Koda}, Jin and {Vanden Bout}, Paul and {Barnes}, Joshua and {Hernquist}, Lars and {Sheth}, Kartik and {Yun}, Min and {Sanders}, David and {Armus}, Lee and {Cox}, Pierre and {Thompson}, Todd and {Robertson}, Brant and {Zschaechner}, Laura and {Tacconi}, Linda and {Torrey}, Paul and {Hayward}, Christopher C. and {Genzel}, Reinhard and {Hopkins}, Phil and {van der Werf}, Paul and {Decarli}, Roberto},
        title = "{ALMA Resolves the Nuclear Disks of Arp 220}",
      journal = {\apj},
         year = 2017,
        month = feb,
       volume = {836},
       number = {1},
          eid = {66},
        pages = {66},
          doi = {10.3847/1538-4357/836/1/66},
archivePrefix = {arXiv},
       eprint = {1605.09381},
 primaryClass = {astro-ph.GA},
       adsurl = {https://ui.adsabs.harvard.edu/abs/2017ApJ...836...66S}
}
\bibliographystyle{aasjournal}

\begin{figure*}[t]
    \figurenum{A1}
    \centering
	\includegraphics[width=0.92\textwidth]{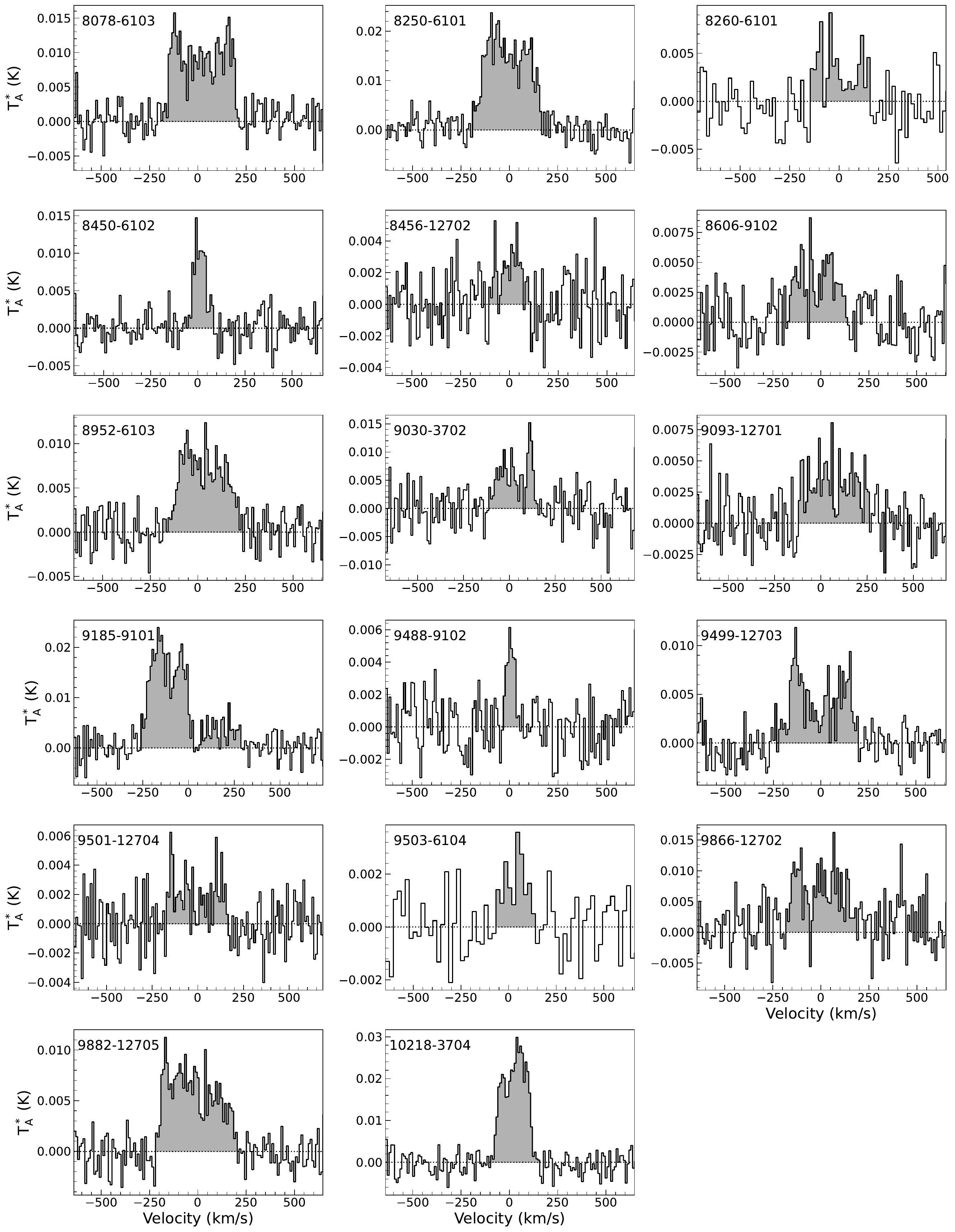}
	\caption{CO(2$-$1) spectra of the detected galaxies from JCMT observations. Most of the spectra were rebinned to a velocity resolution of $\sim$10 $\text{km s}^{-1}$, and some spectra were rebinned to a velocity resolution of $\sim$20 $\text{km s}^{-1}$ to achieve detections. The velocity of each spectrum is Doppler corrected and converted to the barycentric frame, and the zero-point is set based on the optical spectroscopy redshift of the galaxy.
		\label{fig:jcmt_co_2-1}
	}
\end{figure*}

\begin{figure*}[t]
    \figurenum{A2}
    \centering
	\includegraphics[width=1\textwidth]{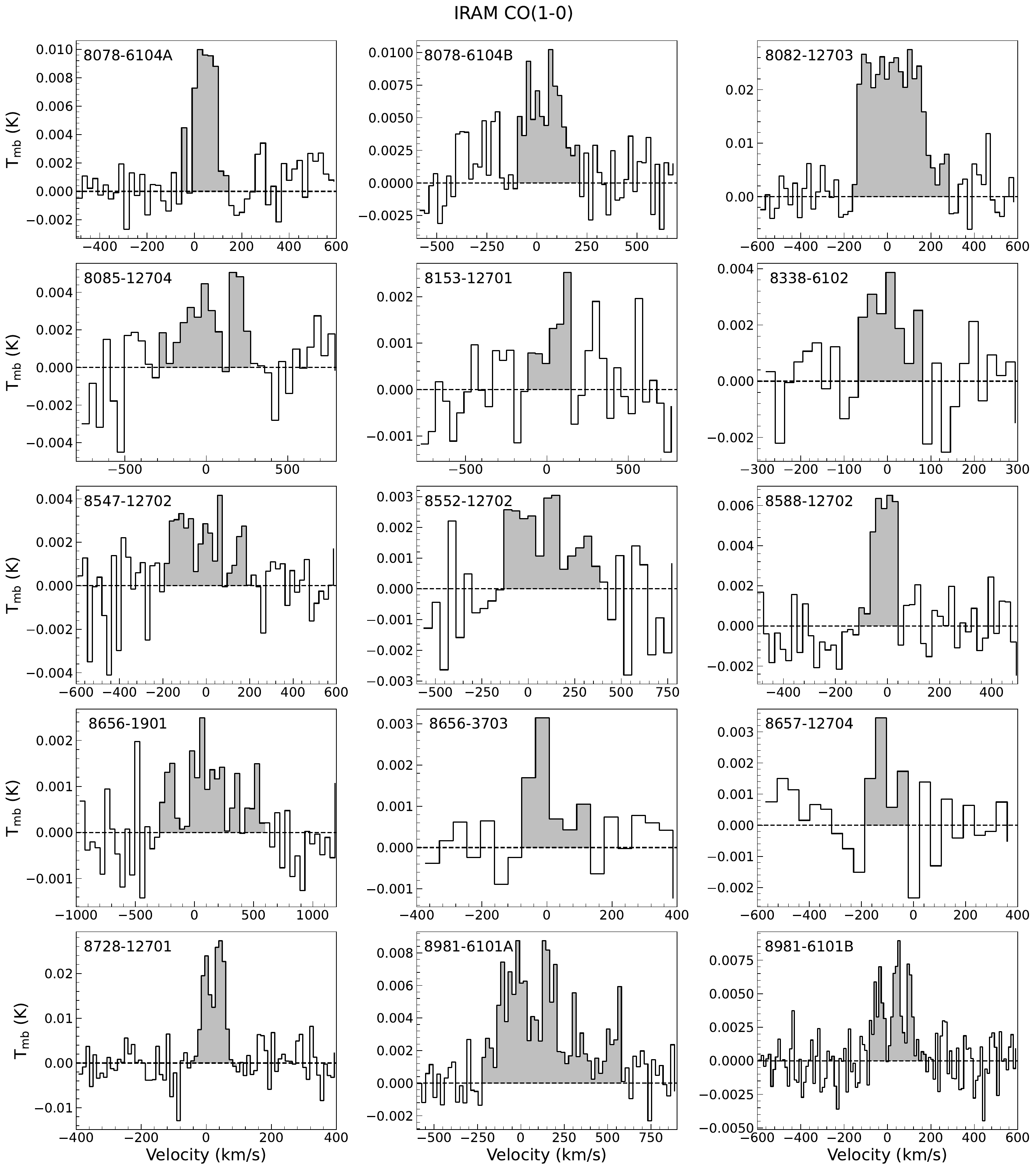}
	\caption{CO(1$-$0) spectra of the detected galaxies from IRAM 30\,m observations. Most of the spectra were rebinned to a velocity resolution of $\sim$20 $\text{km s}^{-1}$, and some spectra were rebinned to a velocity resolution of $\sim$40 $\text{km s}^{-1}$ to achieve detections. The velocity of each spectrum is Doppler corrected and converted to the barycentric frame, and the zero-point is set based on the optical spectroscopy redshift of the galaxy.
		\label{fig:iram_co_1-0}
	}
\end{figure*}

\begin{figure*}[t]
    \figurenum{A2}
    \centering
	\includegraphics[width=1\textwidth]{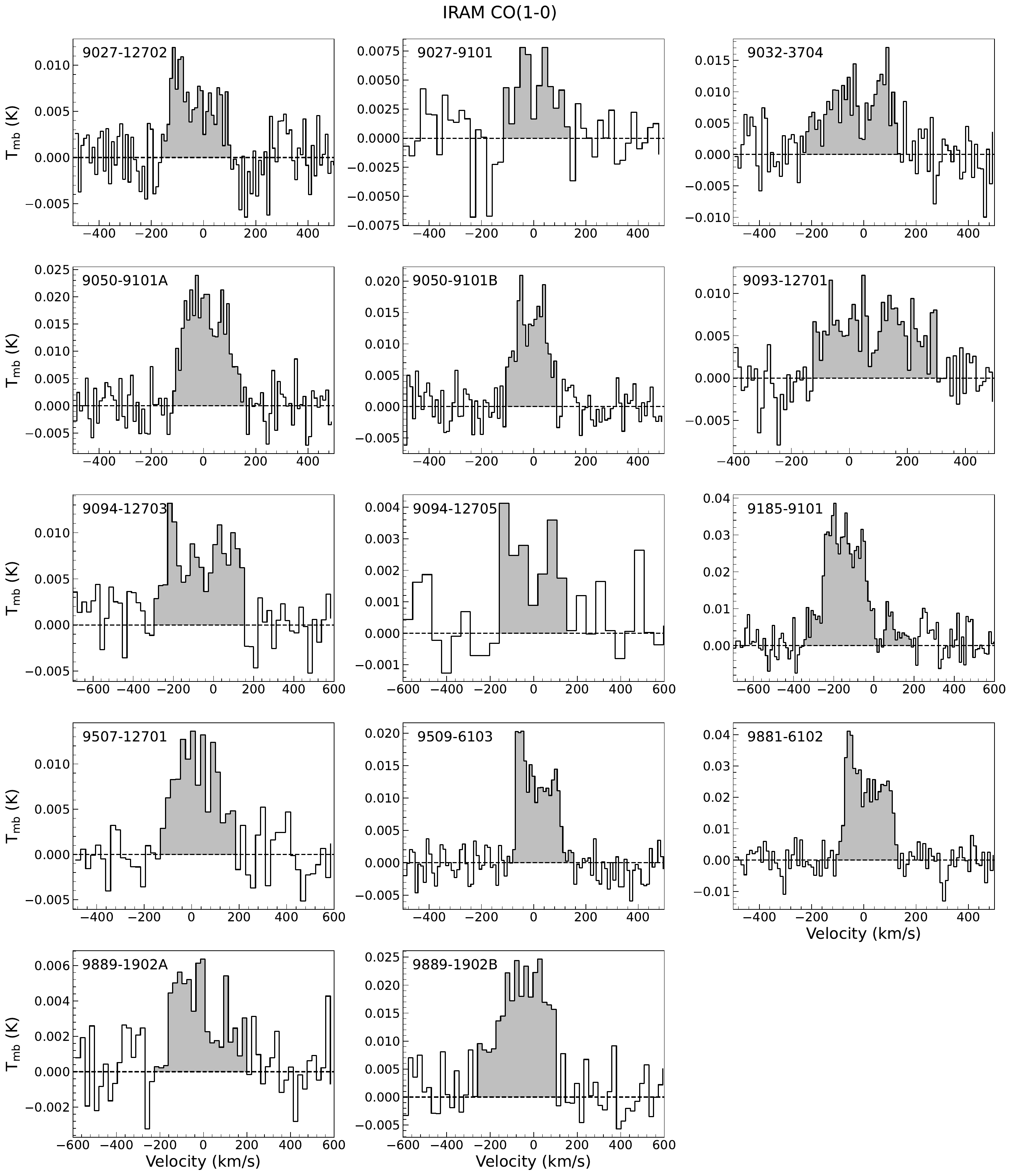}
	\caption{Continued.
		\label{fig:iram_co_1-0_b}
	}
\end{figure*}

\begin{figure*}[t]
    \figurenum{A3}
    \centering
	\includegraphics[width=1\textwidth]{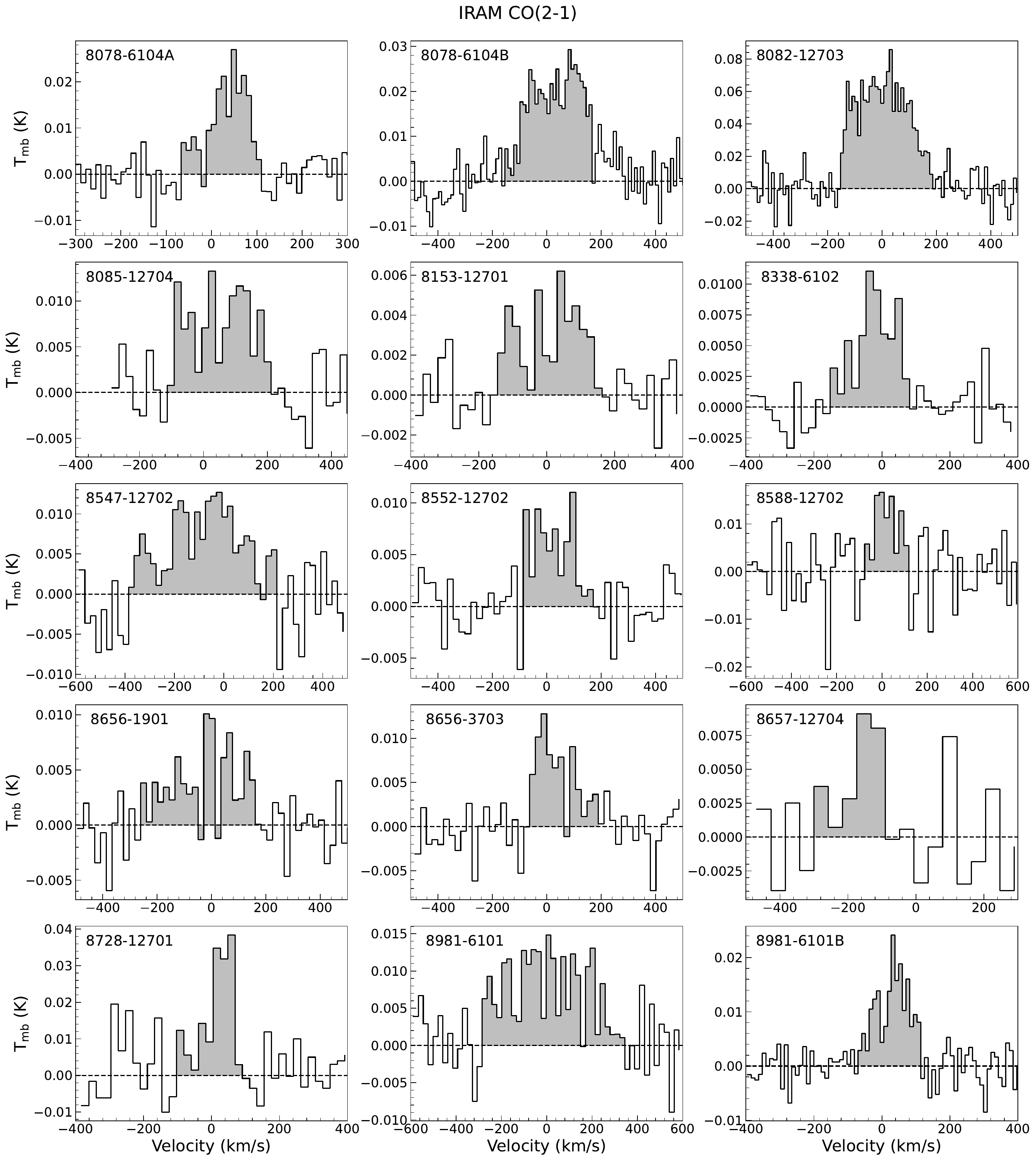}
	\caption{CO(2$-$1) spectra of the detected galaxies from IRAM 30\,m observations. Most of the spectra were rebinned to a velocity resolution of $\sim$20 $\text{km s}^{-1}$, and some spectra were rebinned to a velocity resolution of $\sim$40 $\text{km s}^{-1}$ to achieve detections. The velocity of each spectrum is Doppler corrected and converted to the barycentric frame, and the zero-point is set based on the optical spectroscopy redshift of the galaxy.
		\label{fig:iram_co_2-1}
	}
\end{figure*}

\begin{figure*}[t]
    \figurenum{A3}
    \centering
	\includegraphics[width=1\textwidth]{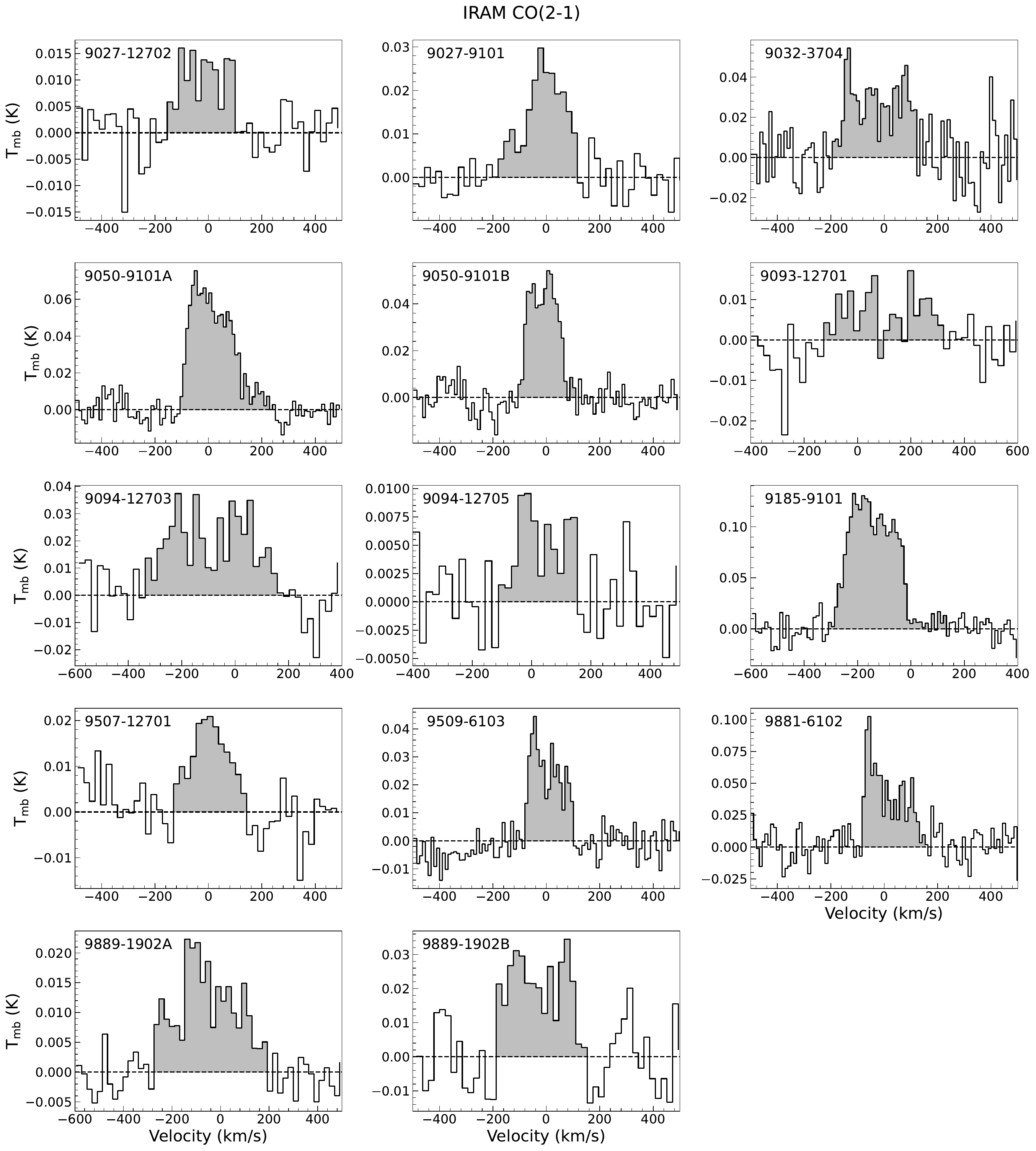}
	\caption{Continued.
		\label{fig:iram_co_2-1_b}
	}
\end{figure*}

%\vspace{5mm}


\end{CJK*}
\end{document}